\documentclass[twoside,12pt]{article}
\usepackage{epsfig}

\usepackage[bookmarks=false]{hyperref}
\usepackage{color}
\usepackage[normalem]{ulem}
\usepackage{amsmath}
\usepackage{amssymb}
\usepackage[numbers,sort&compress]{natbib}
\usepackage[perpage]{manyfoot}
\DeclareNewFootnote{B}[alph]
\usepackage{alphalph}

\newcommand{\be}{\begin{equation}}
\newcommand{\ee}{\end{equation}}
\newcommand{\bea}{\begin{eqnarray}}
\newcommand{\eea}{\end{eqnarray}}

\begin{document}

\title{Hydrodynamic Approaches in Relativistic Heavy Ion Reactions}

\author{
R.\ Derradi de Souza,$^{1,2}$ T.\ Koide,$^{1}$ T.\ Kodama$^{1}$ \\
\\
$^1$Instituto de F\'{\i}sica, UFRJ, CP: 68528, Rio de Janeiro 21941-972 Brazil\\
$^2$Instituto de F\'{\i}sica Gleb Wataghin, UNICAMP, 13083-859 Campinas SP, Brazil
}

\date{}

\maketitle

\begin{abstract} 
We review several facets of the hydrodynamic description of the relativistic heavy ion collisions, starting from the historical motivation to the present understandings of the observed collective aspects of experimental data, especially those of the most recent RHIC and LHC results.
In this report, we particularly focus on the conceptual questions and the physical foundations of the validity of the hydrodynamic approach itself.
We also discuss recent efforts to clarify some of the points in this direction, such as the various forms of derivations of relativistic hydrodynamics together with the limitations intrinsic to the traditional approaches, variational approaches, known analytic solutions for special cases, and several new theoretical developments.
Throughout this review, we stress the role of course-graining procedure in the hydrodynamic description and discuss its relation to the physical observables through the analysis of a hydrodynamic mapping of a microscopic transport model.
Several questions to be answered to clarify the physics of collective phenomena in the relativistic heavy ion collisions are pointed out.
\end{abstract}

\section{Introduction}

\label{sec:Intro}

Hydrodynamics is a theoretical framework to describe the motion of fluids based on their local property and the conservation laws of energy and momentum, and other conserved quantities, for example mass in non-relativistic hydrodynamics.
The main advantage of hydrodynamics resides in the fact that a huge number of degrees of freedom contained in the microscopic composition of the fluids is drastically reduced to a few macroscopic hydrodynamic variables which represent the local property of the fluid. We refer to this property of hydrodynamic description as ``locality''.
The local property is usually represented by thermodynamic relations among the hydrodynamic variables frequently called as equation of state (EoS), plus transport coefficients.
That is, it is assumed that the fluid is in the state of local thermal equilibrium (LTE), or very close to it (see Sec.\ \ref{sec:RelativisticHeavyIonCollisionsAndHydro}).

Hydrodynamics is one of the oldest classical phenomenological theories which has played an important role in the development of science and technology.
Its relativistic form was already developed in the early stages of the theory of relativity \cite{ODarrigol.WorldsofFlow,LDLandau.FluidMechanics} mainly in the context of astrophysics and cosmology.
As we will report in this review, applications of hydrodynamics to the microscopic system such as nuclear collective motion and heavy ion collisions have been shown to be successful.
It is amazing to observe that such a simple scheme of dynamics is able to cover various phenomena from cosmological to hadronic scales.

When we think of typical dynamics of a fluid, we immediately imagine its flow profile as we usually observe in daily life, such like waves of sea, ripples of wine in a glass, vortices in a water-sink, turbulences in winds, etc.
Indeed, in most of these phenomena, the hydrodynamic picture is known to be well established since the validity of the concept of locality in describing the properties of fluids through LTE seems obvious due to the huge number ($\sim$$10^{20}$) of particles involved in a visible dynamics of these fluids. 
In such cases, if the property of the matter is given we can study its response as an initial value problem.
Sometimes we can use the hydrodynamic description to determine the initial condition which leads to the specified final state after the hydrodynamic evolution.
On the other hand, if the properties of the matter is not known, then we can introduce a model to represent them in terms of a few hydrodynamic parameters, and infer them by comparing the predictions with experiments, provided that the hydrodynamic picture is in fact valid.
Of course, hydrodynamic responses of the matter in principle can be nonlinear, and sometimes they are extremely complex.
In such cases the above mentioned hydrodynamic modeling to deduce the initial condition or properties of the matter in question is not trivial at all.
In particular, for example in the presence of turbulence, the time evolution becomes chaotic and the one-to-one correspondence of the initial and final states is lost \cite{WDMcComb.ThePhysicsofFluidTurbulence}.

As mentioned, the hydrodynamic picture is applicable for a vast class of different scales.
However, it should be kept in mind that the meaning of the locality changes depending on the scale of the system in question.
For example, if we try to study hydrodynamics of the atmosphere for weather forecasting purpose, larger-scale simulations are implemented to cover the significant area of the earth.
There, the locality means that any inhomogeneity of the air is completely neglected within the volume, say, of the order of $m^{3}$, and in fact, too precise information is not required.
However, this does not necessarily mean that small scale dynamics is completely neglected.
For example, in the presence of turbulences in a smaller volume than the required precision, they are smeared out and counted as the internal degrees of freedom of the air in our observational scale.
In these situations, the transport coefficients might be modified and some effective quantities, e.g. the so-called eddy viscosity should be used \cite{WDMcComb.ThePhysicsofFluidTurbulence}.
Such a situation occurs frequently when we perform numerical simulations where a certain scale should eventually be introduced due to discretization of space as mentioned above in the example of large scale numerical simulations with cutoff scales, which reproduce the observable phenomena.
This happens also in astrophysical applications such as supernova, galaxy formation and cosmological problems.
This means that the hydrodynamic parameters determined from microscopic theories or laboratory experiments are not necessarily the same as those used in real simulations.

It is more than decades ago since the hydrodynamic modeling in relativistic heavy ion collisions has been shown to be very successful in understanding the nature of so-called collective flow parameters.
In particular, after establishing the almost ideal fluid scenario and the picture of the strongly coupled deconfinement phase of quarks and gluons (quark and gluon plasma - QGP), the hydrodynamic description is considered to be indispensable for some stages of the dynamics of the expanding produced matter in the relativistic heavy ion collisions.
These successes lead to the expectation that some important bulk properties of the QGP and its initial state just after the collisions can be determined to a quantitative level using the hydrodynamic analysis.

However, the situation of relativistic heavy ion collisions is quite different from the aforementioned macroscopic phenomena.
First, we are unable to observe directly the time evolution of flow profile of the matter produced by the collisions in event-by-event basis (EbyE) but we only observe free-streaming final state particles, instead.
In fact, the two scenarios, the hydrodynamic flow and the collection of free-streaming particles are essentially different.
The transition from one scenario to the other is a very complex transport process and referred to as ``freeze-out''.
Second, in the study of relativistic heavy ion physics, the initial condition and the property of the matter are exactly what we want to determine.
In the hydrodynamic approach, these are inputs and are not observable directly determined from the experimental data.
Thus we have to resort to theoretical models to specify the initial condition and the property of the matter, which are out of the scope of the hydrodynamic formalism.
Finally, at the present stages, we are not able to be sure on why and when a hydrodynamic description becomes valid in such an extremely explosive dynamics in the scale of subatomic systems.
All of the above three aspects of the hydrodynamic approach constitute very challenging problems, and extensive studies have been done both experimentally and theoretically.

For the sake of bookkeeping, in this report, we classify these studies roughly in the following three categories.

\begin{enumerate}
\item[1)] Assume the validity of hydrodynamic scenario as working hypothesis, construct a unified description of relativistic heavy ion collision, with the help of certain models of initial conditions and also the freeze-out processes.
By applying such a description to the analysis of experimental data, try to determine quantitatively the parameters in the model.

\item[2)] Focus mainly in qualitative aspects of the hydrodynamic picture.
Look for robust and less model-dependent hydrodynamic signals to explore the initial state dynamics of the collision.
Also, complementary to the approaches classified in 1), study analytic or simplified solutions to clarify the role of hydrodynamic signals.

\item[3)] Establish the foundation and validity of the hydrodynamic approach in the scenario of relativistic heavy ion collisions, and also look for possible alternative (or variant) interpretation of collective phenomena.
\end{enumerate}

Of course the above three categories are complementary and also interrelated so that they are often not quite separable.
The category 1) is presently the main stream of hydrodynamic studies of relativistic heavy ion collisions.
In fact, it is the most urgent and important task to push forward the known hydrodynamic approaches and establish the method of analysis of newly coming experimental data.

On the other hand, this approach deals with the precise quantitative comparison to experimental data so that it necessarily involves quite complex modelings for the initial condition and for the final state interactions, in addition to rather sophisticated numerical techniques.
Thus sometimes the effects of physical parameters in the model are not quite evident for those who do not have these resources at their disposal to study the individual technical issues.
Furthermore, the space of parameters to be determined is quite large so that the uniqueness of the solution is not guaranteed.
In this aspect, the approaches of the category 2) and 3) are helpful to characterize the flow dynamics from more general insights giving feedback to the approach 1).
They are also useful to extend the hydrodynamic picture in different situations such as the beam energy-scan program (BES) at RHIC/BNL, FAIR/GSI and NICA/JINR \cite{Hohne:2013pqa,Schmidt:2014lva}.
In addition, through these approaches we can clarify the limitations and applicability of relativistic hydrodynamics.
This is a fundamental point for the sound understanding of physics of relativistic heavy ion collisions, because the reproduction of experimental data is of course not the ultimate objective of the hydrodynamic approach.
Extensive studies on the role of the hydrodynamic approach in relativistic heavy ion collisions have been done, such as these comprehensive reviews and books \cite{Kolb:2003dz,Heinz:2009xj,Ollitrault:2008zz,Huovinen:2006jp,Sorensen:2009cz,Teaney:2009qa,Hama:2004rr,WFlorkowski.PhenomenologyOfURHIC,Sarkar2009}.
In particular, these recently published reviews \cite{Hirano:2012kj,Int.J.Mod.Phys.A28.11.1340011,Heinz:2013th,Jia:2014jca,Snellings:2014kwa,Petersen:2014yqa} report the present status, giving an overview of the state-of-art of the hydrodynamic approach mainly focusing on 1), including LHC results.
Therefore, in this work we will not repeat in detail the works already reported and refer readers to these review articles and books as well as the original references therein.
Instead, we focus complementary studies on hydrodynamics in relativistic heavy ion collisions, mainly those to be classified in the categories 2) and 3), with the exception of some key point studies which characterize category 1).
As mentioned, they are equally important for a deep understanding of the physics of collective signals in relativistic heavy ion collisions.

The present paper is organized as follows.
In Sec.\ \ref{sec:RelativisticHeavyIonCollisionsAndHydro}, we give a brief overview on the historical motivation and developments of the hydrodynamic approach in relativistic heavy ion collisions.
In Sec.\ \ref{sec:StructureOfHydro}, we show the basic structure of relativistic hydrodynamics and discuss the physical concepts of the approach, both for the ideal and dissipative cases.
In Sec.\ \ref{sec:Practical}, we discuss how the hydrodynamic approach is applied in practice to relativistic heavy ion collisions.
There, we focus on several specific physical problems necessary to connect the hydrodynamic variables to the physical observables of relativistic heavy ion collisions, such as the initial condition, equation of state and transport coefficients and the freeze-out process.
In Sec.\ \ref{sec:Overview} we give an overview on how these questions are dealt with in the present studies and give a concise summary on the present status and accomplishments of the hydrodynamic approach in heavy ion collisions.
In Sec.\ \ref{sec:MoreAboutHydro}, we report on the theoretical efforts from a more general point of view, for example, to formulate the dissipative relativistic hydrodynamics from the various methods, to find the analytic classes of solutions, to establish foundations of the approach in the scenario of relativistic heavy ion collisions, inclusion of some new dynamical effects, etc.
In Sec.\ \ref{sec:CoarseGraining}, we focus our attention on the aspect of coarse-graining in physical observables and investigate its effects on the origin of hydrodynamic description taking an example of event generator code based on a transport theory.
Sec.\ \ref{sec:DiscussionAndFuture} is dedicated on the discussion of future perspectives of the hydrodynamic approach in relativistic heavy ion collisions.

\section{Relativistic Heavy Ion Collisions and Hydrodynamics}

\label{sec:RelativisticHeavyIonCollisionsAndHydro}

In this section, we briefly review the historical background of the studies of relativistic heavy ion collisions.
Research in this area started with theory-driven motivations to search for new, theoretically predicted states of matter.
Since then, it has developed into an experimentally driven science that works not only just to confirm these new states of matter but also to quantitatively determine the properties of these states, even leading to new thought provoking questions.

\subsection{Early Days}

\label{SecII-1}

The use of thermodynamical concepts in high energy elementary particle collisions was introduced for the first time by Fermi to model the multi-particle productions \cite{Fermi:1950jd}.
Landau extended Fermi's fireball model to analyze the effects of collective motion of the produced particles in the framework of relativistic hydrodynamics \cite{Landau:1953gs,Belenkij:1956cd}.
He showed the importance of the longitudinal expansion of the fireball using the Khalatnikov's analytic solution of (1+1)D relativistic hydrodynamics \cite{Khalatnikov}.
At that time, of course, there was no idea of the mechanism of high energy hadronic collisions which leads to multiple pion productions.
Thus, instead of modeling the production mechanism, Landau simply assumed that the created matter could be regarded as a hot gas of massless pions in thermal equilibrium as Fermi did (which is not much far from the present idea of the parton gas).
But he argued that the fireball created in a proton-proton (p--p) collision suffers a strong longitudinal, rather than isotropic, expansion due to the large pressure gradient in this direction because of the Lorentz contraction in the initial state.
This affects naturally the energy dependence of the multiplicity of final particles on the incident energy and generates the characteristic rapidity distribution.

Fermi and Landau's thermodynamic/hydrodynamic approaches have been pursued by many authors to understand the nature of multiple particle production in hadronic collisions observed in cosmic ray phenomena.
These works served as the conceptual origin in the studies of bulk aspects of the strongly interacting matter at very high energies.
In the middle of 60's Hagedorn predicted the existence of the limiting temperature for the hadron resonance gas (HRG) \cite{Hagedorn:1965st}, and later this behavior was discussed with the deconfinement of quarks \cite{Cabibbo:1975ig} (see also Ref.\ \cite{Ericson:2003ya}).
Recently, his statistical bootstrap approach was revisited from the modern perspective of QGP to investigate the hadronization mechanism and transport properties of HRG \cite{NoronhaHostler:2007jf,NoronhaHostler:2008ju,NoronhaHostler:2012ug}.
Also a theoretical framework of HRG was developed in Ref.\ \cite{Dashen:1969ep}.
In his hydrodynamic studies on multi-particle production in high energy hadronic collisions, Carruthers pointed out in 1974 \cite{Carruthers} that the matter which obeys hydrodynamics does not need to be a fluid of known particles.
Instead, he emphasized the importance of using the coarse-grained macroscopic quantities such as energy-momentum tensor and conserved densities to describe the possible new states of matter, which he calls pre-matter.
At that time, the concept of freeze-out to derive the observable particles spectra from the hydrodynamic picture was already studied and the well-known Cooper-Frye formula \cite{Cooper:1974mv} was developed in this context for the multiple particle productions in high energy hadronic collisions (see the discussion in Sec.\ \ref{freezeout}).

The hydrodynamic approach has also been used in nuclear physics since the early days, although before the 70's the main interests have been concentrated on non-relativistic energies.
This is because the principal object of nuclear physics at that time was to investigate the properties of nuclei and their reactions from the point of view of many-body quantum systems.
The underlying nuclear Hamiltonian is given by two-body nucleon-nucleon interactions, represented by potentials which were not established.
Therefore, for the nuclear physics community, scientific interests on highly compressed or excited nuclear matter were not popular yet.
On the other hand, for the elementary particle physics community, high energy proton-nucleus (p--A) or nucleus-nucleus (A--A) collisions were considered to be too complicated, since they certainly provoke excitations of many degrees of freedom and the study of yet unknown hadronic interactions through such complex processes seemed to be almost hopeless.

Nevertheless several pioneering studies emerged suggesting possible existence of exotic states in highly excited hadronic gas as mentioned above, or very high density nuclear isomer states \cite{Lee:1974ma}.
The use of high energy A--A collisions to investigate the nuclear EoS was also discussed \cite{Phys.Rev.Lett.21.1479,Scheid:1974zz}.
In the early 70's, the first relativistic nuclear beam was realized at LBL \cite{Relativistic:1730719}.

At the same time, in 1968, the discovery of the first pulsar \cite{Hewish:1968bj,Pilkington:1968bk} and its identification as a neutron star, which was predicted long before by Landau \cite{Landau-1932} as the form of giant nucleus bound by its gravitational force (more precisely postulated by Baade and Zwicky \cite{Baade-Zwicky-PR-46-76-1934} after the discovery of neutrons), provoked strong interest in the properties of highly compressed nuclear matter in the nuclear physics community (see for example Ref.\ \cite{NKGlendenning.CompactStars}).
The possibility of a quark matter star was first suggested in 1970 by N.\ Itoh \cite{Itoh:1970uw}.
The program of relativistic heavy ion collisions to study the properties of matter at extreme conditions was advocated by T.D.\ Lee in 1974, which is now grown up to one of the contemporary frontiers of sciences.
For more details of these developments, we refer readers to Refs.\ \cite{LarryTDLee} and \cite{Baym:2001in}.
It should be noticed that all these happened at the time when Quantum Chromodynamics (QCD) was emerging \cite{Fritzsch:1972jv,Fritzsch:1973pi} (see also Ref.\ \cite{Fritsch:2012sw} for the development of QCD) and asymptotic freedom was just discovered \cite{Gross:1973id,Politzer:1973fx}, but well before the experimental discovery of jets \cite{Hanson:1975fe,Hanson:1981em}.
Of course now the well-known concept of Quark and Gluon Plasma (QGP) \cite{Shuryak:1978ij} was even not existing among the community.

Thus, in the 70's and 80's, several particle accelerators were adopted to accelerate nuclear projectiles, e.g, BEVALAC/LBL, DUBNA, AGS/BNL, SPS/CERN.
Initially the incident energy of the projectiles for fixed target was around 1 GeV per nucleon, and to investigate whether such a state of the compressed nuclear matter can be achieved was one of the main topics.
During this time period, several relativistic effects and methods were developed, such as relativistic mean field theory and relativistic intranuclear cascades \cite{Cugnon:1982qw,Kodama:1983yk}, in addition to the application of relativistic hydrodynamics.
It was first pointed out by St\"{o}cker \textit{et al.} \cite{Stocker:1981zz}, that the side peaked angular distribution of protons favors the hydrodynamic scenario compared to a simple binary intranuclear cascade.
With the theoretical developments of QCD and the discovery of asymptotic freedom, partons were identified with quarks and gluons in this regime.
Thus, the concept of a plasma of quarks and gluons (QGP) became more concrete as a new state of matter at extreme conditions.
Many theoretical investigations on possible QGP signals in relativistic heavy ion collisions were proposed, such as strangeness enhancement \cite{Rafelski:1982pu}, J/$\Psi$ suppression \cite{Matsui:1986dk}, restoration of chiral symmetry \cite{Rapp:1999ej}, jet quenching \cite{Bjorken:1982tu} and so on \cite{Yagi2005,Letessier2002}.
Influences of the possible new state of matter on collective flow dynamics were also discussed \cite{Nix:1979tc,Prog.Part.Nucl.Phys.4.133-195}.
For more details and further reading of the early stage of the relativistic heavy ion collisions, see Refs.\ \cite{Kapusta:1982va,Nagamiya:1982kn,Harris:1996zx} and books \cite{Wong:1995jf,LaszloBook} which convey the excitements and perspectives for the new research area in nuclear physics of these times.

In order to have a clear signal predicted from theories, we need to achieve a state of QGP in thermal equilibrium, or in other words, we need a large space-time scale of reaction processes.
This is why we consider heavy ion collisions the most promising area to study QGP in the lab.
Elementary collisions such as p--p are considered to have space-time scales too small to reach thermal equilibrium in created matter (for recent surprises, see Sec.\ \ref{sec:Overview}).
Heavy ion collisions play the role of a ``pressure cooker'' to maintain matter in a hot and dense enough state to produce QGP.
In contrast to elementary collisions, the geometry of the initial configuration plays an essential role in heavy-ion collisions.
Of course, the initial collision geometry is not a direct observable but can be inferred from the final state multiplicity of produced particles.
Nowadays, developments of experimental techniques and findings of new methods and observables make it possible to determine the centrality class of the collision in terms of global event observables, such as particle multiplicity and forward energy measured by the zero degree calorimeter (ZDC).
Also techniques of particle identification and methods of particle correlation measurements allow us to obtain more exclusive data, which enables precise comparisons with theoretical models at the quantitative level.
The recent review papers \cite{Ritter:2014uca,Csernai:2014cwa} by the authors working directly on collective flow phenomena in relativistic heavy ion collisions summarize these developments through the last four decades, from the very early days to the present.

\subsection{QCD Phase Diagram}

\label{SecII-2}

\begin{figure}[ptb]
\centering
\includegraphics[height=4.9cm,keepaspectratio]{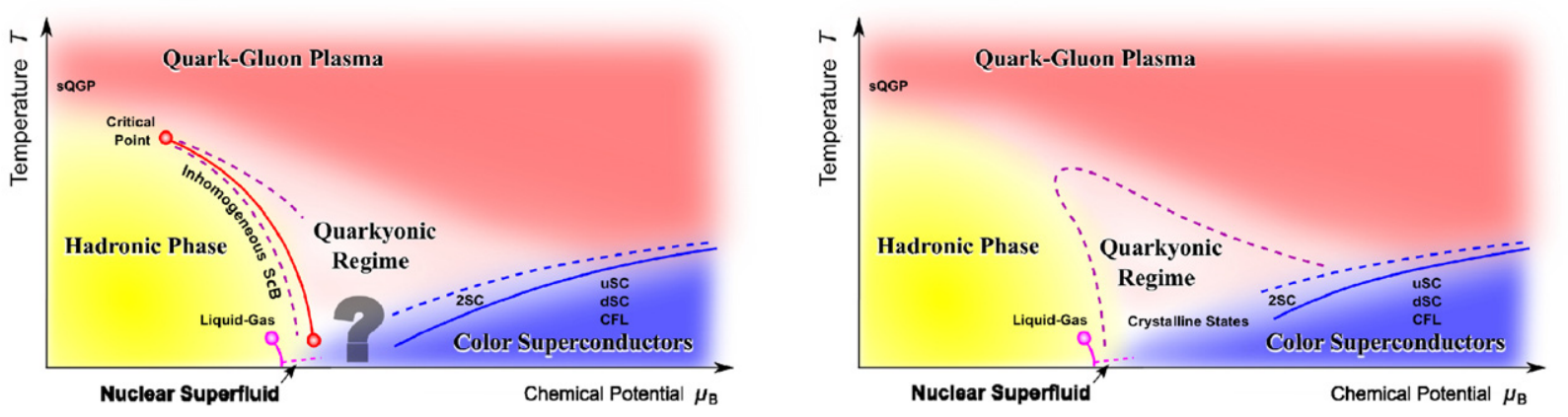}
\caption{(color online) QCD Phase Diagram conjectured from QCD. The left and right panels represents the two distinct predictions, with and without the presence of critical points, respectively. Figure\protect\footnotemarkB\ taken from Ref.\ \cite{Fukushima:2013rx}.}%
\label{fig:QCD_phase_diagram}%
\end{figure}

As mentioned above, one of the main objectives of the relativistic heavy ion program is to investigate the properties and dynamics of the QCD matter at extreme conditions of temperature and density.
The present image of the phase diagram of QCD is summarized in Fig.\ \ref{fig:QCD_phase_diagram}.

\footnotetextB{Reprinted from \textit{Prog.\ Part.\ Nucl.\ Phys}.\ 72, K.\ Fukushima and C.\ Sasaki, \textit{The phase diagram of nuclear and quark matter at high baryon density}, 99, Copyright \textcopyright\ 2013, with permission from Elsevier.}

As seen in Fig.\ \ref{fig:QCD_phase_diagram}, the QCD phase diagram has a much more complex structure compared to the early speculations deduced from the bag model.
First, the lattice QCD (lQCD) calculations showed that the transition from the hadronic phase to the QGP phase at the vanishing baryonic chemical potential ($\mu_{B}=0$) is a smooth cross-over and not a first order transition \cite{Aoki:2006we}.
This behavior is now considered to be well established and we will come back to this point later with respect to the QCD EoS.
Second, the order of the phase transition depends on the baryon chemical potential, leading to a possible critical point (left panel in Fig.\ \ref{fig:QCD_phase_diagram}).
The existence of this kind of critical point was first suggested in 1989 \cite{Asakawa:1989bq} in the context of chiral symmetry breaking.
Although some finite chemical potential results also show the existence of a critical point \cite{Fodor:2004nz,Borsanyi:2012cr}, this is still an open ended question \cite{Philipsen:2012nu}, as shown in the right panel of Fig.\ \ref{fig:QCD_phase_diagram}.
There, the critical point does not even exist anywhere \cite{Fukushima:2013rx}.
Third, the chiral phase transition does not necessarily go with the confinement phase transition.
Possibilities of different behavior of the two phase transitions have been discussed in lQCD and in some theoretical models, and the possible existence of a new phase called quarkyonic matter was suggested in Ref.\ \cite{McLerran:2007qj} where the confined state with the chiral symmetry is realized.
For more detailed information, see review papers \cite{deForcrand:2010ys,Philipsen:2012nu}.
For lower temperatures and high baryonic chemical potentials, quark matter as conjectured by Itoh \cite{Itoh:1970uw} appears, but still exhibits complicated structures such as various color-superconducting phases.
For more detailed discussions on the theoretical investigations of the QCD phase diagram, see the recent reviews, Refs.\ \cite{Fukushima:2010bq,Fukushima:2013rx}.
See also Ref.\ \cite{Tawfik:2014eba} for a general review on statistical properties of strongly interacting matter.

\begin{figure}[ptb]
\centering
\includegraphics[height=7.0cm,keepaspectratio]{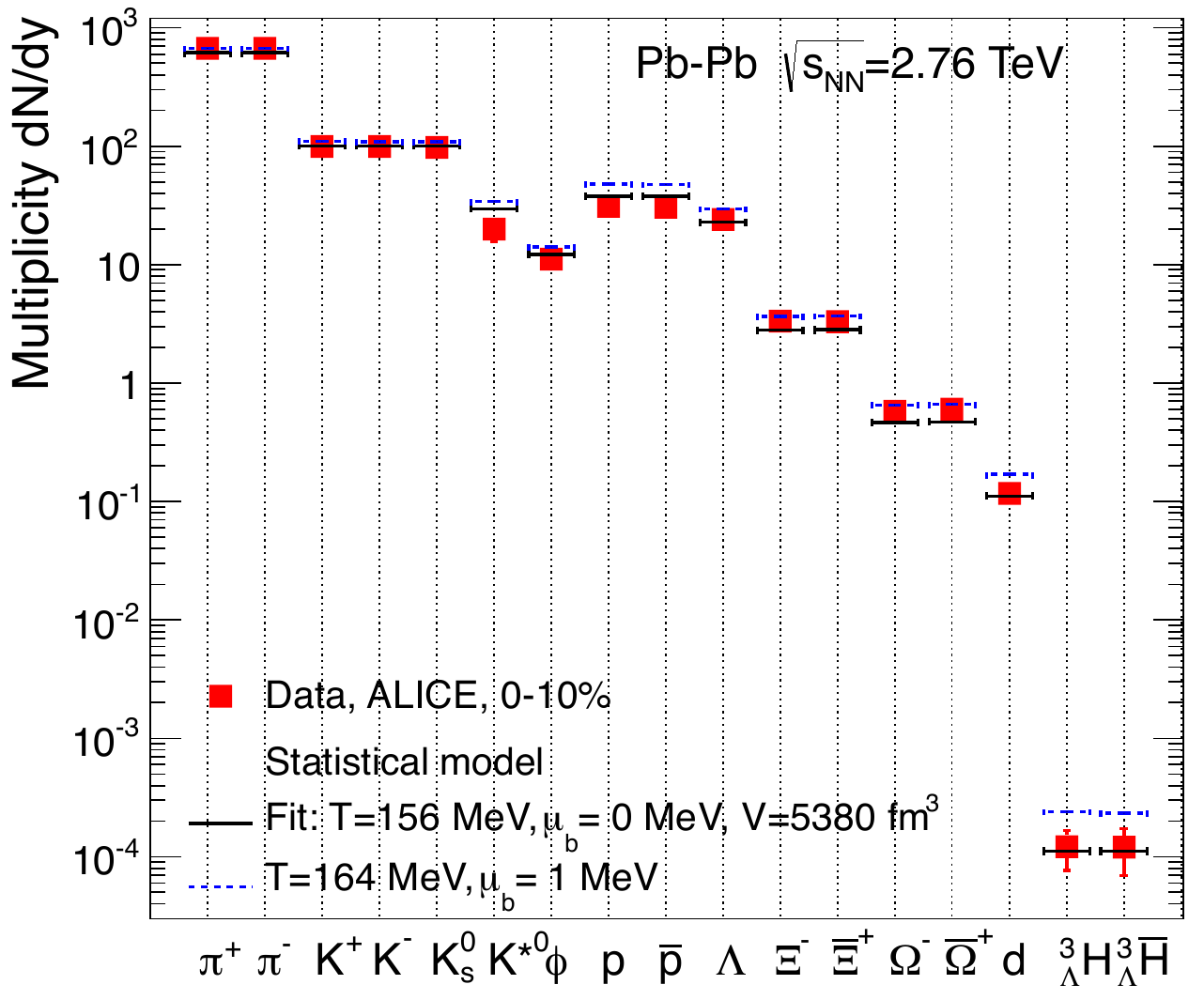}
\caption{Thermal Model Fit of data from Pb--Pb at $\sqrt{s_{\mathrm{NN}}}=2.76$ TeV. Figure\protect\footnotemarkB\ taken from Ref.\ \cite{Stachel:2013zma}.}
\label{fig:ALICE_thermalFits}
\end{figure}

One of the basic motivations of relativistic heavy ion collisions is to experimentally investigate the presumed rich structure of the QCD phase diagram.
By definition, the phase diagram represents the states of a homogeneous matter in thermal equilibrium with infinite volume.
As mentioned, it is not obvious whether the produced matter achieves a thermally equilibrated state in relativistic heavy ion collisions.
One of the experimental indications of thermal equilibrium is observed in the relative abundances of produced particles.
In the thermal models, the relative abundances of various hadron species are given by the thermal equilibrium distribution at the so-called chemical freeze-out stage, specified only by the two parameters, the temperature $T$ and the baryonic chemical potential $\mu_{B}$ \cite{BraunMunzinger:1995bp,Becattini:1997uf,BraunMunzinger:2003zd,Becattini:2002en,Xu:2001zj,Torrieri:2004zz,Wheaton:2004qb}.
In more sophisticated models, effects of the chemical non-equilibrium are also considered by introducing fugacities for quarks \cite{Rafelski:1995vp}.
In Fig.\ \ref{fig:ALICE_thermalFits} we illustrate the recent results of particle ratios with the typical thermal model fits for the RHIC and LHC energies.
\footnotetextB{Reprinted from J. Stachel \textit{et al.}, \textit{Confronting LHC data with the statistical hadronization model}, \textit{J.\ Phys.\ Conf.\ Ser}.\ 509 (2014) 012019, doi:10.1088/1742-6596/509/1/012019, under the terms of the Creative Commons Attribution 3.0 license.}

Whether the perfect chemical equilibrium is achieved or not, it is a fact that quite good reproductions of the data of particle ratios are obtained in terms of the thermal approach.
This indicates that the chemical freeze-out process occurs as if the system is in a state very close to the thermal equilibrium in relativistic heavy ion collisions.
On the other hand, it is intriguing that the thermal models are shown to work also well for the particle ratios in high energy p--\={p} and e--\={e} reactions \cite{Becattini:1995if,Becattini:1996gy}.
Note that in these examples, the thermodynamic model is applied to the system averaged over many events.

The thermal model above is applied to the final HRG.
Thus, this does not imply directly that the system reaches thermal equilibrium at the QGP stage.
Furthermore, if the particle ratios reflect the complete statistical equilibrium of a pure hadron gas, then the information of QGP should be washed out.
The signals of QGP in flavor enhancement/suppression such as strangeness enhancement or J/$\Psi$ suppression should stand out from this thermal model fit of the pure HRG.
See more detailed discussion, Chap.\ 24 of Ref.\ \cite{WFlorkowski.PhenomenologyOfURHIC} and references therein.

One of the QGP signals which may survive throughout final state hadron interactions is collective flow.
Loosely speaking, if the QGP phase emerges, the corresponding pressure is much larger than that of a hadron gas for a given energy density, so that in the presence of QGP phase any spatial inhomogeneity enhances an acceleration of the expansion of the matter. Such an acceleration should manifest in the final particle angular distribution.

In order to quantify the above expectation, we have to model the collision process in terms of a hydrodynamic description.
In the next section, we show the basic equations of relativistic hydrodynamics in the most concise form as possible.
More detailed physical aspects involved in the hydrodynamic formulation will be discussed in the following sections.

\section{Structure of Relativistic Hydrodynamics}

\label{sec:StructureOfHydro}

In this section we introduce the basic concepts and equations of relativistic hydrodynamics.
For the sake of a pedagogical purpose, we start from the familiar non-relativistic Euler equation for the ideal fluid to clarify the concept of coarse-graining used in the hydrodynamic description.

\subsection{Non-relativistic Euler equation}

The non-relativistic Euler equation for an ideal fluid has the following form,
\begin{equation}
\mathbf{\dot{v}}+\left(  \mathbf{v\cdot\nabla}\right)  \mathbf{\ v}=-\frac{1}{\rho_{m}}\mathbf{\nabla}P,\label{NonRelEuler1}
\end{equation}
where $\mathbf{v}$ is the velocity field, $\rho_{m}$ is the mass density and $P$ is the pressure.
In basic physics text books this equation is usually derived by applying Newton's equation of motion to a small cell of the fluid and using Pascal's law which claims that the pressure is isotropic.
Here, we derive it from the conservation of momentum in order to clarify the physical structure of the relativistic hydrodynamics.
To do this, we have to introduce several physical assumptions: first, states of the matter, independent of the details of internal states of microscopic degrees of freedom, can approximately be described in terms of a few number of \textit{macroscopic} classical fields.
We refer to those as hydrodynamical variables, and they are constructed at an arbitrary space-time point by taking the averages over the microscopic degrees of freedom within the vicinity of the point.
The size of \textit{vicinity} should be specified by a certain scale, called the scale of \textit{coarse-graining} (C-G).
We thus represent the system as a continuum medium, but remind that for each space-point we have always to associate a small volume determined by C-G size, to which we refer as ``fluid element''.
By definition these fields should be insensitive to a microscopic dynamics which occurs inside the fluid element, so that any hydrodynamic variables in the co-moving frame should be almost constant, and isotropic in the absence of external forces or internal degrees of freedom such as spin.
Quantitatively speaking, C-G scale should be less than the minimum of $1/\left\vert \nabla\ln(Q)\right\vert $ where $Q$ is the typical hydrodynamic variables.
The condition of the continuum limit for the case of a gas of particles, is achieved when the mean-free path is much less than the C-G scale.
The ratio of these two numbers is called Knudsen number $K$.
Second, forces exert on one fluid element come only from the adjacent fluid elements and expressed in the form of stress tensor.
Third, there should be no tangential stress for a fluid at rest.
This condition is crucial to distinguish the two possible continuum media, say, fluid and solid.
In particular for the ideal fluid where we discuss here, this condition is valid for any fluid element.
To constitute a self-contained dynamical system under the above conditions, we have to construct a closed system of dynamical equations with these classical fields.
This last condition implies that the scales for macroscopic dynamics defined by C-G should be clearly separated from the microscopic ones.

The important question is how to choose the hydrodynamic variables.
Not every physical quantity, even after C-G is necessarily appropriate for the description of dynamics.
Normally, conserved densities are good candidates.
To see this, let $\hat{n}$ and $\mathbf{\hat{J}}$ be, respectively, some conserved density and its current, which are expressed in microscopic degrees of freedom and satisfy the continuity equation.
Then the corresponding C-G density and current, $n$ and $\mathbf{J,}$ should also satisfy the continuity equation,
\begin{equation}
\partial_{t}n=-\nabla\cdot\mathbf{J}.\label{dndt}
\end{equation}
Suppose the C-G scale is given by $L$.
Then from the first condition we mentioned above, the right hand side of Eq.\ (\ref{dndt}) is at most the order of $\left\vert \mathbf{J}\right\vert /L$.
In such a case, a significant change of $n$ occurs only for the time-scale much larger than that of the microscopic motion.
This means that the dynamical time scale of $n$ becomes also that of macroscopic scale compatible with the C-G size.
Therefore, at least $n$ is appropriate variable to characterize the macroscopic motion of the system (see a more detailed discussion in Secs.\ \ref{sec:boltz} and \ref{sec:proj}).
In general, the current $\bf{J}$ can still contain microscopic motions.
In the case of the mass conservation in the non-relativistic hydrodynamics, it can be shown that the corresponding current is also a macroscopic quantity associated with the momentum conservation.

In the non-relativistic case, the mass is a conserved quantity and the corresponding continuity equation for the mass density $\rho_{m}$ and the current $\mathbf{j}=\rho_{m}\mathbf{v}$ is given by,
\begin{equation}
\frac{\partial\rho_{m}}{\partial t}=-\mathbf{\nabla}\cdot(\rho_{m}\mathbf{v}),
\label{mass continuity}
\end{equation}
where $\mathbf{v}$ is the velocity field.
For the later purpose, we express this continuity equation in a Lorentz covariant form, $\partial_{\mu} j^{\mu}=0,$ where $\left(  j^{\mu}\right)  =\left(
\begin{array}
[c]{cc}
\rho_{m}c & \rho_{m}\mathbf{v}
\end{array}
\right)$, with $\left(  x^{\mu}\right)  =\left(  ct,\mathbf{r}\right)  $ and $c$ is the speed of light.

The above continuity equation can readily be generalized for conserved vector quantities.
For example, the total momentum of the medium, $\int dV\ \rho_{m}\mathbf{v}$ should be conserved in the absence of any external forces.
Thus in terms of the continuity equation, we have
\begin{equation}
\partial_{t}(\rho_{m}v^{i})+\partial_{k}(\rho_{m}v^{i}v^{k})=-\partial_{i}P,
\label{momentum cons}
\end{equation}
where the right-hand side accounts for the interaction among the fluid elements through the surfaces with $P$ being pressure. With the use of Eq.\ (\ref{mass continuity}), we arrive at Eq.\ (\ref{NonRelEuler1}).

Equations (\ref{NonRelEuler1}) and (\ref{mass continuity}) constitute a system of partial differential equations for $\rho_{m}$ and $\mathbf{v}$ but due to the presence of the pressure $P$, the system is not closed yet and we need another condition.
For the case of a \textit{barotropic} fluid where $P$ is given by a function of $\rho_{m}$, we can solve the system without any other knowledge of the matter, such as internal energy.

To be complete, let us consider the energy conservation of the fluid element.
The time derivative of the energy can also be expressed as the sum of the energy flux through the surface and the rate of the work done by the external forces and other fluid elements.
The energy density is given by $\rho_{m}\mathbf{v}^{2}/2+\varepsilon_{int}$, where the first term is the translational kinetic energy density and the second term comes from the internal energy of the fluid element.
The conservation of the energy is then written as
\begin{equation}
\partial_{t}\left(  \rho_{m}c^{2}+\frac{\rho_{m}}{2}\mathbf{v}^{2}
+\varepsilon_{int}\right)  +\nabla\cdot\mathbf{v}\left(  \rho_{m}c^{2}
+\frac{\rho_{m}}{2}\mathbf{v}^{2}+\varepsilon_{int}+P\right)  =0,
\label{energy cons}
\end{equation}
where the rest mass energy conservation is added.
Finally, we can combine Eq.\ (\ref{momentum cons}) and Eq.\ (\ref{energy cons}) and rewrite them as a ``relativistic covariant'' expression,
\begin{equation}
\partial_{\mu}T_{\ \ \ NR}^{\mu\nu}=0,
\label{dmuTmunu-nonrel}
\end{equation}
where
\begin{equation}
T_{\ \ ~NR}^{\mu\nu}=\left(
\begin{array}
[c]{cc}
\rho_{m}c^{2}\left(  1+\mathbf{\beta}^{2}/2+\varepsilon_{int}/\rho_{m}
c^{2}\right)  & \rho_{m}c^{2}\mathbf{\beta}^{T}\\
\rho_{m}c^{2}\left(  1+\mathbf{\beta}^{2}/2+\left[  \varepsilon_{int}
+P\right]  /c^{2}\right)  \mathbf{\beta} & \rho_{m}c^{2}\mathbf{\beta\ \beta
}^{T}+P\hat{I}
\end{array}
\right)  . \label{NonRel-T}
\end{equation}
and $\mathbf{\beta}=\mathbf{v}/c$.
Note that the above matrix can be written in the symmetric form as
\begin{equation}
T_{\ \ \ NR}^{\mu\nu}=\left(
\begin{array}
[c]{cc}
\rho_{m}c^{2}\left(  1+\mathbf{\beta}^{2}/2+\varepsilon_{int}/\rho_{m}
c^{2}\right)  & \rho_{m}c^{2}\left(  1+\mathbf{\beta}^{2}/2+\left[
\varepsilon_{int}+P\right]  /\rho_m c^{2}\right)  \mathbf{\beta}^{T}\\
\rho_{m}c^{2}\left(  1+\mathbf{\beta}^{2}/2+\left[  \varepsilon_{int}
+P\right]  /\rho_m c^{2}\right)  \mathbf{\beta} & \rho_{m}c^{2}\mathbf{\beta\ \beta
}^{T}+P\hat{I}
\end{array}
\right)  \label{Tmunu-NREL}
\end{equation}
without altering the energy and momentum conservation equations in the non-relativistic limit.
Equation (\ref{dmuTmunu-nonrel}), which gives the correct Euler equation, looks like a covariant expression.
However, differently from the conserved charge density $j^{\mu}$ in the continuity equation, the above $T_{\ \ ~NR}^{\mu\nu}$ is not yet a covariant tensor under the Lorentz transformation.
The correct covariant quantity is called energy-momentum tensor which is defined in the next subsection.

\subsection{Relativistic Euler equation}

\label{sec:RelEulerEq}

Having in mind the physical structure of the non-relativistic case, we derive the relativistic Euler equation in a more mathematical way to manifest the Lorentz-covariant structure.
We first consider the hydrodynamic variables associated with energy-momentum conservation.

The three-velocity field $\mathbf{v}$ now should be extended to the four-velocity $u^{\mu}$ with the normalization condition $u^{\mu}u_{\mu}=1$.
At a given space-time point, we can always find the frame where $u^{\mu}\rightarrow\left(  1,0,0,0\right)  $.
We call such a frame \textit{local rest frame} (LRF).
As we saw in the non-relativistic derivation, the energy-momentum conservation is described in terms of Eq.\ (\ref{dmuTmunu-nonrel}).
So we expect that the relativistic Euler equation can be derived by rewriting it in a covariant form.
The corresponding quantity, $T^{\mu\nu}$ to $T_{NR}^{\mu\nu}$ should be a second rank tensor which transforms under the Lorentz transformation.
Furthermore, in the ideal case the only non-scalar quantities are the four-velocity field $u^{\mu}$ and the metric tensor $g^{\mu\nu}$, so that the tensor structure of this $T^{\mu\nu}$ should be constructed only from these.
Then, the most general expression of a second rank tensor is written as
\begin{equation}
T^{\mu\nu}=Au^{\mu}u^{\nu}+Bg^{\mu\nu},\label{Tmunu-rEuler}
\end{equation}
where $A$ and $B$ are scalar functions.
In LRF, $u^{\mu}\rightarrow\left(1,0,0,0\right)  ,$ we have
\begin{equation}
T^{\mu\nu}\rightarrow T_{LRF}^{\mu\nu}=\left(
\begin{array}[c]{cc}
A+B & 0\\
0 & -B\mathbf{\hat{I}}
\end{array}
\right)  .\label{TmunuLRF}
\end{equation}
Since we always interpret $T^{00}$ as the energy density (including the rest mass and kinetic energy) in any reference frame, it should reduce to the proper energy density $\varepsilon$ in the local rest frame\footnote{Proper density of any thermodynamic quantity defined in LFR in relativistic hydrodynamics is by definition an invariant quantity. But differently from the non-relativistic case, the corresponding quantity in the observational frame has different value. In this paper, to distinguish this observational value from the proper density, we denote it with $^{\ast}$. For example, the proper charge density in LRF, say $n$ is denoted as $n^{\ast}$ in the observational frame.}.
\begin{equation}
T_{LRF}^{00}=A+B=\varepsilon.
\end{equation}
Furthermore, comparing Eq.\ (\ref{TmunuLRF}) with Eq.\ (\ref{Tmunu-NREL}) for $\mathbf{\beta}=0,$ it is natural to interpret that the proper scalar function $-B$ as the thermodynamic pressure $P$ in LRF.
With this identification, we have the energy-momentum tensor in a general frame as
\begin{align}
\left(  T^{\mu\nu}\right)   &  =\left(
\begin{array}[c]{cc}
\left(  \varepsilon+P\right)  \gamma^{2}-P & \left(  \varepsilon+P\right)  \gamma^{2}\mathbf{\beta}\\
\left(  \varepsilon+P\right)  \gamma^{2}\mathbf{\beta} & \left(
\varepsilon+P\right)  \gamma^{2}\mathbf{\beta\beta}^{T}+P\mathbf{\hat
{I}}
\end{array}
\right)  =  \left(  \varepsilon+P\right)  u^{\mu}u^{\nu}-Pg^{\mu\nu }  .
\label{Tmunu-Matrix}  
\end{align}
Since $\mathbf{\beta}$ is the (three-)velocity of the fluid element, it determines the Lorentz boost from LRF to the observational frame.
The associated Lorentz factor is given by $\gamma=1/\sqrt{1-\mathbf{\beta}^{2}},$ as usual.
The energy and momentum conservations are then expressed in terms of this energy-momentum tensor as
\begin{equation}
\partial_{\nu}T^{\mu\nu}=0.\label{dmuTmunu=0}
\end{equation}
This equation constitutes four constraints among the five unknown hydrodynamic variables, $\varepsilon$, $P$, and $\mathbf{\beta}$. If $P$ is expressed as a function of $\varepsilon,$ then Eq.\ (\ref{dmuTmunu=0}) forms a closed system for four variables, $\varepsilon$ and $\mathbf{\beta}$, as in the case of the barotropic fluid of the non-relativistic fluid.

When the fluid has a conserved charge, we need to include it as another hydrodynamic variable.
The conserved current for this charge is a vector and the most general form is expressed as
\begin{equation}
n^{\mu}=n u^{\mu},\label{Nmu-ideal}
\end{equation}
where $n$ is a new proper scalar density.
In any frame, $n^{\mu=0}$ should be identified with the net charge density.
The charge conservation is expressed in the form of the continuity equation as is the case of the energy-momentum tensor,
\begin{equation}
\partial_{\mu}n^{\mu}=0.\label{ConsN}
\end{equation}
If there are more than one charge, we should introduce Eq.\ (\ref{Nmu-ideal}) for each charge density.
But for the ideal case, the velocity which defines LRF is universal for all of them, so that the continuity equation (\ref{ConsN}) is valid for each charge density.
Therefore, for a system with any number of charges, the continuity equations plus one equation which specifies the relation among $\varepsilon$, $P$, and $\left\{  n_{i}\right\}  $ close the system of equations.
Normally, as is the case of a non-relativistic fluid, we assume that the thermodynamic relation is satisfied for any fluid elements in its rest frame.
Then $P$ is expressed as a function of $\varepsilon$ and $n$ through EoS.
Note that, in the non-relativistic limit, identifying $\rho_{m}=mn$, where $n$ is the number density of particles of mass $m$, and taking $\beta
\ll1,$ $\gamma\rightarrow1+\beta${}$^{2}/2,$
\begin{equation}
\left(  \varepsilon+P\right)  \gamma\rightarrow\rho_{m}c^{2}\left\{
1+\left(  \varepsilon_{int}+P\right)  /\rho_{m}c^{2}\right\}  .
\end{equation}
Here an extra $\gamma$-factor for the energy density is necessary to account for the Lorentz contraction effect on the density \cite{LandauLifshitzBook,Koide:2013nia}.
We see that Eq.\ (\ref{Tmunu-Matrix}) reduces to Eq.\ (\ref{Tmunu-NREL}).

To obtain the relativistic version of Euler equation Eq.\ (\ref{NonRelEuler1}), we first write
\begin{equation}
u_{\nu}\partial_{\mu}T^{\mu\nu}=\frac{dh}{d\tau}+h u^{\mu}u_{\nu
}\partial_{\mu}u^{\nu}+h\partial_{\mu}u^{\mu}=0, \label{udtTmunu=0}%
\end{equation}
where $d/d\tau\equiv u_{\mu}\partial^{\mu}=\gamma d/dt$ is the covariant proper time derivative of the fluid element, and $h=\varepsilon+P$ is the enthalpy density.
Since $u_{\nu}u^{\nu}=1,$ we have $u_{\nu} \partial_{\mu}u^{\nu}=0$.
Furthermore, if we introduce the thermodynamic relations,
\begin{equation}
h  = Ts+\mu n, \qquad d\varepsilon = Tds+\mu dn,
\end{equation}
where $s$ and $n$ are the proper entropy and charge densities, and $T$ and $\mu$ are the temperature and chemical potential, respectively, we can re-express Eq.\ (\ref{udtTmunu=0}) as
\begin{equation}
T\partial_{\mu}\left(  s u^{\mu}\right)  +\mu\partial_{\mu}\left(
n u^{\mu}\right)  =0.
\end{equation}
Since $n$ is a proper scalar density satisfying Eq.\ (\ref{ConsN}), we conclude that $\partial_{\mu}\left(  s u^{\mu}\right)  =0$, that is, the entropy is also conserved.
This also holds when the chemical potential $\mu=0$.

Now, the space components of Eq.\ (\ref{dmuTmunu=0}) in three-vector form is expressed as
\begin{equation}
\partial_{\mu}\mathbf{T}^{\mu}=\frac{d}{d\tau}\left(  h \mathbf{u}
\right)  +h\mathbf{u\partial}_{\mu}u^{\mu}+\mathbf{\nabla}P=0,
\label{diT}
\end{equation}
where we introduced the three-vector notation $\mathbf{T}^{\mu}$ for $\left\{T^{\mu i}\right\}  $ and $\mathbf{u}$ for $\left\{  u^{i}\right\}$, with $i=1,2,3$.
Using the continuity equation for the conserved charges, we can cast Eq.\ (\ref{diT}) into a more familiar form,
\begin{equation}
\frac{d}{d\tau}\left(  \frac{\varepsilon+P}{n}\mathbf{u}\right)
=-\frac{1}{n}\mathbf{\nabla}P, \label{RelEuler}
\end{equation}
which is the relativistic Euler equation.
For a system where the contribution to the energy density from the chemical potential is negligibly small as in the central rapidity region of ultra-relativistic heavy ion collisions, we can use the proper entropy density $s$ in stead of $n$ \cite{Kodama:2001qv}.

In the application to the physics of $\sqrt{s_{NN}}>100$ GeV in RHIC and LHC, we often ignore the existence of conserved charges and discuss only the behaviors of the energy-momentum tensor when we investigate the dynamics of the produced matter at the central rapidity domain.
However, in the forward rapidity domain, or in the future planned experiments such as BES/BNL, CBM/FAIR and NICA/JINR where effects of the large stopping power of nucleus are expected, we cannot ignore the contribution from conserved charges in analyses.

For the sake of the later convenience, we discuss the property of the fluid velocity $u^{\mu}$.
By using Eq.\ (\ref{Tmunu-rEuler}), we find
\begin{equation}
T^{\mu\nu}u_{\nu}=\varepsilon u^{\mu} \label{def-vel-ener}
\end{equation}
where $u^{\mu}$ is an eigenvector of $T^{\mu\nu}$ and its eigenvalue is $\varepsilon$.
Physically, this can be interpreted as $u^{\mu}$ being parallel to the energy flow.
From Eq.\ (\ref{Nmu-ideal}), the charge current is proportional to $u^{\mu}$.
Thus, the fluid velocity defined here is parallel to both of the energy and charge flows.
Moreover, because of this coincidence, $\varepsilon$ coincides with the energy density when $n$ agrees with the charge density.
This is a benefit of the ideal fluid approach because we do not have any ambiguity for the choice of LRF and hence the application of EoS, differently from the case of viscous fluids.

\subsection{Relativistic Dissipative Hydrodynamics}

For the ideal fluid, we have considered an idealized situation where $T^{\mu\nu}$ and $N^{\mu}$ are expressed by only three quantities, $\varepsilon$, $n$ and $u^{\mu}$ (five dynamical parameters) due to the assumption of LTE.
When we do not have these constraints, the general expressions of $T^{\mu\nu}$ and $N^{\mu}$ contains nine independent components more.
We expressed them as
\begin{align}
T^{\mu\nu} &  =(\varepsilon+P)u^{\mu}u^{\nu}-Pg^{\mu\nu}+\Pi^{\mu\nu
},\\
N^{\mu} &  =nu^{\mu}+\nu^{\mu},
\end{align}
introducing the additional quantities, a symmetric tensor $\Pi^{\mu\nu}$ and a four vector $\nu^{\mu}$.
In order to avoid the double counting from already included in the ideal part, these new quantities should satisfy some constraints.
Conventionally, to be consistent with the usual hydrodynamic form of the Navier-Stokes-Fourier (NSF) equation, $\Pi^{\mu\nu}$ is decomposed as
\begin{equation}
\Pi^{\mu\nu}=-\Delta^{\mu\nu}\Pi+h^{\mu}u^{\nu}+u^{\mu}h^{\nu}+\pi^{\mu\nu},
\end{equation}
where $\Pi,$ $h^{\mu}$ and $\pi^{\mu\nu}$ are new variables to characterize the tensor $\Pi^{\mu\nu},$ and $\Delta^{\mu\nu}$ is the projection operator orthogonal to $u^{\mu},$ defined by $\Delta^{\mu\nu}=g^{\mu\nu}-u^{\mu}u^{\nu}$.
The vectors and tensors appearing here are defined to satisfy the following orthogonal conditions,
\begin{equation}
u_{\mu}\pi^{\mu\nu}   =0, \qquad
h^{\mu}u_{\mu}   =\nu^{\mu}u_{\mu}=\pi_{\ ~\mu}^{\mu}=0.
\end{equation}

In the ideal fluid, we can choose the scalar functions $\varepsilon$ and $n$ so as to agree with the energy density and the conserved charge density in LRF, respectively, so that LRF can be defined uniquely.
Unfortunately, this natural identification is not applicable for the above equations.
For example, in LRF of the fluid flow, where by definition, $u^{\mu}\rightarrow(1,0,0,0)$, $N^{\mu}$ still has the spatial components if$\ \nu^{\mu}$ is not identically null:
\begin{equation}
N^{\mu}\rightarrow(n,\nu^{x},\nu^{y},\nu^{z}).
\end{equation}
That is, in LRF of $u^{\mu}$, there is still a flow of the conserved charge.
Although $n=u_{\mu}N^{\mu}$ is the charge density observed in the LRF of $u^{\mu}$, it is not the same as the density in the LRF of $N^{\mu}$.
It is also true for $\varepsilon$, if $h^{\mu}$ is not identically null and we cannot identify $\varepsilon$ with the net energy density measured in LRF of the energy flow.

However, we still have the freedom to choose $u^{\mu}$ to obtain a physically convenient LRF.
For example, when we observe the fluid velocity with respect to the energy flow, it is natural to employ Eq.\ (\ref{def-vel-ener}) as the definition of $u^{\mu}$.
In this case, we should set $h^{\mu}=0$ and then $\varepsilon$ can be identified with the energy density of the fluid element in its LRF.
However if we do so, then in general we cannot set $\nu^{\mu}=0$ simultaneously, and $n$ in the energy LRF defined above does not coincide with the charge density in its own flow (current) in LRF of its own flow.
The charge density in its own LRF, say $n_{\mathrm{proper}}$, is given by
\begin{equation}
n_{\mathrm{proper}}=\sqrt{N_{\mu}N^{\mu}}=\sqrt{n^{2}-\boldsymbol{\nu}^{2}}.
\end{equation}
Physically, this modification comes from the Lorentz contraction of the LRF of the charge current with respect to the energy LRF since the charge current is not in equilibrium with the energy flow.
This identification of the fluid flow as that of the energy is proposed by Landau and Lifshitz \cite{LandauLifshitzBook} and referred often to as Landau frame.

On the other hand, we can set $\nu^{\mu}=0$, instead. Then $N^{\mu}$ is parallel to the fluid velocity and hence $n$ can be identified with the proper charge density in LRF, while $\varepsilon$ is not.
This definition of the fluid velocity is proposed by Eckart \cite{Eckart:1940te} and referred to as Eckart frame (see for another choice, called $\beta$ frame, Ref.\ \cite{Becattini:2014yxa}).

From the view point of relativity, the difference among Landau, Eckart, $\beta$, or any other frames should not cause any physical effects if the theory is truly relativistically covariant.
However, since hydrodynamics is an effective theory, some approximations or truncations involving thermodynamics are implicitly included as we develop more in detail later in Sec.\ \ref{sec:MoreAboutHydro}.
In particular, the thermodynamical relations are only defined in a privileged frame, so that its use in a some particular system may lead to physically different approximations.
In other words,  a general C-G method consistent with relativity to extract the fluid behavior from a microscopic dynamics has not yet been estabilished as will be discussed later in Sec.\ \ref{sec:MoreAboutHydro}.

In this review, we mostly adopt the Landau frame of the fluid velocity unless mentioned otherwise.

\subsection{Relativistic Navier-Stokes-Fourier model}

In non-relativistic hydrodynamics, an unknown tensor is determined by employing linear irreversible thermodynamics \cite{groot1962non}.
When we apply the same argument to define the corresponding non-relativistic quantities to $\Pi$, $\pi^{\mu\nu}$ and $\nu^{\mu}$ and write them down in a covariant form, we obtain
\begin{equation}
\Pi   =-\zeta\partial_{\mu}u^{\mu}, \qquad
\pi^{\mu\nu}    =2\eta\Delta^{\mu\nu\alpha\beta}\partial_{\alpha}u_{\beta}, \qquad
\nu^{\mu}    =\kappa\left(  \frac{nT}{\varepsilon+P}\right)  ^{2}\Delta
^{\mu\nu}\partial_{\nu}\frac{\mu}{T}, \label{NSVIS}
\end{equation}
where
\begin{equation}
\Delta^{\mu\nu\alpha\beta}=\frac{1}{2}(\Delta^{\mu\alpha}\Delta^{\nu\beta
}+\Delta^{\nu\alpha}\Delta^{\mu\beta})-\frac{1}{3}\Delta^{\mu\nu}
\Delta^{\alpha\beta}
\end{equation}
is a symmetric direct product of projection operators orthogonal to the velocity field $u^{\mu}$.
The $\zeta$, $\eta$ and $\kappa$ are the coefficients of the bulk viscosity, shear viscosity and charge diffusion, respectively. The above expressions are obtained from Ref.\ \cite{LandauLifshitzBook}.

This is the relativistic generalization of the NSF theory.
In fact, we can derive the NSF theory by taking the non-relativistic limit of this model.
However, it is known that this model allows a propagation which is larger than the speed of light and is essentially unstable.
This is discussed in Sec.\ \ref{sec:stability}.

Sometimes, the above NSF approach is called the first-order theory, since it corresponds to the first order derivative expansion of the energy-momentum tensor (see Sec.\ \ref{AdS}) or Boltzmann equation as is done in Chapman-Enskog formalism (see \ref{sec:boltz}).
Similarly, the causal relativistic hydrodynamics described in the next section is the second order theory or Israel-Stewart type theory, since the corresponding thermodynamics is extended to account for the second order deviation from the thermodynamic equilibrium.
As we will see later, it is not known yet that dissipative hydrodynamics can be derived from a microscopic dynamics by some systematic expansion from the equilibrium state of the system.
To avoid misleading, we will not use the terms such as first order and second order theories.

\subsection{Causal models}

\label{sec:causal}

In the relativistic NSF theory, the dissipative terms given in Eq.\ (\ref{NSVIS}) are treated as dynamically dependent variables and this is the origin of the inconsistency with relativistic kinematics. In other formulations of dissipative hydrodynamics, the number of independent hydrodynamic variables should be increased from that of the ideal case given by $\varepsilon$, $n$ and $u^{\mu}$. The new variables are the bulk viscosity $\Pi$, the shear viscosity $\pi^{\mu\nu}$ and the charge diffusion $\nu^{\mu}$.
In contrast to relativistic NSF theory, we need to construct the evolution of these variables and so far there is no established theory for these dynamical equations. There are several variations (see Sec.\ \ref{sec:MoreAboutHydro}).

As an example, we write down the expressions obtained by Israel and Stewart \cite{Israel:1976tn,Israel:1979wp}.
They generalized the traditional argument of the linear irreversible thermodynamics and assumed that the entropy (four current) has additional contributions attributed to irreversible currents.
Following the Israel-Stewart theory, let us consider that the entropy four current $S^{\mu}$ is expressed as \cite{Israel:1976tn,Israel:1979wp}
\begin{equation}
S^{\mu}=S_{0}^{\mu}+\frac{\mu}{T}\nu^{\mu}-Q^{\mu},
\end{equation}
where $S_{0}^{\mu}$ is the entropy four current for the ideal fluid and
\begin{equation}
Q^{\mu}=\frac{1}{2}u^{\mu}\left(  \frac{\zeta}{\tau_{\Pi}}\Pi^{2}+\frac{\eta
}{\tau_{\pi}}\pi_{\mu\nu}\pi^{\mu\nu}+\frac{\kappa}{\tau_{\kappa}}\nu_{\mu}
\nu^{\mu}\right)  .
\label{eq:Qmu}
\end{equation}

To satisfy the algebraic positivity of the entropy production rate, $T\partial_{\mu}S^{\mu}=\sigma\geq0$, we find the evolution equations of $\Pi$, $\nu^{\mu}$ and $\pi^{\mu\nu}$ as
\begin{align}
\Pi &  =-\zeta\partial_{\mu}u^{\mu}-\tau_{\Pi}\frac{d\Pi}{d\tau}-\frac{1}
{2}\tau_{\Pi}\Pi\partial_{\mu}u^{\mu}-\frac{T\zeta}{2}\Pi\frac{d}{d\tau
}\left(  \frac{\tau_{\Pi}}{T\zeta}\right)  ,\nonumber\\
\nu^{\mu}  &  =\Delta^{\mu\nu}\left[ -\kappa\partial_{\nu}\frac{\mu}{T}
-\tau_{\kappa}\frac{d\nu_{\nu}}{d\tau}-\frac{1}{2}\tau_{\kappa}\nu_{\nu
}\partial_{\mu}u^{\mu}-\frac{T\kappa}{2}\nu_{\nu}\frac{d}{d\tau}\left(
\frac{\tau_{\kappa}}{T\kappa}\right)  \right]  ,\nonumber\\
\pi^{\mu\nu}  &  =\Delta^{\mu\nu\alpha\beta}\left[  \eta\partial_{\alpha
}u_{\beta}-\tau_{\pi}\frac{d\pi_{\alpha\beta}}{d\tau}-\frac{1}{2}\tau_{\eta
}\pi_{\alpha\beta}\partial_{\lambda}u^{\lambda}-\frac{T\eta}{2}\pi
_{\alpha\beta}\frac{d}{d\tau}\left(  \frac{\tau_{\pi}}{T\eta}\right)  \right]
, \label{Eq.relNSF}
\end{align}
where $d/d\tau=u^{\mu}\partial_{\mu}$.
As seen from the above equations, the time derivatives of the newly introduced variables are contained in a complicated manner, forming a coupled dynamical system together with the original hydrodynamic variables.
In addition, besides $\eta$, $\zeta$ and $\kappa$ which have already appeared in the relativistic NSF model, other transport coefficients , $\tau_{\Pi}$, $\tau_{\kappa}$ and $\tau_{\pi}$ appear in this model.
These parameters are called relaxation times which originated from relatively rapid dynamics compared with that of the relativistic Euler equation, and they play crucial roles to cure the problem of internal inconsistencies in the relativistic NSF theory.

To see this, let us consider an irreversible process of a charged density $n$ satisfying the continuity equation, $\partial_{t}n(\mathbf{x},t)+\nabla\cdot\mathbf{J}(\mathbf{x},t)=0$.
Phenomenologically, the irreversible part of the current $\mathbf{J}$ is known to obey Fick's law,
\begin{equation}
\mathbf{J}(\mathbf{x},t)=-D\nabla n(\mathbf{x},t).
\end{equation}
When we consider the retardation effects which are expected from a microscopic dynamics, such a current has a structure given by
\begin{equation}
\mathbf{J}(\mathbf{x},t)=\int^{t}ds G(t-s)\nabla n(\mathbf{x},s), \label{jdf}
\end{equation}
where $G(t)$ is the memory function which is expressed as the time correlation function of microscopic currents.
When the memory of microscopic correlations vanishes rapidly, $G(t)$ approximately behaves as a Dirac delta function and then we recover Fick's law.
However, when there is no clear separation between microscopic and macroscopic time scales, we should not ignore completely the memory effects.
For this purpose, let us assume that $G(t)$ is approximately expressed as
\begin{equation}
G(t)=\frac{D}{\tau_{D}}e^{-t/\tau_{D}},\
\end{equation}
where we introduced the relaxation time $\tau_{D}$, which represents the time scale of the memory effects.
Substituting this into Eq.\ (\ref{jdf}), we have
\begin{equation}
\mathbf{J}(\mathbf{x},t)=-D\nabla n(\mathbf{x},t)-\tau_{D}\partial
_{t}\mathbf{J}(\mathbf{x},t). \label{mcveq}
\end{equation}
This equation is called the Maxwell-Cattaneo-Vernotte (MCV) equation.
If we take the vanishing limit of $\tau_{D}$, Fick's law is reproduced (see Refs.\ \cite{Koide:2006ef,Koide:2010wt} for more details).
Comparing the above equation with equations (\ref{Eq.relNSF}), one can see that $\tau_{\Pi}$, $\tau_{\pi}$ and $\tau_{\kappa}$ should be interpreted as relaxation times of bulk viscosity, shear viscosity and charge diffusion, respectively.
The last term corresponds to the second terms of the right-hand sides of Eq.\ (\ref{Eq.relNSF}).
It is also possible to understand the role of the third terms in Eq.\ (\ref{Eq.relNSF}) qualitatively considering that these irreversible currents are densities of corresponding extensive quantities (see Ref.\ \cite{Denicol:2009zz}).
For a more precise derivation of the MCV equation from microscopic theory, see Sec.\ \ref{sec:proj}.

Note that in this model, it becomes clear that the values of relaxation times are intimately related to suppress superluminal propagation modes.
From a linear analysis, we can calculate the propagation speeds and in order to maintain relativistic causality, the following relations should be satisfied \cite{Denicol:2008ha,Pu:2009fj}
\begin{equation}
\frac{\zeta}{\tau_{\Pi}(\varepsilon+P)}    \leq1-c_{s}^{2} , \qquad
\frac{\eta}{\tau_{\pi}(\varepsilon+P)}    \leq\frac{3}{4}(1-c_{s}^{2}),
\label{sta-con-shear}
\end{equation}
where $c_{s}$ is the sound velocity.
As we will see later, if the causality is not satisfied the hydrodynamic mode becomes unstable, demonstrating that the construction of relativistic fluid dynamics requires a special care to unify thermodynamics and the theory of relativity.
The former needs a finite space-time domain in which a large number of internal degrees of freedom is accommodated, whereas the latter demands the true microscopic locality.
A similar care should be taken when we deal with a local gauge invariance with hydrodynamics.

The importance of the causal models of dissipative hydrodynamics in relativistic heavy ion collisions were first pointed out by Refs.\ \cite{Prakash:1993bt,Muronga:2001zk,Muronga:2006zw}.
Throughout this review, we use dissipative hydrodynamics instead of viscous hydrodynamics, which is commonly used in the literature, since we consider not only viscosity but also diffusion (heat conduction).

\subsection{Causality and instability}

\label{sec:stability}

The instability of relativistic hydrodynamics has been investigated by various authors.
For example, Hiscock and Lindblam discussed the stability of causal and relativistic NSF models and concluded that the relativistic NSF models both in Eckart and Landau frames are unstable for a linear perturbation around a hydrostatic state \cite{Hiscock:1985zz,Hiscock:1987zz}.
As shown in the previous section, the physical reason for this is that the relativistic extension of the NSF theory treats the thermodynamic reaction as a kind of action at a distance within the C-G scale.
Thus, such a theory necessarily generates acausal modes.

In Refs.\ \cite{Denicol:2008ha,Pu:2009fj}, it was pointed out that there seems to exist a relation between causality and stability: if an acausal mode is contained, such a theory becomes unstable at any cost.
For example, it was shown that, as far as the causality requirement, Eq.\ (\ref{sta-con-shear}) is satisfied in causal models, the stability of linear perturbation is guaranteed in any Lorentz boosted frame.
On the other hand, if they are violated, the nature of stability becomes Lorentz non-invariant, and unstable mode emerges when the hydrostatic background is boosted.
That is, the relativistic NSF theory contains acausal modes and is inconsistent with the relativistic kinematics.
The table \ref{table-sta} shows the stability of various models.

\begin{table}[h]
\caption{Stability of various models.}
\begin{center}
\begin{tabular}[c]{rccc}
\hline
& Relativistic & causal model & causal model \\
& NSF model & without Eq.\ (\ref{sta-con-shear}) & with Eq.\ (\ref{sta-con-shear}) \\
\cline{2-4}
hydrostatic state & \textit{unstable} & \textit{stable} & \textit{stable} \\
moving frame & \textit{unstable} & \textit{unstable} & \textit{stable} \\
\hline
\label{table-sta}
\end{tabular}
\end{center}
\end{table}

The stabilities around Bjorken's scaling solution are discussed for the relativistic NSF theory \cite{Kouno:1989ps} and causal model \cite{Denicol:2008ha}.
Both can be unstable under certain conditions.
Moreover, there are various works where the relation between heat conduction and stability is discussed.
Contrary to the above conjecture, there are various proposals to develop stable relativistic NSF theory \cite{Van:2011yn,GarciaPerciante:2008ui}.
However, the consistency of stability for the Lorentz boost is not studied in these works.

\section{Application to Relativistic Heavy Ion Collisions}

\label{sec:Practical}

The hydrodynamics described in the previous section as it is cannot be applied immediately to the process of relativistic heavy ion collisions.
To perform the hydrodynamic approach, we have to specify the initial condition (IC), the equation of state (EoS), the transport coefficients, and the freeze-out (FO) procedure to connect the hydrodynamical variables to the observed final state particles.
Furthermore, it is crucial to extract appropriate observables which clearly reflect the hydrodynamic scenario from the thousands of produced particles.
In below, we describe the necessary ingredients for the application of hydrodynamic approach to relativistic heavy ion collisions.

\subsection{Initial Condition}

\label{sec:INIC}

As shown in the previous section, hydrodynamics is a system of classical field equations with the use of thermodynamic relations for each fluid element in its LRF.
Initial conditions (IC) for the system are the distributions of energy, charge densities, and the velocity fields at an initial time $t_{0}$ appropriately chosen.
For this, we need to know the transition process where the quantum state of two colliding objects (nucleus) is converted into the macroscopic matter that defines an initial distribution of hydrodynamic variables.
This is a hard task since we are not able to perform real \textit{ab initio} description of collisional processes in QCD without introducing models.
Dynamics at early stages in the collision is exactly what we wish to study with the help of the hydrodynamic approach.
As mentioned before, for heavy ion collisions, the centrality (impact parameter) is relatively well-estimated on an event-by-event (EbyE) basis.
Although the centrality is a crucial factor to determine the overlap geometry of the nuclear collisions, this is not enough to determine more detailed features of initial conditions.
What we have to do is to introduce a model which describes the initial collisional stage reflecting the event geometry.
From this we construct the initial condition of the hydrodynamic equations and then compare the final observables in hydrodynamic calculations with the experimental data as function of collision centrality and other final state kinematic parameter dependences. 

Just to understand the qualitative feature of the hadronic collisions at ultra-relativistic energies, note that the incident hadrons are predominantly emitted in the forward direction, with high longitudinal momenta (leading particles).
The inelasticity distribution $P(x_F)$ of the very high energy proton-proton (p--p) collision is approximately constant where $x_F$ is the Feynmann variable, $P(x_F) \sim const$. having relatively small ($\sim300$ MeV) transverse momentum.
This implies that in terms of rapidity $y=\left(  1/2\right)  \ln\left(  E+p_{z}\right) /\left(  E-p_{z}\right)  $, we get $P(y)\sim\cosh y$, so that the incident protons are emitted predominantly in the large rapidity domain.
Nevertheless, since the average inelasticity is on the order of $1/2\sim1/5$, a large amount of the incident energy is lost and the corresponding energy is used to create the particles.
The rapidity distribution of these particles are symmetric at the central rapidity because most of them are created from the excitation of the vacuum.
Therefore, in ultra-relativistic p--p collisions, the predominant contribution to the rapidity distribution of charged particles at the mid-rapidity comes from the produced particles.
See Ref.\ \cite{Wong:1995jf} for details.

For A--A collisions, due to the multiple collisions inside the nucleus, the stopping power increases for heavier systems, and the baryon number density also tends to immigrate to the smaller rapidity domain, especially for lower energies.
In the early years hydrodynamic simulations of relativistic heavy ion collisions at energies below 1 GeV per nucleon were mainly devoted to see how the nuclear compression effects reflect in the observables such as the collective direct flow \cite{Nix:1979tc,Stoecker:1986ci,Clare:1986qj} and full (3+1)D relativistic hydrodynamic codes were developed in these works.
However, in ultra-relativistic collisions, the baryon numbers carried by the colliding nuclei tend to be distributed in the larger rapidity domain while the produced matter dominates around the central rapidity.
In 1983, J.D. Bjorken \cite{Bjorken:1982qr} made an estimate of the energy density of the produced matter in the central rapidity region of A--A collisions based on the boost invariant solution of the (1+1)D longitudinal hydrodynamics.
Many of early works on hydrodynamic properties from the produced QGP used this kind of initial condition \cite{Baym:1984sr}.
Some (3+1)D codes for such purposes also have been developed (see Refs.\ \cite{Ornik:1989jp,Rischke:1995ir}).

Later, phenomenological or theoretical studies on hadronic collisions advanced, and several versions of more realistic models became available.
However, in the early studies of the data from RHIC, which started in 2001, most of hydrodynamic calculations were based on smooth energy and baryonic distributions (smooth IC).
Furthermore, many calculations were concentrated on the analysis of the central rapidity region of RHIC data so that (2+1)D calculation has mainly been applied to study the transverse dynamics.
For these investigations, the smooth IC's are obtained basically from the geometry of the overlapping area of two colliding nuclei.
For comprehensive reviews, see Refs.\ \cite{Huovinen:2006jp,Hirano:2008hy} and references therein.

More elaborated estimates can be obtained by other microscopic methods based on nucleon-nucleon event generators such as HIJING \cite{Wang:1996yf}, PYTHIA \cite{Sjostrand:2006za}, NEXUS \cite{Drescher:2000ha}, UrQMD \cite{Bleicher:1999xi}, EPOS \cite{Werner:2005jf}, AMPT \cite{Pang:2012he} etc.
See Table 3 of Ref.\ \cite{Hirano:2012kj}.
An essentially different approach is to estimate the QCD color field density from the QCD vacuum structure in the high energy limit which is called the Color Glass Condensate (CGC) \cite{CGC,Albacete:2014fwa}.
In the CGC model for hadronic collisions, the two incident hadrons are described as the CGC sheets due to the Lorentz contraction. While they pass through, their interactions create the coherent classical gluon field (glasma) which in turn transforms into QGP.
Using these models, the initial energy-momentum tensor are estimated from the produced gluon number density.
Another form to introduce the gluon saturation with the mini-jet contribution from nucleon-nucleon collision has been developed in the so-called EKRT model.
See the recent results in Refs.\ \cite{Paatelainen:2013eea,Niemi:2015qia}.

Event generator simulations exhibit usually very bumpy structures in transverse energy density distribution.
Implications of such IC in observables were pointed out by Ref.\ \cite{Gyulassy:1996br}.
The first use of an event generator for the calculation of the initial condition for (3+1)D hydrodynamic simulations was performed in 2001 \cite{Osada:2001hw,Aguiar:2000hw} using the Nexus model following the suggestion by K.\ Werner \cite{Drescher:2000ha,Drescher:2000ec}.
There, the importance of EbyE fluctuations and granularities in IC are emphasized \cite{Aguiar:2000hw,Socolowski:2004hw,Andrade:2006yh}.
The EbyE fluctuations in IC became the central issue after G.\ Rolland \cite{Alver:2010gr} pointed out that such fluctuations in IC are crucial for the analysis of higher order harmonics of collective flow parameters (see discussion in below).

Presently there are several approaches to generate IC.
One is the nucleon-nucleon event generator based Monte-Carlo approach which generally called as Monte-Carlo Glauber model (MC-G).
There are several versions depending on the event generators employed.
Instead of using the Glauber approach, it is also possible to utilize the microscopic transport models in parton/string level, for example, the UrQMD model \cite{Bass:1998ca,Bleicher:1999xi,Petersen:2008dd} and the PHSD model \cite{Cassing:2008nn,Cassing:2008sv,Cassing:2009vt,Bratkovskaya:2011wp}.
Another line is to use the CGC-glasma picture, such as MC-KLN \cite{Drescher:2006ca,Drescher:2007ax,Kharzeev:2004if} and IP-Sat/Glasma \cite{Bartels:2002cj,Kowalski:2003hm,Schenke:2012wb} approaches.

There exists a conceptual difference between the CGC-glasma approach and the other event generator type approaches, which leads to very different pictures for the initial state for the hydrodynamics.
In the CGC-glasma picture, the coherent classical gluon field develops longitudinally between the two CGC sheets as a consequence of color charge exchange when the two sheets collide.
Such a state is composed of color flux tubes, which eventually decay into gluons and thermalize, forming the QGP.
In this picture, the flux tubes are boost-invariant, and particles are generated by the instability of such coherent classical color fields.
The momentum distribution of produced partons then becomes anisotropic, having mainly transverse momenta.
On the other hand, the initial condition generated from the partonic collisions has the large longitudinal momentum as will be shown in Sec.\ \ref{sec:CoarseGraining}.
Such an anisotropy in the initial momentum distribution means that the existence of large values of dissipative quantities in the corresponding $T^{\mu\nu}$, otherwise the pressure would be isotropic.
In most of hydrodynamic simulations the initial dissipative quantities are set to null.
This constitutes a fundamental problem with respect to the initial states of the QGP.

Furthermore, depending on the initial condition model, the initial profile of the energy density distribution changes appreciably.
Dumitru and Nara pointed out that fluctuations of multiplicities of produced particles in p--p collisions within the CGC approach substantially increase the initial fluctuation dominated moments, for example $\epsilon_3$ \cite{Dumitru:2012yr}.

\subsection{Equation of State and Transport Coefficients}

\label{sec:EoS-transport}

Another fundamental input to execute hydrodynamic calculations are a set of all the thermodynamic relations (which in general we refer to as equation of state (EoS)) and the transport coefficients which are introduced in Sec.\ \ref{sec:causal}.

The independent parameters which specify EoS are $T$ and $\mu_{B}$.
As mentioned previously, in the central rapidity region of ultra-relativistic heavy ion collisions the predominant amount of deposited energy goes to the matter created from the QCD vacuum, so that particles and anti-particles are equally populated and hence $\mu_{B}$ is very small.
For null baryonic chemical potential, the results of Lattice QCD (lQCD) calculations are considered to be well-established \cite{Karsch:2006sm,Durr:2010vn} as mentioned in Sec.\ \ref{SecII-2}, even at a quantitative level, since some differences among the calculations from different groups, which has been existing for a decade, is now tending to converge (see Refs.\ \cite{Philipsen:2012nu,Borsanyi:2014rza}).
For the temperature well below the pion mass, the lQCD results have been considered less reliable so that the hadronic resonance gas EoS is often employed \cite{Huovinen:2009yb}.
The lQCD calculations for finite baryonic chemical potential are non-trivial due to the so-called sign problem of the fermionic determinant which appears in the partition function, but several lQCD calculations for finite $\mu_{B}$ values have been studied \cite{Borsanyi:2012cr}.
See more details in Refs.\ \cite{Philipsen:2012nu,Petreczky:2012rq}.
A reliable EoS for finite $\mu_{B}$ is crucial to perform a realistic (3+1)D hydrodynamic calculations for the heavy ion collisional events to be investigated in BES/CBM/NICA programs.
In such a finite $\mu_{B}$ domain, effective theories for QCD are often used to obtain EoS \cite{Fukushima:2010bq,Fukushima:2013rx,Tawfik:2014eba} but a full hydrodynamic calculation has not yet been implemented with such an EoS.

Although the importance of viscosity in the application of hydrodynamics in relativistic heavy ion collisions has been pointed out at the early times \cite{Danielewicz:1984ww} and the necessity of using the causal dissipative hydrodynamics \cite{Prakash:1993bt,Muronga:2001zk,Muronga:2006zw}, initially most hydrodynamic analyses of flow data at RHIC was with ideal hydrodynamics, mainly due to the complexities in the implementation of relativistic dissipative hydrodynamics.
Considering the explosive nature of the hot expanding matter in very short time scale, it would be rather surprising if LTE is attained in such a system.
Nevertheless, the behavior of the collective elliptic flow parameter $v_{2}$ and other bulk observables (see the subsection below) are quite well described by ideal hydrodynamics.
The large value of $v_{2}$ can be interpreted as the large pressure gradient at the initial stage of the hot fluid expansion, leading to the formation of a new state of matter, QGP.
Discrepancies between ideal hydrodynamic calculations and $v_{2}$ as a function of centrality and transverse momentum were expected to be attributed to the effects of viscosity, especially those from the hadronic phase where the mean-free path is expected to be much larger than that in QGP.
Therefore in ultra-relativistic heavy ion collisions, the so-called \textit{Little Bang} \cite{Kolb:2003dz,Heinz:2009xj} picture has been accepted, where a hot and dense (almost perfect) fluid of strong interacting plasma of quarks and gluons (sQGP) is created together with a rather thin mantle of viscous hadronic matter (core-corona), and eventually turns into the bundle of free-streaming hadrons (see Sec.\ \ref{freezeout}).

After the conjecture of the existence of a lower bound called KSS bound of the shear viscosity, $\eta/s\ \geq1/4\pi\,$ where $s$ is the entropy density based on the N=4 Supersymmetric Yang-Mills theory \cite{Kovtun:2004de} in 2005, it provoked a great interest in determining quantitatively the coefficient of the shear viscosity of sQGP.
This is because the success hitherto obtained by the ideal hydrodynamic descriptions of many observed collective variables suggests that the shear viscosity should be very close to this conjectured lower bound.
For a viscous case, we need to specify the transport coefficients such as $\eta$ and $\zeta,$ together with the relaxation times, $\tau_{\pi}$ and $\tau_{\Pi}$.
Generally they depend on thermodynamical quantities, but in the most of calculations, such dependencies are simplified or even taken as constant for $\eta/s$ and $\tau_{\pi}$, and bulk viscosity is simply omitted because it is expected to be small.
The analysis of the collective flow parameters using temperature dependent transport coefficients was introduced by Ref.\ \cite{Song:2009rh}.
For the recent studies, see Refs.\ \cite{Niemi:2015qia,Ryu:2015vwa,Gale:2012rq,Molnar:2014zha}.
In addition, the bulk viscosity is expected to be fairly large near the critical point, and for a quantitative analysis of experimental data, its effect should not be neglected \cite{Denicol:2009am,Monnai:2009ad,Ryu:2015vwa,Noronha-Hostler:2013gga,Noronha-Hostler:2014dqa,Gardim:2014tya,Dusling:2011fd,Bozek:2009dw}.

\subsection{Freeze-out and Final State Interactions}

\label{freezeout}

As already mentioned in the Introduction, in relativistic heavy ion collisions it is dificult to observe directly the time evolution of flow profiles of the matter produced.
Like most explosive evolutions, the hydrodynamic description of hot and dense matter eventually fails when the local density and temperature become low enough to the point where the continuum description loses its meaning.
In this stage, the constituent particles gradually decouple from each other and finally turn into a bundle of free-streaming particles which are captured in the detectors.
We refer to this transition process as the \textit{freeze-out} (FO) process.
More precisely, this is called \textit{kinetic freezeout} in contrast to \textit{chemical freeze-out} for which hadronic chemical composition ceases to change due to the suppression of inelastic channels as the temperature and density decrease.
Temperature and chemical potential determined from thermal models in Sec.\ \ref{sec:RelativisticHeavyIonCollisionsAndHydro} refer to this chemical freezeout.
The FO process is a dynamical process and a precise description requires an elaborate transport approach as is discussed later.

In many simplified hydrodynamic calculations, we assume the so-called sudden FO (sFO), where the particle spectra is calculated directly from the hydrodynamic variables at the kinetic FO surface.
However, if we do not employ the chemical FO which occurs earlier than the kinetic FO, abundances of particles associated with the pair production mechanism, such as anti-proton, would be totally underestimated.
The effect of chemical FO is incorporated simply recording the values of temperature and chemical potential for each hadronic species at the chemical FO and their abundances are adjusted by normalization.
However, even in such a simplified picture, changes in the EoS after the chemical FO should also be accounted for by treating separately chemically \textit{frozen-out} particles \cite{Huovinen:2007xh}.

Theoretically speaking, the description of the transition from hydrodynamic variables to a bundle of free-streaming independent particles should necessarily be of a probabilistic nature, since the former is based on macroscopic distribution functions and the precise microscopic information which is required for the later is smeared out from the beginning.
In practice, what we can do is at best to calculate the final state single particle spectra $d^{3}\sigma/d\vec{p}^{3}$ for each observable particle from the hydrodynamic calculation.

In order to introduce a realistic FO in a hydrodynamic calculation we first have to know when FO starts, i.e. determine what is the physical criteria to switch the hydrodynamic evolution to some non-equilibrium transport scheme.
Second, we need to map the hydrodynamic quantities (fluid) to a state of interacting many particle system. Third, a dynamical description (transport equations) of this many particle system should be implemented till all the particles become free-streaming.
After this stage, we still consider processes of decays of unstable resonances down to the observable mesons and baryons.

\begin{figure}[tbh]
\centering
\includegraphics[height=6.4cm,keepaspectratio]{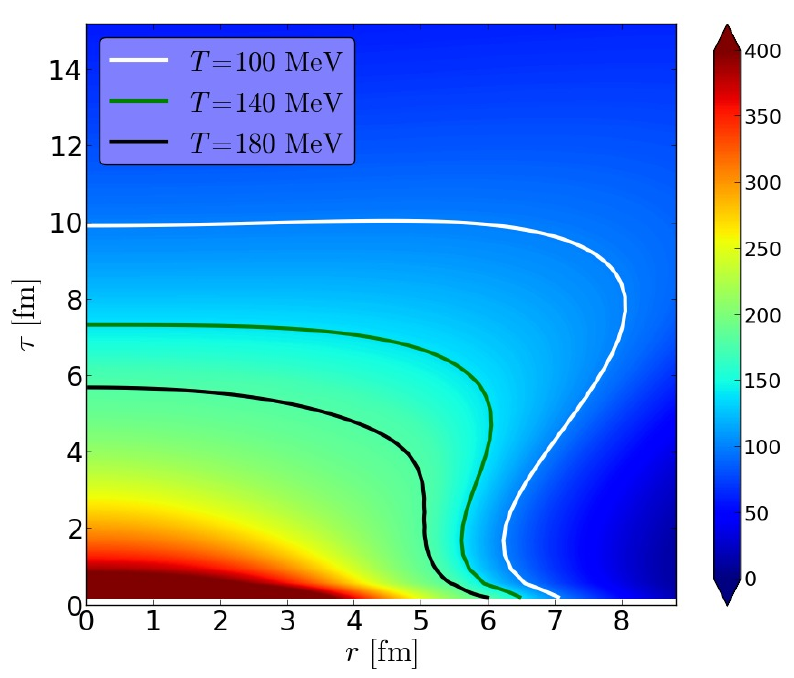}
\caption{(color online) Example of isothermal hypersurfaces. Figure\protect\footnotemarkB\ taken from Ref.\ \cite{Niemi:2014wta}.}
\label{FOS}
\end{figure}

Each of these problems is non-trivial, which constitutes formidable work.
The simplest proposal is the previously mentioned sFO from Cooper and Frye \cite{Cooper:1974mv}, which neglects the third step above.
In the sFO, we first establish the so-called freeze-out surface (FOS), $\Sigma$, using some physical criteria.
Many people uses a constant FO temperate $T_{FO}$ for simplicity but other ways have also been proposed \cite{Bondorf:1978kz,Navarra:1992ej,Schnedermann:1994gc,Bondorf:2000pm,Eskola:2007zc,Schafer:2009dj,Holopainen:2013jna}.
In Fig.\ \ref{FOS}, a typical time evolution of FOS in the transverse direction at central rapidity for a smooth initial condition is shown.
\footnotetextB{Reprinted with permission from H.\ Niemi and G.\ Denicol, \textit{How large is the Knudsen number reached in fluid dynamical simulations of ultra-relativistic heavy ion collisions?}, arXiv:1404.7327 (2014).}

Cooper and Frye proposed to calculate the freeze-out hadronic spectra as
\begin{equation}
E\frac{d^{3}N_{i}}{d\mathbf{p}^{3}}=\int_{\Sigma}d\sigma_{\mu}\ p^{\mu
}f^{\left(  i\right)  }\left(  x,p\right)  ,\ \label{CF-1}
\end{equation}
where $f^{\left(  i\right)  }\left(  x,p\right)  $ is the distribution function of the specie $i$ at FOS.
In the case of an ideal hadronic resonance gas (HRG), the distribution function, $f$, is given by the Maxwell-J\"{u}ttner-type distribution function (see the later section on the Boltzmann Equation) at thermal equilibrium, 
\begin{equation}
f\left(  x,p\right)  \rightarrow f_{TE}\left(  x,p\right)  =\frac{g}{\left(
2\pi\hbar\right)  ^{3}}\frac{1}{e^{\left(  u\cdot p-\mu\right)  /T}\pm
1},\label{CF-2}
\end{equation}
where $g$ is the statistical factor of the particle and the sign $+$ and $-$ is for bosons and fermions, respectively.
Differences in hadronic species enter through $g,\mu$ and its mass \cite{Hama:2004rr}.
This expression is a natural choice for ideal hydrodynamics where LTE is satisfied.
However, in the presence of dissipative effects, the distribution $f$  is not given by Eq.\ (\ref{CF-2}) but there exists a correction,
\begin{equation}
f\left(  x,p\right)  =f_{TE}\left(  x,p\right)  +\Delta f\left(  x,p\right)  .
\end{equation}
When the dissipative corrections are small, $\Delta f$ can be related to $\Pi^{\mu\nu}$ as well as the thermodynamical quantities.
We will discuss more in detail in Sec.\ \ref{sec:boltz} together with the derivation of dissipative hydrodynamics from the Boltzmann equation.

Unfortunately, using the Cooper-Frye formula (CF) one cannot directly calculate the identified particle spectrum, since the FO temperature for chemical FO (that is used for the thermal model for particle ratios) is higher than hydrodynamic FO (kinetic freeze-out) \cite{Arbex:2001vx,Hirano:2002ds}.
However, the dominant part of the produced particles is pions and chemical non-equilibrium effects from heavier baryons are considered to be small.
Thus, the modification of the EoS due to the chemical FO is often neglected.
Even in such a simplified treatment, many hadronic resonances need to be taken into account to calculate the correct multiplicities of observed identified particles through their decays down to the stable hadrons at the kinetic FO temperature, which is usually taken around $T_{FO}\gtrsim120$ MeV.
Several open codes which describe the thermal HRG are available and can be used to obtain the final stable hadrons from the sudden FO hadronic states \cite{Torrieri:2004zz,Wheaton:2004qb,Kisiel:2005hn,Rybczynski:2013yba,Shen:2014vra}.
For more discussion on this point, see Chap.\ 26 of Ref.\ \cite{WFlorkowski.PhenomenologyOfURHIC}.

Another shortcomings of CF is the so-called negative contribution problem.
As seen from Fig.\ \ref{FOS} FOS has two distinct regions: one in which the normal vector to $\Sigma$ is time-like (that is, the surface itself is a space-like 3D volume), and the other space-like (the surface is consisted of 2D space surface expanding in time).
Particle emissions in the domain where the normal is time-like (volume emission) has no conceptual problem in CF.
On the other hand, in the domain where the normal is space-like (surface emission), a portion of particles in the distribution $f$ belongs to the fluid, and should not be counted as emitted particles from the fluid.
However, if we only count the particles which are going out of the fluid surface, then the total energy of the FO particles does not coincide exactly with the total energy of the fluid \cite{LaszloBook}.

In order to improve the problems of sFO, the so-called hybrid models have been developed where the final stage of hydrodynamics is connected to the event generator based hadronic cascade codes or Boltzmann type equations \cite{Petersen:2014yqa}.
Importance of hybrid approach has been demonstrated by various groups.
One of the advantages of these approaches is that the chemical FO processes is naturally incorporated in the scheme.
More detailed description of the hybrid model, see the recent reviews by Hirano \textit{et al.} \cite{Hirano:2012kj} and by H. Petersen \cite{Petersen:2014yqa}.

In most hybrid approaches the transition from the hydrodynamic description to a hadronic transport system is done using CF.
Therefore, the problem of CF mentioned above still persists, since the switching surface has the similar characteristics as the final FOS.
In fact, a small correction originated from the negative contribution may be masked by the fluctuations and uncertainties in the proceeding hadronic cascade processes, but this may not be the case for other observables such as HBT radius (see the next section) especially for Kaons, and particle ratios in smaller systems such as p--p and p--A.
In addition, the negative contribution problem is also a question of theoretical consistency and needs a solution.
Several approaches have been proposed and discussed in \cite{Grassi:1994nf,Grassi:2004dz,Akkelin:2004cp,Csernai:2005ht,Akkelin:2008eh,Sinyukov:2002if,Sinyukov:2002if}.
The method proposed in Ref.\ \cite{Sinyukov:2002if} has been applied to describe these observables and shown to be effective \cite{Karpenko:2009wf,Karpenko:2010te,Shapoval:2014wya,Shapoval:2013bga,Karpenko:2012yf,Shapoval:2013jca}.
One important element of such an approach is that when we switch from the hydrodynamic phase to the hadronic phase, the EoS of lQCD and that of the hadron resonance gas should be consistent otherwise the conservation of energy and momentum is compromised (see Ref.\ \cite{Sinyukov:2013zna}).
It is also pointed out that the hadronic mass spectrum affects the final state collective flow parameters \cite{Noronha-Hostler:2013ria}.

\subsection{Physical Observables}

The signals related to the hydrodynamic evolution in the experimental measurements are those associated to the collective flow profile.
In hydrodynamic regime, if the initial condition has spatial anisotropy, the corresponding pressure gradient will generate the acceleration of the fluid elements, which in turn will affect the angular distribution of the final particles.
On the other extreme, if the produced particles come from just a superposition of independent nucleon-nucleon collisions and there is no interaction among the produced particles, then the final angular distribution is just a sum of single nucleon distribution and the initial collision geometry will not be reflected.
In this sense, an evidence of a systematic dependence of final particle distribution on the initial anisotropy indicates strongly the existence of some collective dynamics.
However, anisotropy in the initial condition is not directly observable.
Even so, if it is related to that created by the non-central collision geometry, it should be strongly correlated to the multiplicity of the produced particles (centrality).

To more quantitative studies, the Fourier expansion of the transverse angular distribution is most commonly applied \cite{Ollitrault:1992bk,Poskanzer:1998yz,Voloshin:2008dg}.
For very high multiplicity events, we can introduce the differential particle distribution for each collisional event as
\begin{equation}
\frac{d^{3}N}{p_{T}dP_{T}dyd\phi}=\frac{1}{2\pi}\frac{d^{2}N}{p_{T}dp_{T}
dy}\left\{  1+2\sum_{n=1}^{\infty}v_{n}\cos n\left(  \phi-\psi_{RP}\right)
\right\}  , \label{Smooth}
\end{equation}
where $\psi_{RP}$ is the angle of the reaction plane (the plane defined by the incident direction and the impact parameter) of the event with respect to the observational frame, i.e. the azimuthal angle of the impact parameter $\vec{b}$ of the collision in the observational frame.
In the above equation, it is assumed an ideal situation where the initial condition has the smooth distribution and symmetric with respect to the reaction plane (RP).
The Fourier coefficients, $v_{n}$ are called collective flow parameters and are functions of $p_{T}$ and $y$.
Sometimes $v_{1}$ is called as directed flow, $v_{2}$ as elliptic flow, and $v_{3}$ as triangular flow.
For smooth initial conditions all the odd flow parameters should vanish at the central rapidity region $\left(  y\simeq0\right)  $, and for the central collision $(\vec{b}\simeq0)$ all the flow parameters should vanish.

Before 2010, most hydrodynamic calculations of elliptic flow for the RHIC experiments were done using smooth initial conditions, except for those mentioned in Sec.\ \ref{sec:INIC}.
In a theoretical calculation we know the event plane a priori, but experimentally we are not able to control the reaction plane in each collisions.
Furthermore, each event does not have smooth and symmetric behavior, so that the comparison of theoretical values to experimental ones requires certain care.
Several methods to determine experimentally $v_{2}$ as function of centrality and $p_{T}$ have been developed \cite{Poskanzer:1998yz,Borghini:2001vi,Bilandzic:2010jr,Voloshin:2008dg,Sorensen:2009cz}.

When we need to consider EbyE fluctuations, Eq.\ (\ref{Smooth}) is not applicable because there is no symmetry on EbyE basis.
The most general form of angular distribution can be parameterized as
\begin{equation}
\frac{d^{3}N}{p_{T}dp_{T}dyd\phi}=\frac{1}{2\pi}\frac{d^{2}N}{p_{T}dp_{T}
dy}\left\{  1+2\sum_{n=1}^{\infty}v_{n}\cos n\left(  \phi-\psi_{n}
^{EP}\right)  \right\}  ,
\label{EbEvn}
\end{equation}
The angles $\Psi^{EP}_n$ should be determined for each value of $p_T$ and $y$ so as to make $v_n$ maximum.
However, due to the insufficient number of particles in each bin of  $p_T$ and $y$, the direct determination of these parameters for one single event is not possible.
Thus, instead of determining directly these parameters, particle correlation measurements are used to obtain the moment distributions of collective flow parameters.
For more detailed discussion on the parameters $v_{n}$ and their experimental determination, see Refs.\ \cite{Voloshin:2008dg,Sorensen:2009cz,Snellings:2014kwa}.
The event angles are also not measurable directly for each single event but correlations among flow parameters can be determined.
Such flow-plane correlations are argued as a good signal of hydrodynamic evolution \cite{Qiu:2012uy}.

Describing the characteristics of collective dynamics generated from the initial state profile is the central issue of the hydrodynamic approach, however, the other basic bulk behaviors of the data such as identified single particle spectra and the size of emission domain should be reproduced simultaneously.
The size of emission domain is measured from the so-called HBT interferometry of two identical particles. Use of the interferometry of two photons emitted from a stellar surface to measure the size of stars was proposed by Hambury Brown and Twiss in 1954 \cite{Philos.Mag.45.1954.663} and this method was first applied in multiple pion production processes in 1960 by G.\ Goldhaber \textit{et al.} \cite{Goldhaber:1960sf}.
Since then HBT interferometry became one of the important methods in hadronic collisions and in relativistic heavy ion collisions in particular.
For comprehensive reviews, see Refs.\ \cite{Baym:1997ce,Wiedemann:1999qn,Weiner:1999th,Heinz:1999rw,Padula:2004ba,Csorgo:2005gd,Humanic:2005ye,Lisa:2005dd} and the references therein.
The application of HBT interferometry for pions produced in relativistic heavy ion collisions determines three parameters, $R_{\mathrm{long}}$, $R_{\mathrm{side}}$ and $R_{\mathrm{out}}$ which characterize the shape of the emission area of the pions as functions of the average momentum of the pion.
Here, $R_{\mathrm{long}}$ represents the longitudinal radius, $R_{\mathrm{side}}$ the side direction radius, and $R_{\mathrm{out}}$ refers to the radius measuring the optical depth of the emission area.
In contrast to the stellar case, the emission surface is dynamical and there exist other mechanisms which create two-particle correlations, thus, special care should be taken to define these parameters and compare the theoretical values to the data \cite{Pratt:2008qv}.

\section{Overview of the Present Status}

\label{sec:Overview}

A large number of experimental and theoretical studies on collective flow phenomena and extraction of QGP properties have been done in the last 40 years \cite{Kolb:2003dz,Heinz:2009xj,Ollitrault:2008zz,Huovinen:2006jp,Sorensen:2009cz,TeaneyQGP4,Hama:2004rr,WFlorkowski.PhenomenologyOfURHIC}.
In particular, the concept of the formation of sQGP and its subsequent hydrodynamic expansion (\textit{Little Bang} picture) of relativistic heavy ion collisions has been established in the last decade as was announced in the BNL news in 2005 \cite{BNLNews}.
It was argued by Hirano and Gyulassy \cite{Hirano:2005wx} that the observed large elliptic flow $v_{2}$ and consequently almost perfect fluidity of the newly created matter can be understood as a sudden increase of the entropy in the QGP phase contrasted to the hadronic gas, thus it is considered as a signal of deconfinement.

As already pointed out, in the hydrodynamic picture the observed flow pattern should reflect the properties of the initial condition.
Therefore, the hydrodynamic description allows for a more detailed study of the nature of the initial conditions using now measurable EbyE fluctuations in flow pattern and their correlations \cite{Gardim:2011xv,Qiu:2012uy,Gale:2012rq,Luzum:2013yya,Huo:2013qma}.
It has been shown that the behavior of these higher harmonic coefficients as functions of particle mass, centrality and transverse momentum are well reproduced by hydrodynamics.
The hydrodynamic modeling to describe the experimental flow parameters on EbyE basis is considered to be a unique tool to study the initial QCD dynamics of nucleus-nucleus (A--A) collision at ultra-relativistic energies.
Additionally, there are attempts to determine the transport properties of QGP from hydrodynamic analysis.
As mentioned in the Introduction, there are already many recent review articles on these subject written by authors who themselves played important roles in these findings \cite{Hirano:2012kj,Gale:2013da,Huovinen:2013wma,Heinz:2013th,Jia:2014jca,Snellings:2014kwa,Petersen:2014yqa}.
Therefore, we refer readers to these references for the development of the model and its understandings.

\subsection{News from recent LHC and RHIC Data}

RHIC has been exploring nuclear matter at extreme conditions over more than a decade now and a huge amount of data have been collected for collisions of light and heavy nuclei at several energies, ranging from below 10 GeV up to the maximum of 200 GeV per nucleon pair in the center of mass frame.
While the understanding of the experimental data has become deeper and many questions concerning the collective aspects of the evolution of the created matter in such collisions have been answered, many other questions have also emerged.
Subsequent data, in particular from the recent results from LHC and the updated RHIC experiments seem to consolidate more firmly the scenario of Little-Bang.
Similarly to the analysis of correlation measurements in modern observational cosmology \cite{Bucher:2015eia}, the flow pattern characterized in terms of higher Fourier components became possible on a real EbyE basis due to large multiplicities in LHC data.

With the start of the heavy ion program at the LHC in 2010, very precise new data has been measured for collision energies of one order of magnitude higher than the maximum achieved at RHIC.
The rich collection of observables measured by the large and complex experiments running at RHIC and
LHC gives us an unique opportunity to study in detail the aspects of the hot and dense matter created in heavy ion collisions.
In the following we highlight the latest experimental results as well as the relevant constraints they are setting for the theoretical developments.

\subsubsection{Anisotropic Flow}

One of the most important signals of collective behavior during the system evolution in heavy ion collisions is the azimuthal anisotropy of the produced particles.
Initially, all the analyses were concentrated on the elliptic flow parameter, the second harmonic of the Fourier decomposition (see equation \ref{Smooth}). Since non-central collisions form the so-called almond shape in the overlap region of the two incident nuclei, it is reasonable to expect the observation of the elliptic anisotropy in the final state if collectivity is present during the evolution \cite{Ollitrault:1992bk,Huovinen:2006jp,Voloshin:2008dg,Heinz:2009xj,Sorensen:2009cz}.

The measurements of the $v_{2}$ coefficient helped to improve our understanding for the collective aspects in heavy ion collisions.
From the experimental side special efforts have been made to develop techniques to compute flow harmonics from the final particles azimuthal distribution and to extract the possible influences from non flow and fluctuations \cite{Wang:1991qh,Poskanzer:1998yz,Voloshin:2008dg,Borghini:2001vi,Adler:2002pu,Bhalerao:2003xf,Selyuzhenkov:2007zi,Bilandzic:2010jr,Aad:2014vba}.
In the last few years higher order flow harmonics have attracted special attention as important observables of the structures of the initial condition \cite{Abelev:2007qg,Adare:2010ux,Alver:2008zza,Qiu:2011iv,Alver:2010gr,ALICE:2011ab}.
There has been a wealth of recent results of $v_{n}$ reported by RHIC and LHC experiments seen in \cite{Adamczyk:2013waa,Adamczyk:2013gw,Abelev:2010tr,Agakishiev:2011eq,Adamczyk:2012ku,Adamczyk:2013gw,Adare:2010ux,Adare:2011tg,CMS:2013bza,Chatrchyan:2013kba,Chatrchyan:2012ta,Chatrchyan:2012xq,Chatrchyan:2012vqa,Aad:2014vba,Aad:2013xma,ATLAS:2012at,Aad:2014eoa,Abelev:2012di,ALICE:2011ab,Aamodt:2010pa,Abelev:2014pua}.

\begin{figure}[tbh]
\centering
\includegraphics[height=6.2cm,keepaspectratio]{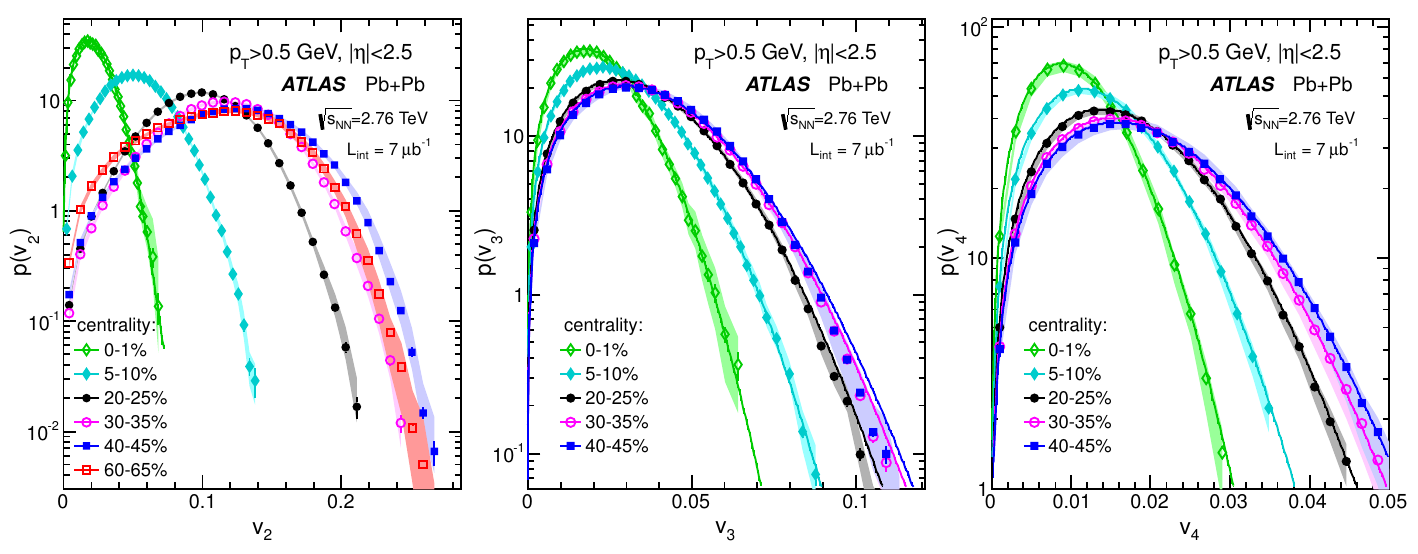}
\caption{Probability density distributions of event-by-event $v_{2}$ (left panel), $v_{3}$ (central panel) and $v_{4}$ (right panel) measured by the ATLAS experiment. Figure\protect\footnotemarkB\ taken from Ref.\ \cite{Aad:2013xma}.}
\label{fig:vn_ebye_dist}
\end{figure}

Regarding the experimental methods used to extract the Fourier coefficients in anisotropic flow analysis, the ATLAS collaboration has recently presented experimental measurements of the EbyE $v_{n}$ distributions  (see Ref.\ \cite{Aad:2013xma} for details).
Such measurements allow to directly extract the mean $\langle v_n \rangle$, the standard deviation $\sigma_{v_n}$, and the relative fluctuation $\sigma_{v_n}/\langle v_n \rangle$ of the $v_n$ distribution, which were previously usually only estimated from second and fourth order cumulants assuming $\sigma_{v_n}\ll \langle v_n \rangle$ \cite{Borghini:2000sa}.
Figure \ref{fig:vn_ebye_dist} shows the probability density distributions of the flow coefficients obtained from EbyE calculations for several centrality windows.
\footnotetextB{Reprinted from G.\ Aad \textit{et al.}, \textit{Measurement of the distributions of event-by-event flow harmonics in lead-lead collisions at $\sqrt{{{s_{NN}}}} $ = 2.76 TeV with the ATLAS detector at the LHC}, \textit{J.\ of High Energy Phys}.\ 2013 (2013) 183, under the terms of the Creative Commons Attribution Noncommercial License. Copyright \textcopyright\ 2013, CERN, for the benefit of the ATLAS collaboration.}
The comparison with results obtained with more traditional methods such as second and fourth-order cumulants ($v_{n}\{2\}$, $v_{n}\{4\}$) and event-plane ($v_{n}\{\textrm{EP}\}$) present a clear systematic ordering between the different techniques, with $v_{n}\{2\} > v_{n}\{\text{EP}\} \geq v_{n}\{\text{EbyE}\} > v_{n}\{4\}$ for the centrality range measured (cf. Fig.\ 10 in Ref.\ \cite{Aad:2014vba}), with $v_{n}\{\text{EbyE}\}$ being the mean value extract from the EbyE $v_n$ distributions. 
Although non-flow correlations may affect more strongly the lower order cumulants (see discussion in \cite{Borghini:2001vi}), the current experimental techniques are able to largely suppress their contribution.
Therefore, the ordering observed may be attributed mainly to flow fluctuations \cite{Miller:2003kd,Drescher:2000ec,Andrade:2006yh,Alver:2006wh,Petersen:2008dd,Flensburg:2011wx}. 
Such effect can be explicitly shown for the second and fourth order cumulants, where the finite variance of the $v_n$ fluctuations contributes positively to $v_n\{2\}$ and negatively to $v_n\{4\}$ \cite{Ollitrault:2009ie}. 

\begin{figure}[tbh]
\centering
\includegraphics[height=6.3cm,keepaspectratio]{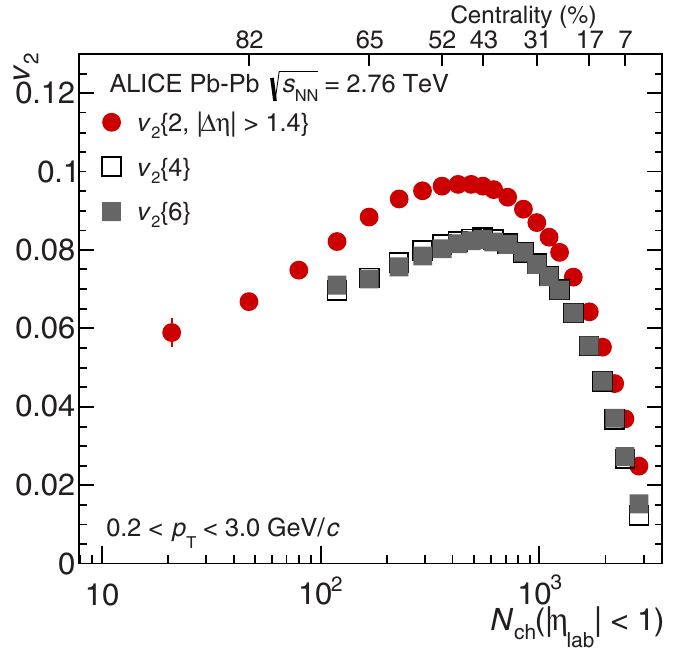}
\includegraphics[height=6.2cm,keepaspectratio]{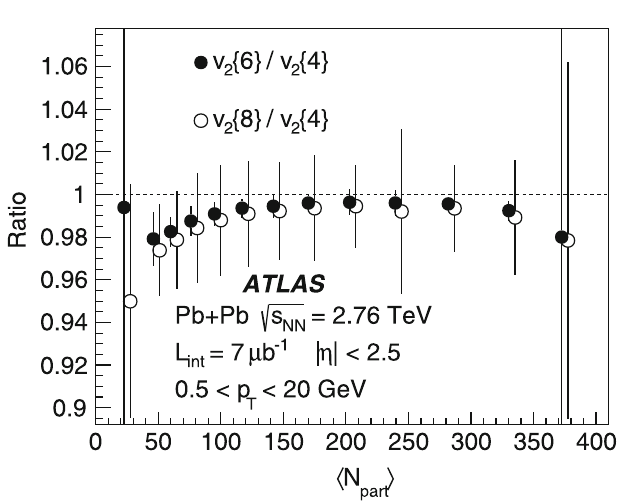}
\caption{Measurements of $v_{2}\{2\}$, $v_{2}\{4\}$ and $v_{2}\{6\}$ from the ALICE\protect\footnotemarkB\ experiment (left panel) and ratios of $v_{2}\{6\}$ and $v_{2}\{8\}$ to $v_{2}\{4\}$ from the ATLAS\protect\footnotemarkB\ experiment (right panel) as a function of centrality in Pb--Pb collisions at $\sqrt{s_{NN}}$ = 2.76 TeV. Figures taken from Refs.\ \cite{Abelev:2014mda} and \cite{Aad:2014vba}, respectively.}
\label{fig:vnFactorisation_ALICE}
\end{figure}

\begin{figure}[htb]
\centering
\includegraphics[height=4.8cm,keepaspectratio]{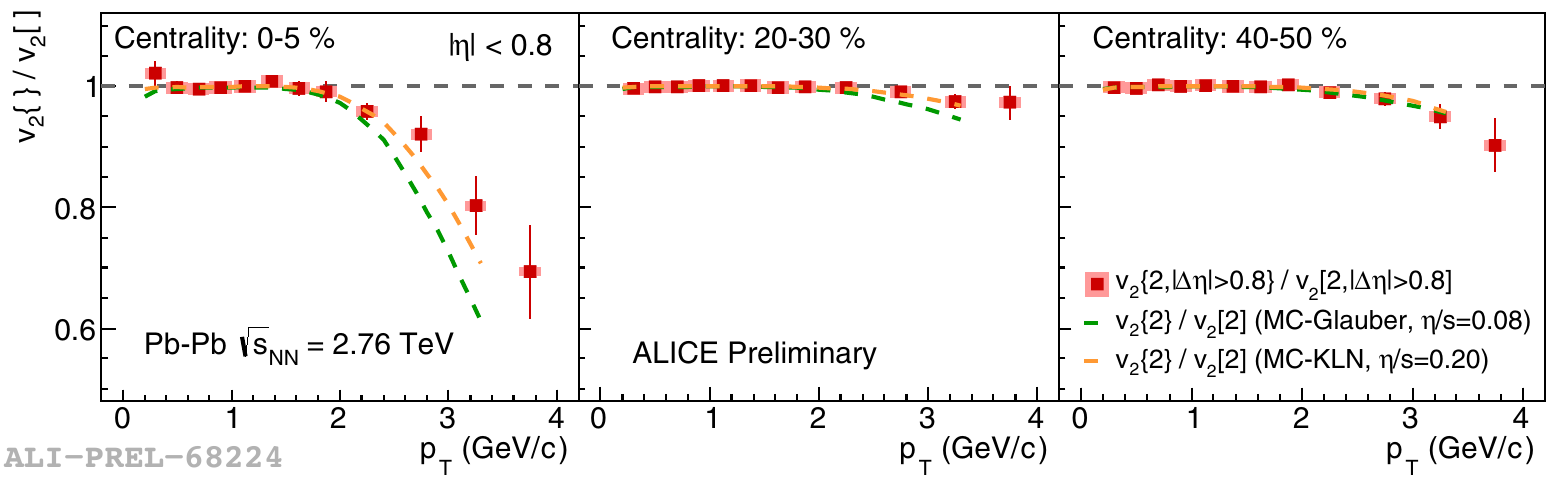}
\caption{(color online) Ratio of $v_{2}\{2\}$ over $v_{2}[2]$ as a function of $p_{T}$ for three centrality
classes measured by the ALICE experiment for Pb--Pb collisions at $\sqrt{s_{NN}}$= 2.76 TeV \cite{Zhou:2014bba}. Comparison with hydrodynamic calculations \cite{Heinz:2013bua} are also shown.
Figure\protect\footnotemarkB\ taken from Ref.\ \cite{Zhou:2014bba}.}%
\label{fig:vnFactorisationRatio_ALICE}%
\end{figure}

Assuming that the development of anisotropic flow is driven by the pressure gradients defined by the structures of the initial density profile, the measurement of flow fluctuations may reveal important properties of the initial state.
However, it is important to remember that the fluctuation spectrum extracted from the measurements of flow coefficients can be distorted by nonlinear effects during the collective dynamic evolution of the system.
The precise connection with the initial fluctuation spectrum can be established only when the non-equilibrium dynamics of the system is understood, but such measurements will help to constrain the developments of initial condition models \cite{Luzum:2013yya}.
In particular, the measurements of flow coefficients using different order cumulants present different sensitivity to flow fluctuations and allows for further exploration of the details of the underlying flow distribution.
Results from PHOBOS \cite{Alver:2010rt} and STAR \cite{Agakishiev:2011eq} Collaborations for Au--Au collisions at $\sqrt{s_{NN}}$ = 200 GeV are consistent with a Bessel-Gaussian functional form for the $v_{2}$ distribution \cite{Voloshin:2007pc}, which implies $v_{2}\{4\} \approx v_{2}\{6\}$.
Furthermore, very precise measurements performed by the ALICE experiment \cite{Abelev:2014mda} also seem to qualitatively follow a Bessel-Gaussian model of flow fluctuations at the LHC (see the left panel in Fig.\ \ref{fig:vnFactorisation_ALICE}).
Although, it was found that $v_{2}\{4\}$ is slightly different than $v_{2}\{6\}$, which raised the possibility of non-Bessel-Gaussian fluctuations, in agreement with the recently proposed power law distribution
\cite{Yan:2013laa}.
The ATLAS Collaboration has also reported measurements of $v_{2}\{6\}/v_{2}\{4\}$ and $v_{2}\{8\}/v_{2}\{4\}$ where significant deviations are observed for most central and peripheral cases (see the right panel in Fig.\ \ref{fig:vnFactorisation_ALICE}) \cite{Aad:2014vba}.
\addtocounter{footnoteB}{-1}\footnotetextB{Reprinted from B.\ Abelev \textit{et al.}, \textit{Multiparticle azimuthal correlations in p-Pb and Pb-Pb collisions at the CERN Large Hadron Collider}, \textit{Phys.\ Rev}.\ C90 (2014) 054901, doi:10.1103/PhysRevC.90.054901, under the terms of the Creative Commons Attribution 3.0 License. Copyright \textcopyright\ 2014, CERN, for ALICE Collaboration.}
\stepcounter{footnoteB}\footnotetextB{Reprinted from G.\ Aad \textit{et al.}, \textit{Measurement of flow harmonics with multi-particle cumulants in Pb+Pb collisions at $\sqrt{s_{\mathrm {NN}}}=2.76$ = 2.76  TeV with the ATLAS detector}, \textit{Eur.\ Phys.\ J}.\ C74 (2014) 3157, under the terms of the Creative Commons Attribution 4.0 License. Copyright \textcopyright\ 2014, CERN, for the benefit of the ATLAS collaboration.}
There have been some extensive investigations on the correlation of the observed flow fluctuations with the initial state profile \cite{Alver:2007qw,Agakishiev:2011eq,DerradideSouza:2011rp,Collaboration:2011yba}
and, as reported by the ATLAS experiment, the eccentricity distributions calculated with Glauber and MC-KLN models \cite{Miller:2007ri,Drescher:2006pi} (see Sec.\ \ref{sec:INIC}) cannot completely describe the observed data.
Another interesting consequence of EbyE fluctuations recently pointed out in Refs.\ \cite{Gardim:2012im,Heinz:2013bua} is the possible factorization breaking of two-particle correlation probability distributions into a product of single-particle distributions.
Figure\ \ref{fig:vnFactorisationRatio_ALICE} shows preliminary\footnote{The effects of non-flow contribution on $v_{2}\{2\}$ and $v_{2}[2]$ are being further investigated.} results of the ratio $v_{2}\{2\}/v_{2}[2]$ for differential ($p_{T}$ dependent)  flow coefficients obtained from Pb--Pb collisions at $\sqrt{s_{NN}}$ = 2.76 TeV by the ALICE experiment \cite{Zhou:2014bba}.
The numerator ($v_{2}\{2\}$) in the above ratio is the usual second order cumulant flow coefficient (see for instance Refs.\ \cite{Borghini:2000sa,Borghini:2001vi,Bilandzic:2010jr}), in which only one of the particles is taken from the $p_{T}$ bin of interest, while the denominator ($v_{2}[2]$) is computed by taking both particles from the $p_{T}$ bin of interest \cite{Heinz:2013bua}.
The factorization breaking (deviations from unity) as a function of the transverse momentum is shown to be the result of decoherence in the angular correlations induced by initial fluctuations and is predicted even for pure hydrodynamic calculations \cite{Gardim:2012im}.
The intensity of the factorization breaking observed in the experimental data is thus of special interest in constraining the initial state and transport properties of different model calculations \cite{Heinz:2013bua}.

With the latest experimental results the interpretation and understanding of the enormous amount of information contained in particle correlation analysis have become a very challenging task.
Therefore, newer and more sophisticated techniques need to be developed in order to help us disentangle the true meanings of our measurements.
In this sense, a very recent work by Bhalerao \textit{et al.} \cite{Bhalerao:2014mua} proposed the application of the statistical technique of principal component analysis (PCA) to study EbyE fluctuations in two-particle correlations (2PC).
\footnotetextB{Reprinted with permission from Y.\ Zhou (for the ALICE Collaboration), \textit{Searches for $p_{\mathrm{T}}$ dependent fluctuations of flow angle and magnitude in Pb-Pb and p-Pb collisions}, \textit{Nucl.\ Phys}.\ A 931 (2014) 949. Copyright \textcopyright\ 2014 The Authors.}
The method approximates the Fourier coefficients of the pair distribution, $V_{n\Delta}$, as a sum of flow fluctuation components (modes),
\begin{equation}
V_{n\Delta}(p_{1},p_{2}) \approx\sum_{\alpha=1}^{k} V_{n}^{(\alpha)}(p_{1})
V_{n}^{(\alpha) *}(p_{2}),\label{eq:Vn_PCA}
\end{equation}
where the coefficients $V_{n\Delta}$ are determined by the statistics of the single-particle complex Fourier coefficient, $V_{n\Delta} (p_{1},p_{2}) = \langle V_{n} (p_{1}) V_{n}^{*}(p_{2}) \rangle$, and $p_{i}$ is a shorthand notation for the coordinates $p_{T}$ and $\eta$ \cite{Bhalerao:2014mua}.
In the absence of flow fluctuations the left hand side of Eq.\ (\ref{eq:Vn_PCA}) factorizes, i.e. only $\alpha=1$ is present, thereby recovering the usual anisotropic flow.
One of the advantages of this new method is the ability to make use of all the information contained in 2PC, therefore, allowing for a more robust observable to study flow fluctuations. Indeed, it has already been shown that previously unknown subleading modes in both rapidity and transverse momentum are found to be present in data measured by the ALICE experiment \cite{Bhalerao:2014mua,Mazeliauskas:2015vea}, which in turn are argued to be responsible for factorization breaking as observed in two-particle correlation analysis \cite{Aamodt:2011by,Gardim:2012im,Heinz:2013bua}.

\subsubsection{Two-Particle Correlations}

The study of two-particle correlations (2PC) has become one of the main topics
of investigation in the last years.
The general idea consists of constructing a correlation function $C(\Delta\phi,\Delta\eta)$ by taking the
distribution of pairs relative differences $\Delta\phi$ and $\Delta\eta$
between a \textit{trigger} and an \textit{associated} particle in the ``same''
event and then dividing by the distribution of pairs of particles from
``mixed'' events, in which the \textit{trigger} is taken from one event and
the \textit{associated} is taken from a similar but different event, which
thus cancels all the effects due to detector acceptance and efficiency. Given
the desired ranges for the transverse momentum of the trigger and associated
particles, the correlation function can be written as,
\begin{equation}
C(\Delta\phi,\Delta\eta) \equiv\frac{N_{\mathrm{mixed}}}{N_{\mathrm{same}}}
\times\frac{N_{\mathrm{same}}(\Delta\phi,\Delta\eta)}{N_{\mathrm{mixed}%
}(\Delta\phi,\Delta\eta)}.
\end{equation}

As a matter of fact, we recall the observation of structures such as the
long-range near-side ($\Delta\phi\sim0$) \textit{ridge} and the away-side
($\Delta\phi\sim\pi$) double-hump in the $\Delta\eta\times\Delta\phi$
distributions obtained from Au--Au collisions at RHIC
\cite{Abelev:2009af,Alver:2009id}. Likewise, recent
measurements at LHC also confirm the observation of such structures in Pb--Pb
collisions \cite{ATLAS:2012at,Aamodt:2011by,CMS:2013bza}.

The complete understanding of the origin of such structures is still under
intense investigation. Initially, the proposed interpretations were based on
the medium response and conical flow induced by jet quenching \cite{CasalderreySolana:2004qm,Armesto:2004pt,Shuryak:2007fu,Stoecker:2004qu} as well as
longitudinal flux tubes in the initial state
\cite{Dumitru:2008wn,Wong:2008yh}. However, it was later shown that
EbyE hydrodynamic simulations with inhomogeneous initial condition
could produce similar structures \cite{Takahashi:2009na} suggesting
that the ridge and the double-hump structures could be explained by the
presence of higher order flow harmonics
\cite{Luzum:2010sp,Alver:2010gr,Teaney:2010vd,Sorensen:2010zq}%
, which appears naturally as the odd harmonics contribute constructively at
$\Delta\phi\sim0$ and destructively at $\Delta\phi\sim\pi$. The Fourier
coefficients in this case are given by,
\begin{equation}
V_{n\Delta} = \langle\cos n \Delta\phi\rangle= \frac{\sum_{m=1}^{N}
\cos(n\Delta\phi_{m})C(\Delta\phi_{m})}{\sum_{m=1}^{N} C(\Delta\phi_{m})},
\end{equation}
where the correlation function $C(\Delta\phi)$ is scaled by the respective
number of pairs in the mixed and same events distributions of the projected
$\Delta\eta$ slice.

\begin{figure}[tbh]
\centering
\includegraphics[height=6.2cm,keepaspectratio]{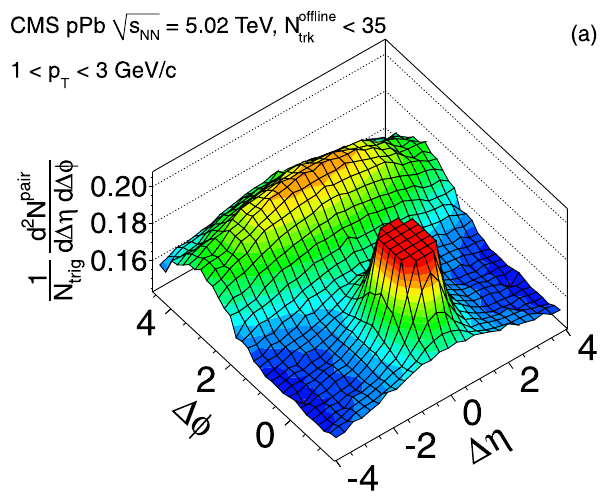}
\includegraphics[height=6.2cm,keepaspectratio]{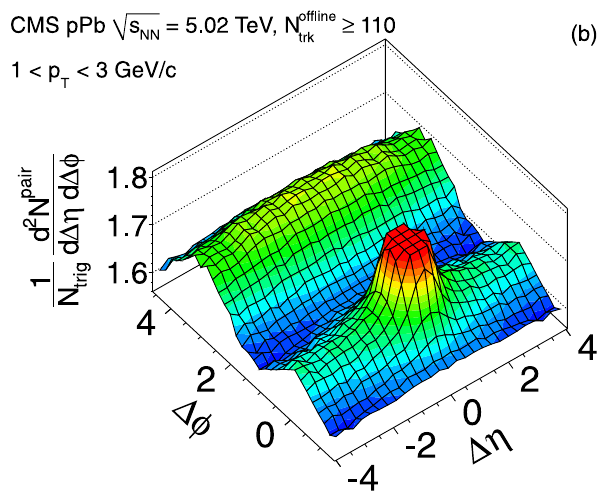}\caption{$\Delta
\eta\times\Delta\phi$ distributions for charged particle pairs measured by the
CMS experiment for low (left panel) and high (right panel) multiplicity events
of p--Pb collisions at $\sqrt{s_{NN}}$ = 5.02 TeV. Figures\protect\footnotemarkB\ taken from Ref.\ \cite{CMS:2012qk}.}%
\label{fig:CMS_2PCpPb}%
\end{figure}

In the earlier studies of 2PC performed at RHIC for p--p
and d--Au collisions, the observed results were very similar
\cite{Adams:2003im} and the structures observed for Au--Au
collisions were not present. Therefore, these results were commonly taken as
reference, where no QGP medium was formed, to compare with Au--Au collisions
data. However, recent measurements at the LHC for high multiplicity events in
p--Pb \cite{CMS:2012qk,Aad:2012gla,ABELEV:2013wsa} and
even in p--p \cite{Khachatryan:2010gv,ATLAS:2012ap,ATLAS:2012as} collisions
have indicated the appearance of the same structures usually observed only in
A--A collisions. Figure \ref{fig:CMS_2PCpPb} shows a comparison between the
two-dimensional distributions of charged particle pairs measured by the CMS
experiment for low (left panel) and high (right panel) multiplicity events
\cite{CMS:2012qk}, where the ridge-like shape can be clearly observed
in the second case.

\footnotetextB{Reprinted from S.\ Chatrchyan \textit{et al.}, \textit{Observation of long-range, near-side angular correlations in pPb collisions at the LHC}, \textit{Phys.\ Lett}.\ B718 (2013) 795, under the terms of the Creative Commons Attribution 4.0 License. Copyright \textcopyright CERN, on behalf of the CMS Collaboration.}

\begin{figure}[tbh]
\centering
\includegraphics[height=11cm,keepaspectratio]{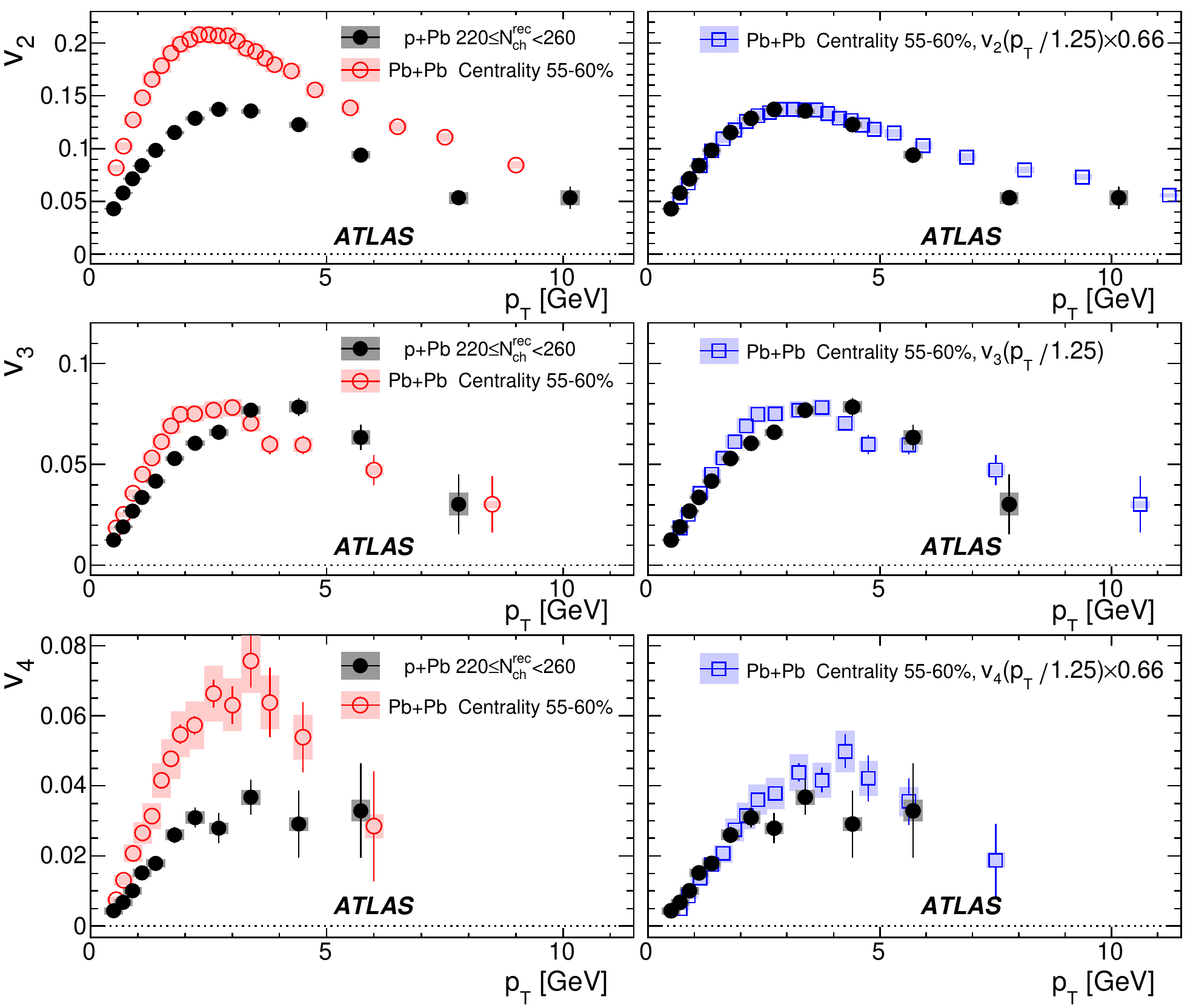}\caption{$v_{n}
$ coefficients as functions of $p_{T}$ for Pb--Pb collisions at $\sqrt{s_{NN}}$ = 2.76 TeV and p--Pb collisions at $\sqrt{s_{NN}}$ = 5.02 TeV. Figure\protect\footnotemarkB\ taken from Ref.\ \cite{Aad:2014lta}.}
\label{fig:2PCvn_pPb_ATLAS}
\end{figure}

In addition to the already surprising similarities between the measurements of peripheral Pb--Pb collisions and the high multiplicity selection in p--Pb collisions for the 2-dimensional $\Delta\eta\times\Delta\phi$ correlation function, comparisons of the first few coefficients extracted from Fourier decomposition show remarkable consistence in shape between the cases.
Figure\ \ref{fig:2PCvn_pPb_ATLAS} shows the comparisons between the results obtained by the ATLAS experiment for peripheral 55-60\% Pb--Pb collisions at $\sqrt{s_{NN}}$ = 2.76 TeV and high multiplicity p--Pb collisions at $\sqrt{s_{NN}}$ = 5.02 TeV \cite{Aad:2014lta}.
The plots on the left show the original values extracted from each analysis and the plots on the right show the results extracted from Pb--Pb collisions rescaled both vertically, by an empirical factor of $0.66$, and horizontally, by a factor of $1.25$ as proposed in Ref.\ \cite{Basar:2013hea} to take into account the
difference between the $\langle p_{T} \rangle$ values in the two systems.
\footnotetextB{Reprinted from G.\ Aad \textit{et al.}, \textit{Measurements of long-range pseudorapidity correlations and azimuthal harmonics in $\sqrt{s_{NN}}$ = 5.02 TeV proton-lead collisions with the ATLAS detector}, \textit{Phys.\ Rev}.\ C90 (2014) 044906, doi:10.1103/PhysRevC.90.044906, under the terms of the Creative Commons Attribution 3.0 License. Copyright \textcopyright\ 2014 CERN, for the ATLAS Collaboration.}

Following the observations made by the LHC experiments for the p--p and p--Pb
collisions, the PHENIX collaboration presented new $v_{2}$ measurements for
central d--Au collisions at $\sqrt{s_{NN}}$ = 200 GeV compatible with the
values reported for p--Pb collisions at $\sqrt{s_{NN}}$ = 5.02 TeV
\cite{Adare:2014keg}.

The unavoidable question is whether the dynamics governing the evolution of the system in each case, namely A--A, p--A and p--p collisions, is similar or not.
If there is collectivity in p--A and p--p collisions, can hydrodynamics still be applied to such small systems?
Indeed, elliptic flow measurements of identified hadrons in p--Pb collisions by the CMS \cite{Khachatryan:2014jra} experiment seem to be consistent with the quark number scaling previously observed in A--A collisions \cite{Adams:2004bi,Adare:2006ti,Abelev:2010tr,Adamczyk:2013gw}.
Thus, these newly produced experimental results have undoubtedly imposed serious challenges to our theoretical understandings.
The assumptions generally used in the hydrodynamic approach for A--A collisions, namely the thermalization time, local thermal equilibration and coarse-graining scale, may need to be revised if one wants to describe small systems such as p--A and p--p.
At the same time, it is a great opportunity to probe the edges between the two worlds usually described independently with macroscopic and microscopic theories (see interesting discussion in Ref.\ \cite{Shuryak:2013ke}).

\subsubsection{Direct Photons and Pion Femtoscopy Measurements}

Another important experimental probe of the properties of the hot and dense medium created in relativistic hadronic collisions is the measurement of direct photons.
Since photons are not affected by strong interactions they can emerge from the fireball almost undisturbed while carrying information about their emission source.

In heavy ion collisions, direct photons can be classified as \textit{prompt photons}, which are mainly produced by leading order pQCD processes in the very initial stages, and \textit{thermal photons}, emitted by radiation of the medium, both in partonic and subsequent hadronic phases.
The thermal photons spectrum should therefore depend on the temperature of the respective production source \cite{Conesa:2007gr,Wilde:2012wc}.
Recent hydrodynamic modeling suggests the possibility of extracting information of the early hydrodynamic evolution of the system (see Ref.\ \cite{Shen:2013vja} and references therein).

Experimentally, the measurement of direct photons is a very complex task that requires careful identification and removal of photons from hadron decays in order to isolate the direct contribution \cite{Adare:2014fwh}.
Thus, experimental techniques have had to develop over detector upgrades and experiments at RHIC and LHC have now been able to present crucial direct photons measurements.

\begin{figure}[ptb]
\centering
\includegraphics[height=6.4cm,keepaspectratio]{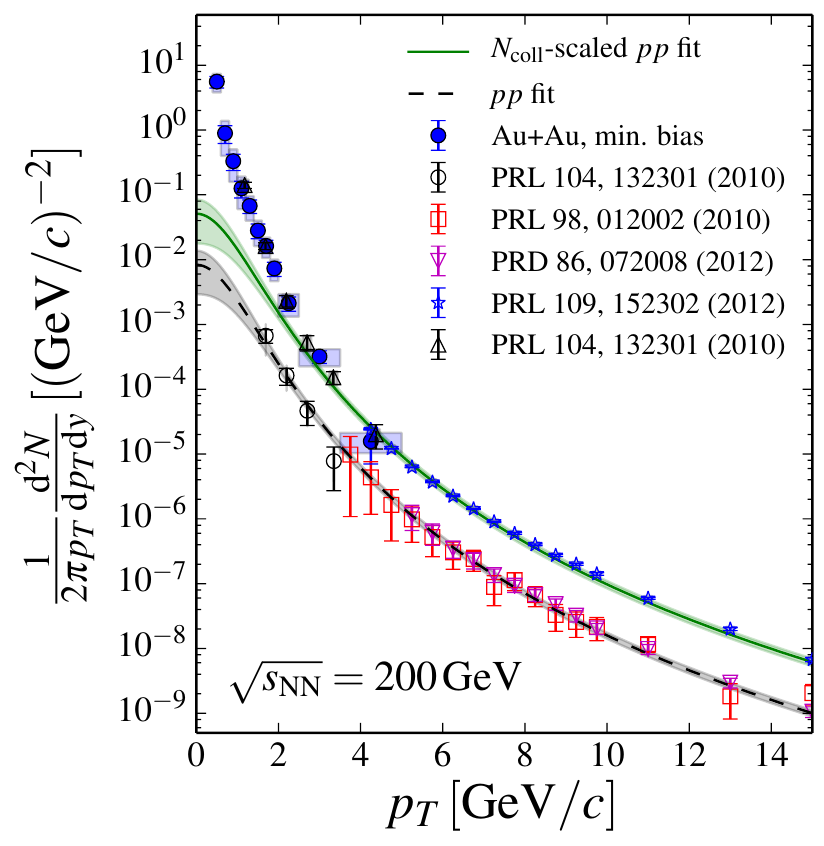}
\includegraphics[height=6.45cm,keepaspectratio]{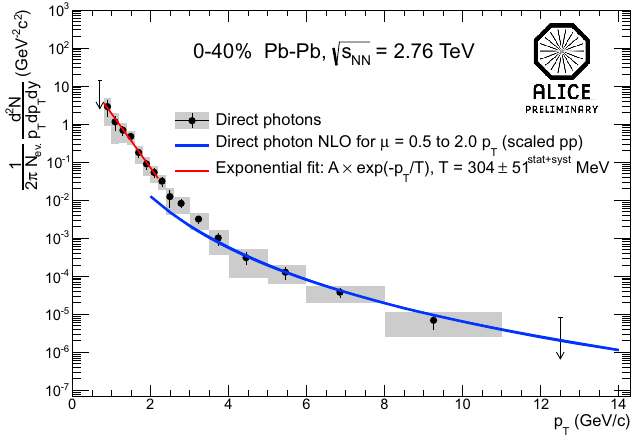}
\caption{$p_{T}$ spectra of direct photons measured by the PHENIX\protect\footnotemarkB\ experiment for Au--Au and p--p collisions at $\sqrt{s_{NN}}$ = 200 GeV \cite{Adare:2014fwh} (left panel) and results from the ALICE\protect\footnotemarkB\ experiment for Pb--Pb collisions at $\sqrt{s_{NN}}$ = 2.76 TeV \cite{Wilde:2012wc} (right panel). Figures taken from Refs.\ \cite{Adare:2014fwh} and \cite{Wilde:2012wc}, respectively.}
\label{fig:direct_photons_PHENIX_ALICE}
\end{figure}

The PHENIX experiment at RHIC recently observed an exponential excess in direct photon transverse momentum yields for Au--Au collisions compared to p--p collisions \cite{Adare:2008ab,Adare:2014fwh}.
The left panel in Fig.\ \ref{fig:direct_photons_PHENIX_ALICE} shows the combined $p_{T}$ spectra for both p--p and Au--Au collisions at $\sqrt{s_{NN}}$ = 200 GeV.
\addtocounter{footnoteB}{-1}\footnotetextB{Reprinted with permission from A.\ Adare \textit{et al.}, \textit{Centrality dependence of low-momentum direct-photon production in Au$+$Au collisions at $\sqrt{s_{_{NN}}}=200$ GeV}, arXiv:1405.3940 (2014).}
\stepcounter{footnoteB}\footnotetextB{Reprinted with permission from M.\ Wilde (for the ALICE Collaboration), \textit{Measurement of Direct Photons in pp and Pb--Pb Collisions with ALICE}, \textit{Nucl.\ Phys}.\ A904-905 (2013) 573c, Copyright \textcopyright\ 2013 CERN.}
The excess obtained from the difference between the Au--Au results and the $N_{\mathrm{coll}}$-scaled p--p results is well described by an exponential distribution with an average inverse slope of 240 MeV in the $p_{T}$ range from $0.6$ to $2$ GeV/$c$ \cite{Adare:2014fwh}.
In a similar analysis, the ALICE experiment at LHC also observed an excess of direct photons in Pb--Pb collisions at $\sqrt{s_{NN}}$ = 2.76 TeV with respect to NLO pQCD predictions scaled by $N_{\mathrm{coll}}$ \cite{Wilde:2012wc} (see right panel in Fig.\ \ref{fig:direct_photons_PHENIX_ALICE}).
The inverse slope parameter extracted from the exponential fit in the $p_{T}$ range from $0.8$ to $2.2$ GeV/$c$ is on the order of 300 MeV \cite{Wilde:2012wc}.

In a hydrodynamic picture, these results can be interpreted as the effective temperature of a thermalized source convoluted over its whole time evolution.
In both cases, i.e. $T_{\mathrm{RHIC}} \sim240$ MeV and $T_{\mathrm{LHC}} \sim 300$ MeV, the values are well above the phase transition temperature $T_{C} $ of approximately 170 MeV predicted by lattice QCD calculations for the transition to the QGP phase \cite{Wilde:2012wc}.
Such high values of temperature may indicate the production of direct photons at very early times during the system evolution \cite{Lohner:2012ct}.
On the other hand, PHENIX and ALICE experiments have also reported measurements of significant elliptic flow of direct photons in the low-$p_{T}$ region for collisions of Au--Au at $\sqrt{s_{NN}}$ = 200 GeV \cite{Adare:2011zr} and Pb--Pb at $\sqrt{s_{NN}}$ = 2.76 TeV \cite{Lohner:2012ct}, respectively.
While one would expect a small elliptic flow if the direct photons are emitted predominantly at the very early stages, since flow develops towards the final stages the observed large values of $v_{2}$ also support relevant photon production later in the evolution.
Therefore, the simultaneous description of the above observations has become truly challenging to the current theoretical developments (see Ref.\ \cite{Gale:2012xq}).

Comparisons from predictions of EbyE hydrodynamic calculations \cite{Chatterjee:2013naa} as well as from microscopic dynamic model simulations using the PHSD model \cite{Linnyk:2013hta} have been presented for ALICE and PHENIX results, respectively.
As shown in these works, the initial condition fluctuation in EbyE hydrodynamic calculations seems to play an important role in the final direct photon $v_{2}$ \cite{Chatterjee:2013naa}.
Investigations using a (3+1)D hydrodynamic model which takes into account viscous corrections suggest that the flow of direct photons may be also especially sensitive to the equation of state and the transport properties of the system \cite{Dion:2011pp}.
Furthermore, it is worth noticing the excellent agreement of the predictions given by the PHSD model with respect to the PHENIX measurements (cf.\ Fig.\ 11 in Ref.\ \cite{Linnyk:2013hta}).
Within the approach employed in this model, the photons produced during the QGP phase account for slightly less than 50\% of the observed spectrum and have small $v_{2}$ (on the order of 2-3\% only), whereas the main contribution to the large $v_{2}$ comes from hadronic channels.
These channels, however, indirectly require the strong interactions of the partonic phase to effectively build up the observed asymmetry of the final particle momentum distribution.
In another recent study employing the hydrodynamic framework \cite{Shen:2015qba} it was suggested that the precise measurement of direct photons in p--A collisions may furnish relevant information on the fluid dynamic profile at early times, in particular, concerning the possible formation of a QGP phase.

\begin{figure}[htb]
\centering
\includegraphics[height=5.5cm,keepaspectratio]{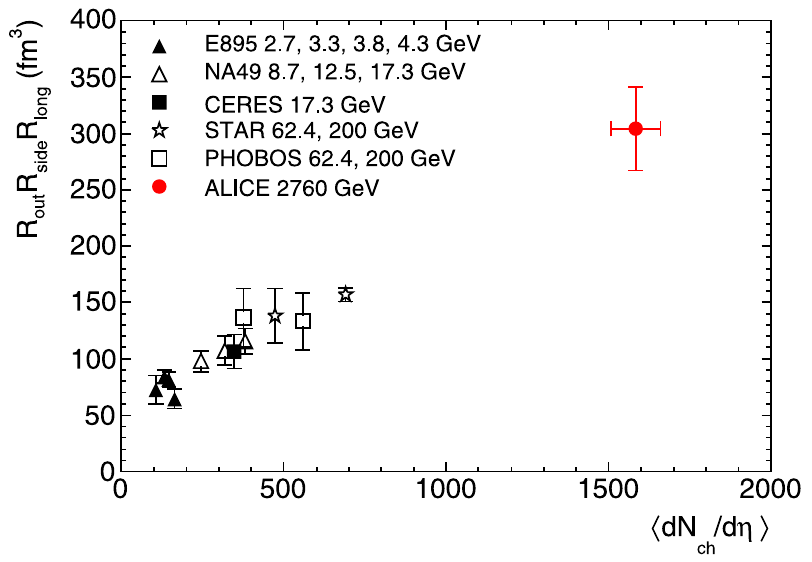}
\includegraphics[height=5.5cm,keepaspectratio]{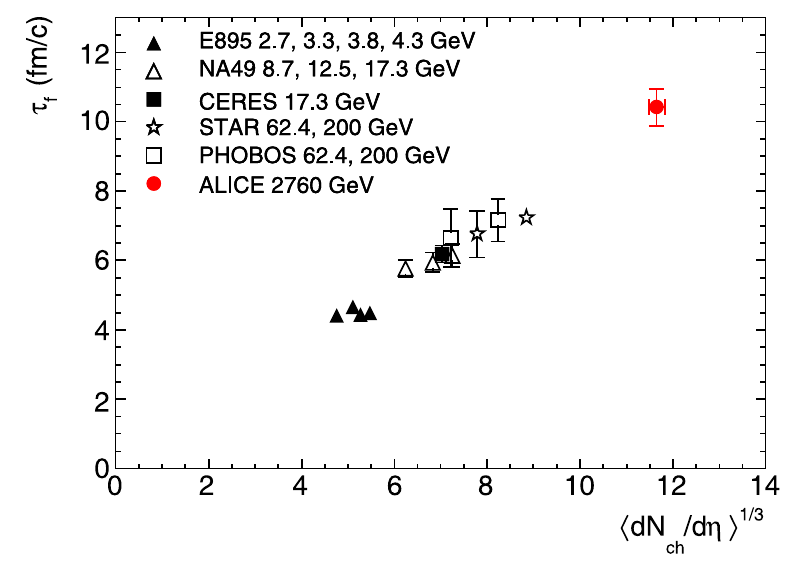}
\caption{Product of HBT radii extracted from pion correlations at $k_{T}$ = 0.3 GeV/$c$ (left panel) and the decoupling time extracted from $R_{\mathrm{long}}(k_{T})$ (right panel) as functions of the charged particle multiplicity for several collision energies. Figures\protect\footnotemarkB\ taken from Ref.\ \cite{Aamodt:2011mr}.}
\label{fig:HBT_ALICE}
\end{figure}

Finally, femtoscopy measurements based on quantum-statistical interferometry of identical bosons as proposed by Hanbury Brown and Twiss \cite{Philos.Mag.45.1954.663,Nature178.1046-1048}, sometimes also called HBT correlations, bring additional constraints to the modeling of the dynamic evolution of the system produced in high energy collisions.
As discussed previously, the main goal of this experimental technique is to infer about the space-time scales of the emitting source \cite{Baym:1997ce,Wiedemann:1999qn,Lisa:2005dd}, which allows one to extract information about the duration of expansion and the system size at the kinetic FO. The details of the analysis performed on the experimental data can be found in the Refs.\ \cite{Adams:2004yc,Abelev:2009tp,Abelev:2013pqa,Aamodt:2011mr}.
Figure \ref{fig:HBT_ALICE} shows the dependence of the product of the HBT radii (left panel) and the decoupling time $\tau_{f}$ (right panel) with the charged particle multiplicity for several femtoscopy measurements at different collision energies.
The later is an estimated quantity obtained by assuming that $R_{\mathrm{long}}$ is proportional to duration of the longitudinal expansion and, therefore, to the decoupling time of the system (see Ref.\ \cite{Aamodt:2011mr} for more details).
The common linear behavior shown in this figure, that extends over such a large range in collision energies (also reported in Ref.\ \cite{Abelev:2009tp}), is consistent with a universal condition for pion FO \cite{Adamova:2002ff} and establishes important constraints for model developments.
Extended measurements using three pion correlations \cite{Abelev:2013pqa} and the cumulant expansion \cite{Abelev:2014pja} have also been recently reported by the ALICE collaboration for Pb--Pb collisions at $\sqrt{s_{NN}}$ = 2.76 TeV as well as for p--Pb at $\sqrt{s_{NN}}$ = 5.02 TeV and p--p at$\sqrt{s_{NN}}$ = 7 TeV.
These improved measurements restrict even further the model developments.
In particular, as concluded in Ref.\ \cite{Abelev:2014pja}, the results obtained seem to be consistent with predictions from CGC initial conditions (IP-Glasma) without a hydrodynamic evolution, which may be key to understand the role between the initial condition and the collective expansion in such small systems.
\footnotetextB{Reprinted with permission from K.\ Aamodt \textit{et al.}, \textit{Two-pion Bose-Einstein correlations in central Pb--Pb collisions at $\sqrt{s_{NN}}$ = 2.76 TeV}, \textit{Phys.\ Lett}.\ B696 (2011) 328. Copyright \textcopyright\ 2010 CERN.}
This is however still controversial, inasmuch as some hydrodynamic calculations are shown to reproduce the observed collective flow nature in p-Pb \cite{Kozlov:2014hya,Kozlov:2014fqa,Werner:2013ipa}.

\section{More about on Relativistic Hydrodynamics}

\label{sec:MoreAboutHydro}

As discussed until now, the success of hydrodynamic modeling in nucleus-nucleus collisions is overwhelming and hydrodynamics is now one of the fundamental tools for the study of the initial state and transport properties of QCD matter in terms of experimentally observed collective flow parameters.
In spite of these successes, one should be aware of the fact that there still exist several questions to be clarified within the hydrodynamic approach used in relativistic heavy ion collisions.

\begin{figure}[tbh]
\centering
\includegraphics[height=6.4cm,keepaspectratio]{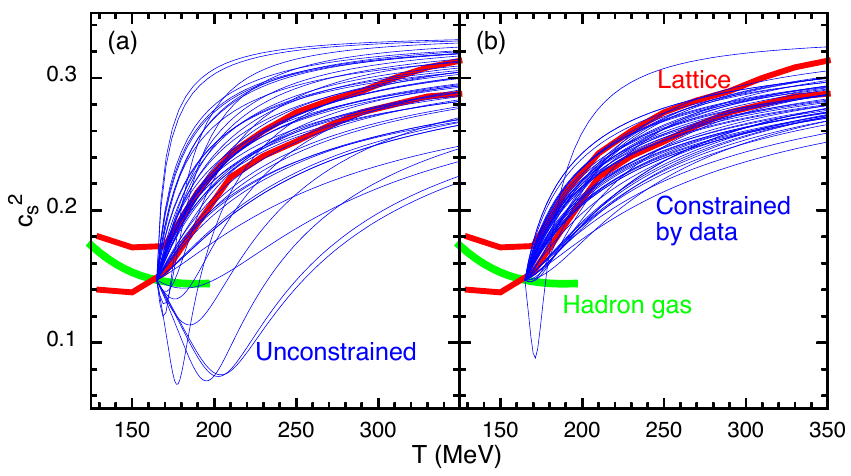}
\caption{(color online) Equation of state calculations from lattice QCD and from models unconstrained (left panel) and constrained (right panel) by data from RHIC and LHC. Figure\protect\footnotemarkB\ taken from \cite{Pratt:2015zsa}.}
\label{fig:EoS_SPratt}
\end{figure}

First, the hydrodynamic fit to experimental data is not unique.
There are still model-dependent systematic uncertainties.
In a recent study, S.\ Pratt \textit{et al.} \cite{Pratt:2015zsa} proposed a method to constrain in a global manner the EoS in hydrodynamic modeling using the conditions imposed by observables, such as particle spectra, HBT radii, and elliptic flow coefficients.
In Fig.\ \ref{fig:EoS_SPratt} we see that the overall shape of the EoS is considerably constrained but there is still a large class of different EoS can be equivalently good with respect to the observables selected.
Of course, not all the constraints available from the data have been considered in this example, such as the higher order Fourier flow components, and the inclusion of additional information may constrain the EoS even further, which makes such kind of analysis very promising based on the state-of-art statistical methods.
But at the same time, in the present analysis, no effects of viscosities, especially their temperature dependence, are taken into account.
Also, several ambiguities which come from the initial condition and the final state hadronic cascade part are absent.
Therefore, we still cannot conclude that the hydrodynamic representation is uniquely determined at the present stage.

Second, according to the Little Bang scenario, in the central rapidity region a hot strongly coupled QGP is formed at a very early time after the collision ($\sim$$10^{-1}$ fm) in almost thermal equilibrium.
Unfortunately we do not know yet what is the exact mechanism behind the thermalization process though several approaches have been proposed \cite{Geiger:1992si,Baier:2000sb,Kharzeev:2005qg,Mueller:2005un,Epelbaum:2014xea,Berges:2013fga}.
Furthermore, the success of the hydrodynamic description does not necessarily imply that the matter in question is in LTE.
For a peculiar example, we may consider an ideal gas of massless particles.
When the system is locally isotropic in the momentum space, then independently of the momentum distribution, the energy-momentum tensor becomes exactly that of an ideal fluid with the ``EoS'', $\varepsilon=3p$ in LRF.
See also Ref.\ \cite{Noronha:2015jia}. 

In addition, at the present stage the great success of the hydrodynamic approach is mainly concentrated in the mid-rapidity region of ultra-relativistic collisions where the baryonic chemical potential is close to zero, i.e., where the matter created from the vacuum (converted from the incident energy) is dominant.
When we extend the hydrodynamic description to the non-central rapidity domain or to lower energies to describe the BES at RHIC, CBM at FAIR, and NICA programs, one needs a hydrodynamic model which accounts for the high baryonic current and the corresponding EoS at finite baryonic chemical potential.
Furthermore, in the domain where the baryonic charge current becomes dominant we expect significant effects of non-equilibrium since these currents strongly preserve the information about the initial incident state.
In order to discuss quantitatively these situations, one still has to clarify the validity and the resolution power of the hydrodynamics.
\footnotetextB{Reprinted with permission from S.\ Pratt \textit{et al.}, \textit{Constraining the Equation of State of Superhadronic Matter from Heavy-Ion Collisions}, \textit{Phys.\ Rev.\ Lett}.\ 114 (2015) 202301. Copyright \textcopyright\ 2015 American Physical Society.}

In this section we first discuss several approaches to relativistic viscous hydrodynamics and the meaning of the physical assumptions behind them.
In particular, we discuss the concept of C-G hidden in the final form, which is essential to understand the resolution power of hydrodynamic approaches when we use them as a tool to extract the initial condition of energy and momentum distributions from the observed flow profiles.

\subsection{Different formulations of relativistic dissipative hydrodynamics}

\label{sec:DifferentFormulations}

There are several approaches used in the formulation of relativistic dissipative hydrodynamics and, differently from the ideal case, the resulting equations are not unique.
Thus, the theory of dissipative hydrodynamics is not yet completely established as we will see below.

\subsubsection{Macroscopic approaches}

In section \ref{sec:causal}, we briefly discussed the phenomenological derivation employed by Israel and Stewart.
In that case, the usual thermodynamics is extended to include $\Pi$, $\pi^{\mu\nu}$, and $\nu^{\mu}$ as additional thermodynamic variables.
Once the number of thermodynamic variables is changed, various thermodynamic laws are affected.
However, there is no established theory for such a generalization of thermodynamics to include irreversible currents.

One example of such a theory is called extended irreversible thermodynamics (EIT) \cite{Rep.Prog.Phys.62.1035}.
In this theory, the first law of thermodynamics near equilibrium is generalized to be
\begin{equation}
T dS = dE + PdV - \mu dN - \frac{\zeta}{\tau_{\Pi}} \Pi d(\Pi V) - \frac{\eta
}{\tau_{\pi}} \pi_{\mu\nu} d(\pi^{\mu\nu}V) - \frac{\kappa}{\tau_{\kappa}}
\nu_{\mu}d(\nu^{\mu}V).
\end{equation}
By calculating the entropy production with this new equation as is done in the linear irreversible thermodynamics, one obtains evolution equations for viscosities.
The results are the same as those in Eq.\ (\ref{Eq.relNSF}).
The internal-variable theory (IVT) proposes another generalization of thermodynamics, but the essential structure of the derivation is the same as in EIT \cite{Ciancio1991}.

In the examples above, $\Pi$, $\pi^{\mu\nu}$, and $\nu^{\mu}$ are considered to be additional thermodynamic variables.
However, this is not the only way to extend thermodynamics. 
As an example, let us consider the divergent-type theory (DTT) \cite{Geroch:1990bw,Heinson:1995xw,Calzetta:1997aj} where thermodynamics is obtained from 
\begin{equation}
T dS^{\mu}= - \mu dN^{\mu}- u_{\nu}dT^{\mu\nu} - T \xi_{\nu\delta} dA^{\mu
\nu\delta},
\end{equation}
where $\xi_{\nu\delta}$ and $A^{\mu\nu\delta}$ are additional thermodynamic variables.
By using these variables, the viscous tensor is expressed as,
\begin{align}
\pi^{\mu\nu} = \eta\xi^{\mu\nu} - \frac{\lambda_{1} \tau_{\pi}T^{4}}{3\eta}
\left(  \xi^{\mu\alpha} \xi_{\alpha}^{\nu}- \frac{1}{3} \Delta^{\mu\nu}
\xi^{\alpha\beta}\xi_{\alpha\beta} \right)  .
\end{align}
The evolution equation of $\xi^{\mu\nu}$ is induced by the velocity gradient as is the case of the shear viscosity.
Another phenomenological method which can be used to derive hydrodynamics is known as GENERIC.
We refer the reader to the Refs.\ \cite{Grmela:1997zz,Ottinger2005} for further details.

There is an important property in phenomenological derivations which is known as Curie's principle.
In linear irreversible thermodynamics \cite{groot1962non}, irreversible currents with different tensor properties cannot be coupled.
In hydrodynamics, we consider three thermodynamic forces,  $\partial_{\mu} u^{\mu}, \partial^{\mu} u^{\nu}$, and  $\partial^{\mu} (\mu/T) $ but these terms are not consider to mix.
For example, we can consider a vector current formed by a linear combination of thermodynamic forces such as 
\begin{equation}
 F^\mu \partial_{\nu} u^{\nu} + G \partial^{\mu} \frac{\mu}{T},
\end{equation}
where $F^{\mu}$ and $G$ are the vector and scalar Onsager coefficients, respectively.
This is obviously a mathematically covariant vector but its inclusion is not allowed from Curie's principle, where any thermodynamic force of a given rank tensor can induce only irreversible currents of lower or equal rank. 
In non-relativistic fluids, this principle is experimentally established within linear irreversible thermodynamics. 

The applicability of Curie's principle in relativistic and/or non-linear regimes is not clear, but assuming that it can be employed, one can reduce the number of terms which can possibly appear just based on Lorentz invariance.
For example, we assume that $Q^{\mu}$ is given by Eq.\ (\ref{eq:Qmu}) in the derivation of the Israel-Stewart theory \cite{Israel:1979wp}.
Therefore, it is possible to extend the form of $Q^{\mu}$ to include the cross terms $\Pi$ and $\nu^{\mu}$.
In fact, such terms are considered in the original argument by Israel and Stewart.
However, those terms lead to a coupling between $\Pi$ and $\nu^{\mu}$, which is then forbidden by Curie's principle.

\subsubsection{Mesoscopic approaches (Truncation of Boltzmann equation)}

\label{sec:boltz}

The relativistic Boltzmann equation for a rarefied non-degenerate gas of classical particles\footnote{Effects of quantum statistics according to fermion or boson can also be treated \cite{Denicol:2014loa}.} of mass $m$, without external forces, can be written as \cite{Cercignani2002}
\begin{equation}
p^{\mu}\frac{\partial}{\partial x^{\mu}}f=\int\left(  f^{\left(  3\right)
}f^{(2)}-f^{\left(  1\right)  } f \right)  \Phi\frac{d\sigma}{d\Omega}%
d\Omega\frac{d^{3}p^{\left(  1\right)  }}{p_{0}^{\left(  1\right)  }},
\label{RelBol}%
\end{equation}
where the right hand side is called the collision term, which describes the changes of the one particle distribution function $f\left(x,p\right)$ in terms of binary collisions,
\begin{equation}
p+p^{\left(  1\right)  }\rightarrow p^{\left(  2\right)  }+p^{\left(
3\right)  }, \label{collision}%
\end{equation}
with the differential cross section denoted by $d\sigma/d\Omega$.
Note that $f^{\left(  i\right)  }=f\left(  x,p^{\left(  i \right)  }\right) $ and $\Phi$ is the invariant flux, $\Phi=\sqrt{p^{\mu}p_{\mu}^{\left(  1\right)}  -m^{2}}$.
In the collision term, energy and momentum are conserved and these conservation laws reduce the integration of the final states specified by $\left\{  p^{\left(  2\right)  },p^{\left(  3\right)  }\right\}  $ from six to two, which corresponds to the integral of the solid angle $\Omega$ between $p$ and $p^{\left(  1 \right)  }$.
Since the integrand of the collision term is Lorentz invariant, we can choose the center of mass system of a given $p, p^{(1)}$ so that the final momenta are trivially expressed as functions of the solid angle.
We also have to integrate over all the colliding momenta, $p^{\left(  1\right)  }$ on $p$.

The Boltzmann equation is an effective theory to describe the non-equilibrium dynamics of a rarefied gas.
This is obtained by, at least in classical non-relativistic case, the lowest order truncation of the exact many-body dynamics called Bogoliubov-Born-Green-Kirkwood-Yvon (BBGKY) hierarchy equation \cite{Reichl2009,Liboff2003}.
That is, the two particle distribution function which appears in the collision term of the BBGKY equation is approximately replaced by the products of the one particle distribution function as is shown on the right hand side of Eq.\ (\ref{RelBol}).
This truncation is called Stosszahlansatz or molecular chaos assumption and, due to this approximation, the time reversal symmetry is violated in the Boltzmann equation.
This is a kind of coarse-graining procedure, leading to the existence of the so-called H-function, which decreases in time.
Furthermore, such a truncation is consistent if the system is dilute.
Therefore, the Boltzmann approach is expected to provide a solid theoretical framework to describe the transport properties involved in the dynamics of a relativistic dilute gas such as, for example, the hadron gas phase present in the kinetic freezeout stage of relativistic heavy ion collisions.

On the other hand, since the microscopic interactions are considered as point-like (which is consistent with molecular chaos), any hysteresis effects in the dynamics are not taken into account.
In fact, it is known that in the non-relativistic case the dynamic evolution of fluids described by the Boltzmann equation and by the Navier-Stokes-Fourier (NSF) theory show some qualitative differences.
The famous example is the velocity correlation functions.
The value obtained from the non-relativistic Boltzmann equation decays exponentially in time \cite{chapman1970mathematical} while that from NSF theory one finds a power law tail \cite{Reichl2009,Liboff2003,Phys.Rev.Lett.18.988,cond-mat/9707146,Dorfman198177}.
In addition, the bulk viscosity identically vanishes in the Boltzmann approach for the non-relativistic case.
Therefore, one has to be aware of the applicability and limitations of the Boltzmann approach.

Nevertheless, when extracting the long-time-scale dynamics described by the linearized Boltzmann equation it is known that one can reproduce the form of NSF theory.
That is, the linearized Boltzmann equation transfers some basic features of the microscopic collisional processes to the macroscopic viscous dynamics.
Because of this, it is important to investigate how hydrodynamics emerges from the Boltzmann equation especially in the relativistic domain where one does not have a firm phenomenological guidance as shown in the previous subsection.

The Chapman-Enskog method \cite{Liboff2003,Cercignani2002} is a typical method employed to obtain macroscopic equations of motion from the Boltzmann equation.
As mentioned above, the Boltzmann equation itself is already an effective theory, but it still contains rapid motions that are irrelevant in the hydrodynamic regime.
Therefore, in this case we are not interested in the exact behavior of the solution and, in fact, we want to extract an \textit{effective} one-particle distribution function of the type $\bar{f}(x,p)=f_{0}\left(  x,p\right) +\delta f$ which characterizes the long-time-scale dynamics of the relativistic Boltzmann equation, where $f_0$ and $\delta f$ are defined below.

How can one determine the macroscopic dynamics?
In quantum mechanics, the time evolution of the wave function is expressed in terms of a linear combination of the energy eigenfunctions.
Then, the scale of time oscillations is characterized by the magnitude of the energy eigenvalues.
Here we want to apply a similar argument to proceed with the coarse-graining reduction scheme of the Boltzmann equation.
For this, we consider the following linearized Boltzmann equation,
\begin{equation}
p^{\mu}\mathbf{\partial}_{\mu}\bar{f}=f_{0}\mathbf{\hat{C}}\phi,
\label{LinearBoltzmann}%
\end{equation}
where
$\phi\equiv\delta f/f_{0}=\bar{f}/f_{0}-1$.
Here we introduced the linear collision operator $\mathbf{\hat{C}}$ defined by
\begin{equation}
\mathbf{\hat{C}}\phi=\int f_{0}^{(1)}\left[  \left(  \phi^{\left(  3\right)
}+\phi^{\left(  2\right)  }\right)  -\left(  \phi^{\left(  1\right)  }%
+\phi\right)  \right]  \Phi\frac{d\sigma}{d\Omega}d\Omega\frac{d^{3}p^{\left(
1\right)  }}{p_{0}^{\left(  1\right)  }}. \label{ColOp-f0}%
\end{equation}
In the derivation of the collision operator above, we required that the basis function $f_{0}$ should satisfy the detailed balance relation (see below),
\begin{equation}
f_{0}\left(  p\right)  f_{0}\left(  p^{\left(  1\right)  }\right)
=f_{0}\left(  p^{\left(  2\right)  }\right)  f_{0}\left(  p^{\left(  3\right)
}\right)  , \label{detailedBallance}%
\end{equation}
or equivalently%
\begin{equation}
\ln f_{0}+\ln f_{0}^{\left(  1\right)  }=\ln f_{0}^{\left(  2\right)  }+\ln
f_{0}^{\left(  3\right)  }. \label{log of detaild ballance}%
\end{equation}
The eigenvalue problem of this operator plays a crucial role.
The collision term of the linearized Boltzmann equation contains various processes whose scales are characterized by the magnitude of the eigenvalues of the collision operator,
\begin{equation}
\mathbf{\hat{C}}\phi_{\lambda}^{\alpha}=\lambda\phi_{\lambda}^{\alpha},
\end{equation}
where $\alpha$ is the degeneracy index of the eigenvalue $\lambda$.
Any function $\phi$ which describes the changes of $\bar{f}$ due to the effect of binary collisions should be expressed in terms of a linear combination of these eigenfunctions, $\phi=\sum_{\lambda,\alpha}c_{\lambda,\alpha} \phi_{\lambda}^{\alpha}$.

It is known that all the eigenvalues are real and non-positive \cite{Cercignani2002}.
The slowest modes have zero eigenvalue $\lambda=0$.
It is obvious from Eq.\ (\ref{ColOp-f0}) that any function $\phi\left(  p\right)$ which is conserved in the collision process Eq.\ (\ref{collision}) is an eigenfunction of $\mathbf{\hat{C}}$ with a null eigenvalue.
In a binary collision, as is well-known, there are five conserved quantities: particle number, energy, and momentum.
Thus, they are degenerated eigenfunctions with $\lambda=0$,
\begin{equation}
\mathbf{\hat{C}}\mathbf{1}=\mathbf{\hat{C}}p^{\mu}=0. \label{conserved}%
\end{equation}

On the other hand, due to the condition Eq.\ (\ref{log of detaild ballance}), $\ln f_{0}$ is an eigenfunction of $\mathbf{\hat{C}}$ with $\lambda=0$.
Since there are no other independent conserved quantities than Eq.\ (\ref{conserved}), the most general form of $f_{0}$ should be expressed as
\begin{equation}
\ln f_{0}=\alpha(x)\mathbf{1}-\beta_{\mu}(x)p^{\mu}. \label{ce-dfexp}%
\end{equation}
As was pointed out before, when there are conserved quantities the time scales of the corresponding densities are very long compared to the microscopic time scale.
Therefore, it is natural to choose such $f_{0}$ as the starting point of the C-G reduction procedure with respect to their time-scales.
Additionally, the above $f_{0}$ is nothing but the Maxwell-J\"{u}ttner distribution and this expansion is equivalent to the expansion around the equilibrium distribution function.

Now, suppose that the amplitude of $\phi$ is very small ($\left\vert \phi\right\vert \ll1$) so that the effective distribution is essentially given by $\bar{f}=f_{0}$.
Then one can show that the dynamics described by Eq.\ (\ref{LinearBoltzmann}) is reduced to ideal relativistic hydrodynamics by identifying
\begin{equation}
T^{\mu\nu}\left(  x\right)  =\left\langle p^{\mu}p^{\nu}\right\rangle
_{\left(  f_{0}\right)  },
\end{equation}
where we defined the notation for the average of any function $O\left( p\right) $ by $f_{0}$ as
\begin{equation}
\left\langle O\right\rangle _{\left(  f_{0}\right)  }\equiv\frac{1}{\left(
2\pi\right)  ^{3}}\int\frac{d^{3}\mathbf{p}}{p^{0}}O\left(  p\right)f_0(x,p)  .
\end{equation}
Using the Landau definition, Eq.\ (\ref{def-vel-ener}) for the fluid velocity $u^{\mu}$, one can in fact show that
\begin{equation}
T^{\mu\nu}\left(  x\right)  =\left(  \varepsilon+P\right)  u^{\mu
}u^{\nu}-g^{\mu\nu}P,
\end{equation}
with
\begin{equation}
\varepsilon  = \langle(u_{\mu}p^{\mu})^{2}\rangle_{(f_{0})},\qquad
P  = -\frac{1}{3}\langle\Delta^{\mu\nu}p_{\mu}p_{\nu}\rangle_{(f_{0})}.
\end{equation}
Note that, for consistency, the time scale of the variation in $\alpha(x)$ and $\beta_{\mu}(x)$ should be much slower than the microscopic ones.
Because $\ln f_{0}$ describes the motion in the functional space spanned only by the five eigenfunctions with $\lambda=0$ the dynamics of the parameters $\alpha$ and $\beta_{\mu}$ are completely disentangled from the change in the momentum states due to the collision term.
Therefore, within the time scale corresponding to the change of $\alpha$ and $\beta_\mu$, detailed balance is considered to be held.
That is, the evolution only through the changes in $\alpha$ and $\beta_{\mu}$ corresponds to those of the quasi-static process that maintains the local equilibrium state of the gas.
In other words, the dynamics is time-reversible, i.e., that of an ideal fluid.

Therefore, to consider the violation of the detailed balance, which is directly related to the effect of dissipation, we need to expand the dynamical functional space by introducing other eigenfunctions of $\mathbf{\hat{C}}$ with finite eigenvalues.
In the Chapman-Enskog method, this extension is implemented by assuming that the first order deviation from the space of $\lambda=0$ is described by the linearized Boltzmann equation (\ref{LinearBoltzmann}) (see Ref.\ \cite{Reichl2009}).
By definition, the deviation of the one-particle distribution should relax at scales quicker than the time scale of $f_{0}$ and thus, in the long-time limit, the right hand side of Eq.\ (\ref{LinearBoltzmann}) is approximately replaced by
\begin{equation}
p^{\mu}\partial_{\mu}\bar{f}\approx p^{\mu}\partial_{\mu}f_{0}.
\end{equation}
Substituting this into the linearized Boltzmann equation, the deviation is given formally by
\begin{equation}
\phi=\mathbf{\hat{C}}^{-1}p^{\mu}\partial_{\mu}\ln f_{0}.
\end{equation}
Now the right-hand side contains a component proportional to $p^{\mu}p^{\nu}$ which is not a collision invariant anymore and the deviation from $f_{0}$ is affected by a component which involves a faster time scale than those from the collision invariants.
Substituting this into the linearized Boltzmann equation, we obtain the relativistic NSF theory.
This is the physics behind the coarse-graining of time scales in the Chapman-Enskog method and the above-mentioned scheme is mathematically implemented as a systematic expansion in powers of the Knudsen number.

Relevant contributions which survive even after taking the long-time limit are sometimes called secular terms in the derivation of coarse-grained dynamics.
For example, it is known that the systematic collection of the secular terms is important to obtain a master equation in other mechanical systems such as celestial mechanics and quantum mechanics \cite{vanHove1954517,DeMasi1984}.
The C-G procedure employed here corresponds to the van Hove limit.
However, the way the secular terms can be resummed is neither trivial nor unique.
In Ref. \cite{Tsumura:2012kp}, for example, the method to reduce nonlinear differential equations is applied to the relativistic Boltzmann equation and another relativistic generalization of the NSF theory is obtained.

In the Chapman-Enskog method the perturbative contribution to the effective distribution is expressed in terms of $f_{0}$ which is a function of the collisional invariants.
On the other hand, to obtain a causal model, one needs to introduce new hydrodynamic variables as mentioned before.
Naively thinking, such additional terms can be obtained by just generalizing the form of $f_{0}$ itself.
However, this is a very difficult task and this constitutes one of the reasons why it is not trivial at all to obtain causal hydrodynamic models using the Chapman-Enskog approach.

The moments method, which was proposed by Grad \cite{Grad1949}, is a representative method to obtain causal hydrodynamic models from the Boltzmann equation.
In this approach, the Boltzmann equation is converted to coupled equations for the moments, which are, for example, defined by
\begin{equation}
\mathcal{M}_{n}^{\alpha_{1}\alpha_{2}\cdots\alpha_{l}}\equiv\int d\Gamma
_{p}(u_{\mu}p^{\mu})^{n}p^{<\alpha_{1}}\cdots p^{\alpha_{l}>}\delta
f\equiv\langle(u_{\mu}p^{\mu})^{n}p^{<\alpha_{1}}\cdots p^{\alpha_{l}>}%
\rangle_{(\delta f)},
\end{equation}
where $p^{<\alpha_{1}}\cdots p^{\alpha_{l}>}=\Delta_{\beta_{1}\cdots\beta_{l}}^{\alpha_{1}\cdots\alpha_{l}}p^{\beta_{1}}\cdots p^{\beta_{l}}$ with $\Delta_{\beta_{1}\cdots\beta_{l}}^{\alpha_{1}\cdots\alpha_{l}}$ being the symmetric traceless projection operator for any of two $\alpha_i$'s or two $\beta_i$'s \cite{Denicol:2012cn}.
One can obtain the exact coupled equations for these moments, which is equivalent to the Boltzmann equation if the expansion converges.

In the Chapman-Enskog method, $\phi$ (or equivalently $\delta f$) is determined dynamically by solving the linearized Boltzmann equation.
On the other hand, in the moments method implemented by Israel and Stewart, a functional ansatz of $\phi$ is introduced,
\begin{equation}
\phi=\lambda_{1}\Pi+\lambda_{2}\nu_{\mu}p^{\mu}+\lambda_{3}\pi_{\mu\nu}p^{\mu
}p^{\nu}, \label{14m}%
\end{equation}
where
\begin{equation}
\Pi   =-\frac{1}{3}\langle\Delta^{\mu\nu}p_{\mu}p_{\nu}\rangle_{(\delta f)},\qquad
\pi^{\mu\nu}    =\langle\Delta^{\mu\nu\alpha\beta}p_{\alpha}p_{\beta} \rangle_{(\delta f)},\qquad
\nu^{\mu}   =\langle p^{\mu}\rangle_{(\delta f)}.
\end{equation}
This ansatz for $\phi$ is called 14 moment approximation \cite{Israel:1979wp}.
By assumption, the time evolution of $\phi$ is restricted in the dynamical functional space spanned by $\Pi$, $\nu^{\mu}$, and $\pi^{\mu\nu}$.
As in the case of Chapman-Enskog's method, one assumes that the deviation from $f_{0}$ is described by the linearized Boltzmann equation (\ref{LinearBoltzmann}).
Note that the above assumption for $\phi$ resembles the Kawasaki relation which gives the relation between equilibrium and non-equilibrium distribution functions \cite{Hoover1991}.
Once dynamical equations for these new variables are obtained in a closed form, one can derive the dissipative hydrodynamics equations.

However, closing the system of dynamical equations is not granted.
In the moments method, one introduces a new ambiguity in the choice of the moments to derive dissipative hydrodynamics which does not have a counterpart in the Chapman-Enskog method \cite{Garcia2008149}.
For example, in the original argument of Israel and Stewart, the equation for the bulk viscosity is obtained from the moment equation of $u_{\nu}u_{\lambda}\partial_{\mu}\langle p^{\mu}p^{\nu}p^{\lambda}\rangle_{(\bar{f})}$.
It is however possible to obtain the equation for the bulk viscosity from another moment equation, for example, $\mathcal{M}_{0}$.
The form of the transport coefficients calculated in the moments method depends on the choice of the equation for the moments, which are used to close the coarse-grained dynamics \cite{Denicol:2010xn}.

It is possible to close the moment equation without using the 14 moment approximation (\ref{14m}).
In this approach, the form of the effective particle distribution function $\bar{f}$ is not specifically calculated.
Instead, the linearized Boltzmann equation is mapped into a set of infinitely coupled equations for the moments and a coarse-graining reduction with respect to the time scale is applied directly to the moment equations by introducing an expansion around a diagonalized basis of moments of the collision operator \cite{Denicol:2014loa}.

The equations for the moments ${\cal M}_{r}$, ${\cal M}_{s}^{μ}$, $\cdots$ have contributions from the collision operator.
The corresponding terms can be expanded as
\begin{equation}
C^{\mu_{1} \cdots\mu_{l}}_{r-1} = \int d\Gamma_{p} (u_{\mu}p^{\mu}%
)^{r}_{\mathbf{p}} p^{< \mu_{1} }\cdots p^{\mu_{l}>} f_{0} \hat{C} \phi= -
\sum^{N_{l}}_{n=0} A^{(l)}_{rn} \mathcal{M}^{\mu_{1} \cdots\mu_{l}}_{n} .
\end{equation}
The above equation is equivalent to the Boltzmann equation.
To obtain C-G dynamics, as is done in the Chapman-Enskog expansion, where the eigenfunctions of the collision operator with the null eigenvalue $\lambda=0$ play an important role, one needs to introduce an appropriate truncation scheme in the system of moments.
For the derivation of the hydrodynamic equation, the moments for $l=0,1,2$ are important.
Each $N_{l}$ is in principle $\infty$, but we introduce the truncation by choosing finite numbers $N_0$, $N_1$ and $N_2$ as three parameters of this approach.
Note that ${\cal M}_1$ and ${\cal M}_2$ for $l=0$ and ${\cal M}_1^{\mu}$ for $l=1$ are excluded from the basis of the expansion.

In the basis where the matrix $A^{(l)}_{rn} $ is diagonal, the moment $\mathcal{M}^{\mu_{1} \cdots\mu_{l}}_{n}$ is changed into $X^{\mu_{1} \cdots\mu_{l}}_{n}$.
Let us assume that the deviation from equilibrium can be spanned only by $X_{0}$, $X^{\mu}_{0}$ and $X^{<\mu\nu>}_{0}$ among all possible moments $X^{\mu_{1} \cdots\mu_{l}}_{n}$.
This corresponds to an expansion of the dynamical functional space in terms of $N_{0}-2$ scalar moments ($\mathcal{M}_{0},\mathcal{M}_{3} \cdots\mathcal{M}_{N_{0}}$), $N_{1}-1$ vector moments ($\mathcal{M}_{0}^{\mu}, \mathcal{M}_{2}^{\mu}, \cdots \mathcal{M}_{N_{1}}^{\mu}$) and $N_{2}$ tensor moments ($\mathcal{M}_{0}^{\mu\nu},\cdots\mathcal{M}_{N_{2}}^{\mu\nu}$).

The moments equations for $X_{0}$, $X_{0}^{\mu}$ and $X_{0}^{<\mu\nu>}$ still couple to other higher order moments.
Then the truncation of the coupled moment equation for $X_{n}^{\mu_{1}\cdots\mu_{l}}$ is carried out by maintaining the first order terms of the expansion in terms of two parameters, the Knudsen number and the inverse Reynolds number.

To express hydrodynamics in terms of $\Pi$, $\nu^{\mu}$ and $\pi^{\mu\nu}$, we change from the diagonalized basis to the original basis of $\mathcal{M}^{\mu_{1} \cdots\mu_{l}}_{n}$.
Then, for example, $\mathcal{M}_{0}$, which is associated with the bulk viscosity, is expressed as a linear combination of $X_{0} ,X_{3} \cdots X_{N_{0}}$.
To eliminate the dynamics of $X_{3} \cdots X_{N_{0}}$, we further assume that the time evolution of these moments is much faster than $X_{0}$ and the values are replaced by the stationary solution of the corresponding equations for the moments.

The hydrodynamic theory which is obtained in this scheme is called transient hydrodynamics.
One can easily see that this hydrodynamic theory leads to a different model from the one obtained from the 14 moment approximation in general because the dynamical functional space used to perform the coarse-grained dynamics is different.
In the 14 moment approximation, the dynamics of $\phi$ is considered to take place in the space spanned by $\Pi$, $\nu^{\mu}$, and $\pi^{\mu \nu}$.
On the other hand, in transient hydrodynamics, $\phi$ is not written explicitly and, thus, a direct comparison is not trivial to be performed.
However, the moment equations are expanded in the basis spanned by $N_{0}-2$ scalar moments, $N_{1}-1$ vector moments, and $N_{2}$ tensor moments and the scale of coarse-graining is different than that obtained in the 14 moment approximation.

Interestingly enough, if we choose $N_{0} =2$, $N_{1} =1$, and $N_{2} = 0$, then the moment equations are expanded in terms of $\mathcal{M}_{0}$, $\mathcal{M}^{\mu}_{0}$ and $\mathcal{M}^{\mu\nu}_{0}$ in transient hydrodynamics, which is similar to the 14 moment approximation.
In fact, transient hydrodynamics leads to the same result as the 14 moment approximation in this case.
There is also another approach based on the Boltzmann equation, cf. Ref.\ \cite{Jaiswal:2013fc}.

It is often claimed that hydrodynamics is obtained from an underlying microscopic theory by expanding in powers of the Knudsen number. 
This view seems to be based on the Chapman-Enskog method for the Boltzmann equation discussed above. 
However, such a perspective is quite suspicious because one cannot improve upon the coarse-graining procedure even if one considers higher order correlations in the Chapman-Enskog method.
In fact, non-relativistic hydrodynamics including such correction terms is called the Burnett and super Burnett equations \cite{Garcia2008149}, which are known to suffer from an instability called Bobylev instability.
Moreover, there are no examples of fluids which are correctly described by either the Burnett or super Burnett equations. 
Therefore, it is not obvious whether one should consider that C-G can be systematically implemented via the Knudsen number expansion.

Finally, we have one more comment on the so-called derivation of hydrodynamics from the Boltzmann equation.
As was mentioned above, the Boltzmann equation is appropriate to describe dilute systems while the hydrodynamic description works better in dense systems.
Therefore, the discussion presented here should not be interpreted as a rigorous derivation of the hydrodynamic equations from the Boltzmann equation itself.
Rather, it is expected that the asymptotic behavior towards equilibrium in the Boltzmann equation correctly describes the deviation from a given LTE and in fact it can reproduce the structure of the NSF theory.
This is because the asymptotic LTE state is an input in the case of linearized Boltzmann equation. 
Furthermore, we note that the dynamics described by the exact Boltzmann equation does not coincide with NSF theory as mentioned before.

\subsubsection{Projection Operator Method}

\label{sec:proj}

Microscopic dynamics has time-reversal symmetry and special care is need to violate it to obtain C-G dissipative dynamics. 
In this subsection, we explain how this is done in the framework of the projection operator method.

The most traditional way to implement C-G in microscopic dynamics is the projection operator method in statistical physics \cite{Zwanzig2001,Breuer2002,Reichl2009,Kubo1983,Koide:2010wt}.
In this approach, we first choose the dynamical space onto which our non-equilibrium dynamics is mapped and define a projection operator $P$ to implement the coarse-graining with the chosen variables.
By using this projection, we can systematically project out irrelevant rapid motions contained in a microscopic dynamics.

Let us define the (time-independent) projection operator satisfying the following general properties, 
\begin{equation}
P^{2}  = P,\qquad
Q  = 1-p.
\end{equation}
From these definitions, one can see that $Q$ extracts the dynamics orthogonal to $P$, that is, the irrelevant rapidly moving degrees of freedom.
By using the properties of the projection operators mentioned above, the Heisenberg equation of motion for an arbitrary operator $O$ is expressed as 
\begin{equation}
\partial_{t}O(t)=e^{iLt}PiLO+\int_{0}^{t}dse^{iL(t-s)}PiLQe^{iLQs} iLO+Qe^{iLQt}iLO,
\label{pom}
\end{equation}
where $L$ denotes the Liouville operator defined by
\begin{equation}
iLO\equiv\left\{
\begin{array}[c]{ll}
i[H,O] & \mathrm{for\ quantum\ systems}, \\
\left\{H,O\right\}_{PB} & \mathrm{for\ classical\ systems}.
\end{array}
\right.
\end{equation}
Here $\left\{H,O\right\}_{PB}$ is the Poisson bracket of $H$ and $O$.
This equation is called time-convolution equation of the projection operator and can be derived without specifying the concrete definition of $P$.
The first term on the right-hand side is called the free-streaming term and corresponds to a collective motion which does not produce entropy.
The second term represents the dissipative part of the time evolution while the third term is interpreted as the noise term.
In the following, we exclusively discuss quantum mechanical systems.

There are several possible choices for $P$, but we consider here the so-called Mori's projection operator in the following discussion.
Let us consider a set of macroscopic variables by an n-dimensional vector, $\mathbf{A}^{T}=(A_{1},\cdots,A_{n})$.
Then the Mori projection operator is defined as
\begin{equation}
PO=\sum_{i=1}^{n}c_{i}A_{i},
\end{equation}
where the coefficient is
\begin{equation}
c_{i}=\sum_{j=1}^{n}(O,A_{j}^{\dagger})(A,A^{\dagger})_{ji}^{-1}.
\end{equation}
The inner product is given by Kubo's canonical correlation, which is defined by
\begin{equation}
(X,Y)=\int_{0}^{\beta}\frac{d\lambda}{\beta}\mathrm{Tr}[\rho_{eq}e^{\lambda
H}Xe^{-\lambda H}Y]\equiv\int_{0}^{\beta}\frac{d\lambda}{\beta}\langle
e^{\lambda H}Xe^{-\lambda H}Y\rangle_{eq},
\end{equation}
where $\rho_{eq}=e^{-\beta H}/\mathrm{Tr}[e^{-\beta H}]$ with $\beta$ being the inverse temperature.
See Refs.\ \cite{Koide:2008nw,Koide:2007ja,Huang:2010sa,Huang:2011ez,Koide:2005qb} for details.

Let us consider some type of microscopic dynamics characterized by the Hamiltonian $H$ and derive its coarse-grained dynamics assuming that the dynamics is described by only one macroscopic variable, say, $B$.
Then the macroscopic evolution equation of this macroscopic variable $B$ is obtained from Eq.\ (\ref{pom}) as 
\begin{equation}
\frac{\partial}{\partial t} B(t) = \int^{t}_{0} ds D(s) B(t-s) + \xi(t),
\end{equation}
where
\begin{equation}
D(t)   = ( \xi(t), \xi(0) ) (B,B)^{-1},\qquad
\xi(t)    = Q e^{iLQt}iL B.
\end{equation}
Here we assumed that $B$ is an Hermitian operator.
This equation is still equivalent to the Heisenberg equation of motion and, to obtain dissipative behavior, we need to introduce the coarse-graining of the time scale.

Note that if the projection operator $P$ is chosen appropriately in order to completely extract macroscopic motions, the time scale of the noise term $\xi(t)$ is very short because it is proportional to $Q$ and we can ignore the characteristic time scale of $(\xi(t),\xi(0))$ in $D(t)$.
Then we can assume that the time dependence of $D(t)$ is approximately proportional to $\delta(t)$ and the exact equation above is reduced to 
\begin{equation}
\frac{\partial}{\partial t}B(t)=\gamma B(t)+\xi(t), \label{beq}%
\end{equation}
where $\gamma=\int_{0}^{\infty}dsD(s)$.
This is called the time-convolutionless limit (TCL), which is equivalent to a Markov approximation \cite{Huang:2010sa,Huang:2011ez}.
We shall soon see a concrete example of this equation.
As seen in Eq.\ (\ref{beq}) the equation obtained in the projection operator method contains a noise term that is related to the viscosity through the fluctuation-dissipation theorem.
The relativistic hydrodynamic model with noise terms is discussed in Ref.\ \cite{Kapusta:2011gt} and references therein.
However, this type of approach is known to present difficulties such as Lorentz covariance and stability \cite{fox1978gaussian}.

The choice of the hydrodynamic variables is closely related to the applicability of the TCL, that is, the consistent violation of the time reversal symmetry.
One can easily see that the TCL is not applicable when the time correlation function of the noise converges to a finite value in the long time limit,
\begin{equation}
\lim_{t \rightarrow\infty} D(t) = const,
\end{equation}
because then $\gamma$ diverges.
This means that macroscopic dynamics has not yet been extracted and we need to extend the definition of the projection operator $P$.

For example, in dynamics of diffusion, it is known that the projection operator is defined with two macroscopic variables, $n(\mathbf{x})$ and $\mathbf{J}(\mathbf{x})$, which satisfy the continuity equation, to apply the TCL limit\footnote{In the textbooks of statistical physics \cite{Reichl2009,Zwanzig2001,Fick2012}, it is written that the diffusion equation can be obtained in the projection operator method.
However, in the derivation, an artificial approximation is introduced.
If this approximation is not used, we encounter the singular behavior of the diffusion coefficient and the definition of the projection operator should be generalized.
This fact was first pointed out in Ref.\ \cite{Koide:2005qb}.}, $\partial_{t} n (\mathbf{x}) = - \nabla\cdot\mathbf{J} (\mathbf{x})$.
Then the equation corresponding to Eq.\ (\ref{beq}) is given by
\begin{align}
\frac{\partial}{\partial t} n(\mathbf{x},t) + \nabla\mathbf{J}(\mathbf{x},t)
&  = 0,\\
\tau_{D} \frac{\partial}{\partial t} \mathbf{J}(\mathbf{x},t) + \mathbf{J}%
(\mathbf{x},t)  &  = -D \nabla n(\mathbf{x},t) , \label{eq_j}%
\end{align}
where the diffusion coefficient $D$ and the relaxation time $\tau_{D}$ are, respectively, given by
\begin{equation}
D    = -\left(  \frac{\partial\mu}{\partial n} \right)  _{T} \lim_{\omega\rightarrow0}\lim_{\mathbf{k} \rightarrow0} G^{R}_{\mathbf{J}} (\omega,\mathbf{k}),\label{diff_coe} \qquad
\frac{D}{\tau_{D}}    = \frac{1}{3} \frac{\sum_{i} \int d^{3} \mathbf{x}(\mathbf{J}^{i}(\mathbf{x}),\mathbf{J}^{i} (\mathbf{0}))} {\int d^{3}\mathbf{x} (\delta n(\mathbf{x}),\delta n(\mathbf{0}))},
\end{equation}
where $\delta A \equiv A - \langle A \rangle_{eq}$ and the retarded Green function is
\begin{equation}
G^{R}_{\mathbf{J}} (\omega,\mathbf{k}) = - \frac{i}{3}\sum_{i} \int^{\infty
}_{0} dt \int^{\infty}_{-\infty} d^{3}\mathbf{x} e^{-i\omega t}e^{i\mathbf{kx}%
} \langle[\mathbf{J}^{i} (\mathbf{x,t}), \mathbf{J}^{i} (\mathbf{0},0)]
\rangle_{eq}.
\end{equation}
Here the noise term has been dropped.

The projection operator approach provides information on how one should violate the time reversal symmetry.
For example, one may consider that the TCL is the lowest order truncation of the following expression of the time-convolution integral term, 
\begin{equation}
\int^{t}_{0} ds D(s) B(t-s) = \sum_{n=0}^{\infty}\int^{t}_{0} ds D(s)
\frac{\partial^{n} B(t)}{\partial t^{n}} (-s)^{n} , \label{markov-exp}%
\end{equation}
and consider that the solution can be improved by considering higher order terms but that is not true.
First of all, the TCL is an artificial operation to violate the time reversal symmetry and this is done not for the purpose of the simplification of dynamics.
In fact, it is known that the TCL is the most appropriate procedure for the violation of time-reversal symmetry in the projection operator method because it can pick up all the secular terms correctly and an exact relation, called f-sum rule, is maintained \cite{Huang:2011ez}.
Alternatively, if we consider a correction term to the TCL, secular and non-secular terms are considered on an equal footing and the f-sum rule is violated \cite{Huang:2011ez,Koide:2005qb}.
If we want to take into account more microscopic dynamics associated with $D(t)$, we need to extend the number of dynamical variables to define the projection operator as was done in the example above.

The problem of the projection operator method is that it is difficult to maintain the manifest Lorentz covariance.
The attempt to formulate relativistic hydrodynamics is, thus, so far limited to the derivation of linearized equation \cite{Minami:2012hs}.
It is however still possible to utilize it to define transport coefficients because these are Lorentz scalar quantities and, in this case, it is enough to consider linear deviations from equilibrium.
In fact, it is known that this approach leads to the correct transport coefficients that appear in NSF theory \cite{Koide:2007ja}.

To obtain, for example, the shear viscosity in a causal hydrodynamic model the projection space is defined by the two variables, $\mathbf{A} = T^{yx}$ and $T^{0x}$.
Then $\eta$ and the corresponding relaxation time $\tau_{\pi}$ are defined by
\begin{equation}
\frac{\eta}{\beta(\varepsilon+ P)}    = \frac{\eta_{GKN}}{\beta^{2} \int d^{3} \mathbf{x} (T^{0x}(\mathbf{x}),T^{0x}(\mathbf{0}))} ,\qquad
\frac{\eta}{\tau_{\pi}(\varepsilon+ P)}    = \frac{\int d^{3} \mathbf{x}(T^{yx}(\mathbf{x}),T^{yx}(\mathbf{0}))}{\int d^{3} \mathbf{x} (T^{0x}(\mathbf{x}),T^{0x}(\mathbf{0}))} . 
\label{shear-etatau}
\end{equation}
where $\eta_{GKN}$ is the usual result of the shear viscosity coefficient in the relativistic NSF theory \cite{Koide:2009sy,Huang:2011ez,Koide:2010wt}.
This expression is called Green-Kubo-Nakano (or simply Kubo) formula, but, exactly speaking, this result was obtained using Zubarev's non-equilibrium statistical operator method \cite{Ann.Phys.154.229}.
Note that the results above are universal and are thus valid not only at finite temperature but also at finite temperature and nonzero density \cite{Huang:2011ez}.
The results for the bulk viscosity coefficient are shown in Refs.\ \cite{Huang:2010sa,Huang:2011ez,Koide:2010wt}.

The shear viscosity result has an interesting relation to other kinetic theory calculations and also AdS/CFT calculations.
In fact, the lowest order calculation of Eq.\ (\ref{shear-etatau}) without including the pair creation-annihilation effect gives the exact same result as the moments method within the 14 moment approximation obtained in Ref.\ \cite{Denicol:2010br}.
However, the contributions from the pair creation-annihilation effect can be as large as 20 $\%$ in Eq.\ (\ref{shear-etatau}) and 80 $\%$ for the corresponding quantity for the bulk viscosity.
Thus, one cannot ignore this quantum effect.

The transport coefficients of transient hydrodynamics assume different values as the number of moments in the expansion is increased.
However, this difference is reasonable because the expansion basis of the dynamical functional space is now different.
In the definition of the projection operator above, the deviation from thermal equilibrium is expressed by the $T^{0x}$ and $T^{yx}$, the former corresponds to $(\varepsilon+ P )u^{x}$ and the latter $\pi^{yx}$.
This corresponds to expand the dynamical functional space using the 14 moments as was done in the moments method for the Boltzmann equation.
To reproduce the result of, for example, the 23 moments approximation done in transient hydrodynamics, one needs to expand the definition of the projection operator.
For such a systematic generalization of the projection operator, see Ref.\ \cite{Koide:2008nw}.

On the other hand, if we consider the next order correction of the expansion by Eq.\ (\ref{markov-exp}), the form of Eq.\ (\ref{shear-etatau}) is modified as was pointed out in Eq.\ (65) of Ref.\ \cite{Huang:2011ez,Huang:2010sa}, and it coincides with the same expression obtained from AdS/CFT.
See also Ref.\ \cite{Denicol:2011fa} where it was shown how one can reproduce the result of 14 moments approximation by modifying the AdS/CFT calculation.

It is also worth mentioning that we need to use the modified definition of the energy-momentum tensor for the scalar field theory in calculating the bulk viscosity consistent with conformal symmetry \cite{Huang:2010sa,Huang:2011ez}.
It should also be noted that there are two philosophically different concepts involving the transport coefficients.
When an irreversible current is induced by the application of an external field, as is the case of electric conductivity, such a perturbation can be expressed in the form of the interaction Hamiltonian.
On the other hand, phenomena like heat conduction and viscosity are induced by the changes of the boundary conditions and, thus, these effects cannot be expressed as the interaction term.
The importance of this difference was already pointed out by Kubo himself.
See also the discussions in Ref.\ \cite{Zwanzig1965}.

As we have discussed, there are two main factors needed to obtain the C-G dynamics from a given microscopic theory; one is the choice of macroscopic variables and the other is the Markov limit. 
These are, however, not independent. To employ the Markov limit, we need to choose a complete set of macroscopic variables. 
The difficulty of the C-G procedure is attributed to the absence of the established criterion to choose macroscopic variables, which is a common problem not only in the macroscopic and mesoscopic derivations which we have discussed, but also in the derivation of conformal fluids which will be discussed soon later.
In the present case we chose conserved densities and the corresponding currents as macroscopic variables, but there are situations where one may have to extend this definition.
For example, if there is a continuous phase transition, 
because of the critical slowing down the time scale of the order parameter field becomes sufficiently slow and we need to consider it as another macroscopic variable
\cite{Koide:2004yn}.

\subsubsection{Conformal hydrodynamics and transport coefficients with AdS/CFT correspondence}

\label{AdS}

Another method to derive relativistic hydrodynamics from the microscopic degrees of freedom in the strong coupling limit has been developed within the framework of AdS/CFT correspondence.
There are various studies involving conformal hydrodynamics \cite{Baier:2007ix} (in which the hydrodynamic equations change covariantly under Weyl transforms of the metric) and recently this approach has been extended to study nonconformal theories as well \cite{Finazzo:2014cna}.
The derivation of hydrodynamics itself is phenomenological, but one can calculate the transport coefficients from a microscopic theory using the AdS/CFT correspondence, which means the equivalence between a conformal field theory such as $\mathcal{N}=4$ supersymmetric Yang-Mills theory and superstring theory in anti-de Sitter spaces \cite{Baier:2007ix}.

First we discuss the derivation of hydrodynamics in this case.
The philosophy of this approach is based on the interpretation that hydrodynamics can be obtained from kinetic theory within the Chapman-Enskog method which can be regarded as the Knudsen number (or equivalently derivative) expansion of the linearized Boltzmann equation as we have discussed.
As it has been discussed so often in other approaches, one first specifies the hydrodynamic variables.
We choose slow variables associated with the conservation of energy and momentum, $\varepsilon$ and $u^{\mu}$.
If hydrodynamics can be expressed only by those variables, one obtains the ideal hydrodynamic case which we have already discussed.
To obtain a dissipative theory, one needs to include the deviation from equilibrium.
In this approach, this is assumed to be given as a function of $\varepsilon$ and $\partial^{\mu}u^{\nu}$ and consider up to the second order terms in derivatives.
Another method used in this derivation can be found in Ref.\ \cite{Bhattacharyya:2008jc}.

The advantage of considering a conformal fluid is that the continuity equation of the energy-momentum tensor should transform covariantly under Weyl transformations and, thus, the number of second order terms are reduced.

The hydrodynamic theory derived this way has a term associated with the vorticity,
\begin{equation}
\Omega^{\mu\nu} = \frac{1}{2} \Delta^{\mu\alpha} \Delta^{\nu\beta}
(\partial_{\alpha}u_{\beta}- \partial_{\beta}u_{\alpha}).
\end{equation}
Such a term does not appear in the kinetic derivation of the Chapman-Enskog method and the 14 moment approximation of the moments method and, thus, has been considered as a quantum effect.
However, it was found that the same term can be naturally obtained in kinetic theory using transient hydrodynamics \cite{Denicol:2012cn}.

One can easily see that this is similar to the second order correction in the Chapman-Enskog method and then what we obtain is the relativistic generalization of the Burnett equation.
In fact, it is known that this equation contains an unphysical propagation mode \cite{Baier:2007ix}.
To obtain a causal model, one needs to replace some of the terms in the equation by the time derivative of $\pi^{\mu\nu}$ by hand.

Another advantage of discussing a conformal fluid is that one can compute the transport coefficients directly from the AdS/CFT correspondence.
The retarded Green's function of the energy-momentum tensor can be calculated from two different ways; the first one involves the energy momentum tensor equation which was derived phenomenologically above while the other one involves the microscopic theory, such as supersymmetric Yang-Mills theory, via the AdS/CFT correspondence.
Expressing the retarded Green's function in the Fourier variables $\omega$ and $k$ and comparing both results at the same order one can find the desired transport coefficients.
This method is essentially the same as the indirect Kubo method Ref.\ \cite{Zwanzig1965}.
For example, the shear viscosity coefficient obtained is given by the famous ratio
\begin{equation}
\frac{\eta}{s} = \frac{1}{4\pi}.
\end{equation}
This result was first obtained in Refs.\ \cite{Policastro:2001yc,Kovtun:2004de} and it was considered to be a lower bound for the value of the shear viscosity coefficient, called the KSS bound.

There are several attempts to generalize the approach mentioned above to non-conformal fluids, see Refs.\ \cite{Kanitscheider:2009as,Buchel:2007mf,Bigazzi:2010ku,Finazzo:2014cna}.

As was mentioned above, this derivation of hydrodynamics is inspired by the success of the kinetic derivation.
However, as was also emphasized before, there is a critical difference between the kinetic derivation and the present microscopic derivation.
That is, in the Boltzmann equation time reversal symmetry is already broken but the microscopic theory is still symmetric.
In the above argument involving the AdS/CFT correspondence, it is implicitly assumed that the symmetry is appropriately violated by the truncation of the derivative expansion.
The validity of the assumption should be carefully investigated.

\subsection{Variational Principle}

\label{sec:vp}

In this subsection we present a quite different view from what has been discussed until now.
As mentioned before in this text, hydrodynamics is a classical effective theory for interacting matter, which emerges from the underlying microscopic degrees of freedom through some kind of coarse-graining procedure.
There are many ways to introduce a given C-G scheme.
In order to obtain a closed system of dynamical equations only in terms of these C-G hydrodynamic variables, one usually needs to introduce further some approximations or truncations.
As discussed in the previous section, depending on how one introduces these truncations, many different forms of dissipative hydrodynamics can be obtained.

A different way to obtain a closed system of dynamic equations for a set of given C-G variables is the variational method.
It has been shown in the very early days that relativistic hydrodynamics can also be formulated in terms of a variational principle \cite{Taub:1954zz,Andersson:2006nr}.
Thus, if we express the model action in terms of the selected hydrodynamic variables, the variational principle leads to the optimal dynamics for these variables.
Below we show how the effective dynamics of C-G variables can be described in variational form.

\subsubsection{Coarse Graining in Variational Principle}

In order to perform the C-G explicitly, let us consider for example a classical microscopic system which contains a large number of quickly moving point-like particles.
We can define the density
\[
\tilde{\rho}\left(  \mathbf{r},t\right)  =\sum_{i}\delta\left(  \mathbf{r}-\mathbf{r}_{i}\left(  t\right)  \right)
\]
where the sum is done over all the particles in the system and $\mathbf{r}_{i}\left(  t\right)  $ refers to the position of $i$-th particle at time $t$.
The symbol $^{\sim}$ is used to show that the variable is spiky (delta-function like) in space and rapidly changing in time.
The corresponding current is%
\[
\mathbf{\tilde{j}}\left(  \mathbf{r},t\right)  =\sum_{i}\frac{d\mathbf{r}_{i}}{dt}\delta\left(  \mathbf{r}-\mathbf{r}_{i}\left(  t\right)  \right)  ,
\]
and we can easily see that they satisfy the continuity equation,
\begin{equation}
\partial_{t}\tilde{\rho}+\mathbf{\nabla}\cdot\mathbf{\tilde{j}}=0.\label{Continuidade-n0}
\end{equation}
However, we usually do not require a very precise resolution both in space and in time to describe collective flow behavior.
Thus, we introduce an averaged smooth density distribution $\rho\left(  \mathbf{r},t\right)  $ and current $\mathbf{j}$ from the original distributions using a four-dimensional smoothing kernel \cite{Mota:2012qv} as
\begin{align}
\rho\left(  \mathbf{r},t\right)   &  =\int dt^{\prime}\int d^{3}
\mathbf{r}^{\prime}U\left(  t^{\prime}-t\right)  W\left(  \mathbf{r}
-\mathbf{r}^{\prime}\right)  \tilde{\rho}\left(  \mathbf{r}^{\prime}
,t^{\prime}\right)  ,\label{nsmooth}\\
\mathbf{j}\left(  \mathbf{r},t\right)   &  =\int dt^{\prime}\int
d^{3}\mathbf{r}^{\prime}U\left(  t^{\prime}-t\right)  W\left(  \mathbf{r}
-\mathbf{r}^{\prime}\right)  \mathbf{\tilde{j}}\left(  \mathbf{r}^{\prime
},t^{\prime}\right)  ,
\end{align}
where
\begin{equation}
\int_{-\infty}^{\infty}dt\ U\left(  t\right)  =\int d^{3}r\ W\left(
\mathbf{r}\right)  =1,\mathbf{\ }\label{NormalizationUW}%
\end{equation}
and
\begin{equation}
U(t), W(\mathbf{r}) \rightarrow 0, \ \ \ |t|,|\mathbf{r}| \gg \tau, h
\label{SupportLimit}
\end{equation}
with $\tau$ and $h$ are given constants which characterize the time and space C-G scale, respectively.
We can take, for example, $U$ and $W$ as Gaussian distributions with width $\tau$ and $h$.
It is easy to verify that $\partial\rho/\partial t+\nabla\cdot\mathbf{j}=0$.

In an analogous manner, we can construct the smoothed (coarse-grained) energy-momentum tensor $T^{\mu\nu}$ as a convolution of the original very spiky and quickly changing $\tilde{T}^{\mu\nu}$.
It is easy to show that such an energy-momentum tensor again satisfies the continuity equation, $\partial_{\mu}T^{\mu\nu}=0$, as far as the original $\tilde{T}^{\mu\nu}$ does.
Through the above C-G procedure, the two macroscopic fields, $j^{\mu}\left(\mathbf{r},t\right)$, and $T^{\mu\nu}\left(  \mathbf{r},t\right)  $ can be constructed and, by definition, they are smooth and mildly changing within the scales of $W$ and $U$.
Now, instead of constructing the equation of motion from the microscopic dynamics for the original variables (in the present example, the set of large number of trajectories $\left\{  \vec{r}_{i}\left( t\right)  \right\}  $ ), we ask what is the optimal time evolution for $j^{\mu}\left(  \mathbf{r},t\right)$, and $T^{\mu\nu}\left(  \mathbf{r},t\right)$, given that we know the initial and final values of them.

Usually, the variational principle describes time-reversal symmetric problems.
Thus, we first consider the ideal fluid case.
In this case, the number of independent variables are reduced to five and they are specified by the velocity field $u^{\mu}$ with $u^{\mu}u_{\mu}=1$ and the two proper scalar densities, $\varepsilon$ and $n$.
They are related to $j^{\mu}\left(  \mathbf{r} ,t\right)$, and $T^{\mu\nu}\left(  \mathbf{r},t\right)  $ as $n=\sqrt{j^{\mu}j_{\mu}}$, and $u^{\mu}$ and $\varepsilon$ are determined by the definition of Landau frame as the eigenvector and eigenvalue of $T^{\mu\nu }\left(  \mathbf{r},t\right)$.
The number density $n^{\ast}$ in an observing frame is given by $n\gamma$, where $\gamma=u^{0}$ is the Lorentz factor.
Now, one needs to determine the form of the functional to be optimized.

\subsubsection{Hydrodynamic Action and the Ideal Fluid}

Now we would like to construct the variational principle to describe the dynamics of these C-G variables.
By construction, the $\left(  0,0\right)  $ and $\left(  0,i\right)  ,i=1,2,3$ components of energy-momentum tensor $T^{\mu\nu}$ represent the energy and the momentum densities carried by the fluid element, respectively.
To stress this, let us denote them as
\begin{equation}
\left(
\begin{array}[c]{c}
\mathcal{H}\\
\mathbf{\pi}
\end{array}
\right)  \equiv\left(
\begin{array}[c]{c}
T^{00}\\
T^{0i}
\end{array}
\right)  .
\end{equation}
Since $E=\int d^{3}\mathbf{r}\mathcal{H}$ is the total energy of the system, we can identify $\mathcal{H}$ with the Hamiltonian density.
Now, from Landau's definition of the local rest frame $T^{0\nu}u_{\nu}=\varepsilon u^{0}$ we obtain
\begin{equation}
-\varepsilon=\mathbf{\pi\cdot v}-\mathcal{H},
\end{equation}
where $\mathbf{v}$ is the velocity of the fluid element.
This equation indicates that the quantity $-\varepsilon$ should be the Lagrangian density of the system ($L=p\dot{q}-H$).
Therefore, the optimal dynamics will be given by modeling the action,
\begin{equation}
I=-\int d^{4}x\ \varepsilon\left(  x\right)  .
\label{HydroAction}
\end{equation}
in terms of hydrodynamic variables, i.e., the local proper thermodynamic variables and the velocity field $\mathbf{v}$.
For example, for an adiabatic motion of fluid element, we can specify $\varepsilon$ as a function of any of the conserved proper densities, say $\rho$\footnote{For instance, $\rho$ can be chosen to be the entropy density, $s$ in ideal hydrodynamics \cite{Kodama:2001qv}.}.
In terms of the density in the observational system of reference $\rho=\rho^{\ast}/\gamma$ and the action is
\begin{equation}
I\left[  \rho,\mathbf{v}\right]  =-\int d^{4}x\ \varepsilon\left(
\frac{\rho^{\ast}}{\gamma}\right)  ,
\label{ModelAction}
\end{equation}
where $\gamma=1/\sqrt{1-\mathbf{v}^{2}}$ is the usual Lorentz factor.
The variation should be taken with respect to the density $\rho\left(  x\right)  $ and the velocity field $\mathbf{v}\left(  x\right)  $.
In fact, it is known that this procedure reproduces the equations of motion ideal relativistic hydrodynamics \cite{Elze:1999kc}.

Usually the variational procedure is performed in Eulerian coordinates and a constraint among $\rho$ and $\mathbf{v}$ has been introduced using the Lagrange multiplier method.
Below we sketch how to directly derive the equation of motion from the action Eq.\ (\ref{ModelAction}) using Lagrange coordinates.

We consider the situation where there is no chaotic motion and the time extension is finite so that we can follow the fluid motion for each infinitesimal fluid elements from the beginning.
Let $\mathbf{r}_{\mathbf{R}}\left(  t\right)  $ be the space coordinate of a fluid element at time $t$ whose initial position was $\mathbf{R}$.
We assume that $\mathbf{r}_{\mathbf{R}}\left(  t\right)  $ and $\mathbf{R}$ have an one to one correspondence for any $t$.
The fluid velocity is $\mathbf{v}_{\mathbf{R}}\left(  t\right)  =d\mathbf{r}_{\mathbf{R}}\left(  t\right)  /dt$.
Let $\rho_{0}\left(  \mathbf{R}\right)  $ be the initial distribution of the fluid elements.
Without any loss of generality, we can always choose the distribution of initial fluid elements to be $\rho_{0}=const$.
Then, the fluid density distribution $\rho\left(  \mathbf{r},t\right)  $ is determined as
\begin{equation}
\rho^{\ast}\left(  \mathbf{r},t\right)  =\frac{1}{J}\rho^{\ast}_{0} \label{densities}
\end{equation}
where $J=\det\left\vert \partial\mathbf{r/\partial R}\right\vert$,
is the Jacobian of the transformation $\mathbf{R}\rightarrow\mathbf{r}=\mathbf{r}_{\mathbf{R}}\left(  t\right)  $.
Equation (\ref{densities}) is a direct consequence of the conservation of the number of particles in the fluid element, $\rho\left(  \mathbf{r},t\right)  d^{3}\mathbf{r}=\rho_{0}d^{3}\mathbf{R}$.
The action (\ref{ModelAction}) can be re-expressed as
\begin{equation}
I\left[  \left\{  \mathbf{r}_{\mathbf{R}},\mathbf{v}_{\mathbf{R}}\right\}
\right]  =\int dt\int d^{3}\mathbf{R}~\mathcal{L}\left[  J,\gamma\right]  ,
\label{ActionLagrange}
\end{equation}
with the Lagrangian density $\mathcal{L}$
\begin{equation}
\mathcal{L}\left[  J,\gamma\right]  =-J\ \varepsilon\left(  \frac
{\rho^{\ast}_{0}}{\gamma\ J}\right)  .
\end{equation}

Note that the Lagrangian above contains only $t$ and $\mathbf{R}$ derivatives of the dynamical variables $\mathbf{r}_{\mathbf{R}}\left(  t\right)  $ through the Lorentz factor $\gamma$ and the Jacobian $J.$ Using the properties of the determinant, the term which appears in the Euler-Lagrange equation can be calculated as
\[
\sum_{k=1}^{3}\frac{\partial}{\partial\mathbf{R}_{k}}\left[  \frac
{\partial\mathcal{L}}{\partial J}\frac{\partial J}{\partial\left(
\partial\mathbf{r}_{\mathbf{R}}/\partial\mathbf{R}_{k}\right)  }\right]
=J\ \nabla_{\mathbf{r}_{\mathbf{R}}}\left(  \frac{\partial\mathcal{L}
}{\partial J}\right)  .
\]
Furthermore,
\begin{align}
\left(  \frac{\partial\mathcal{L}}{\partial J}\right)   &  =-\frac{1}{\rho^{\ast}_{0}}\frac{\partial}{\partial\left(  1/\rho\right)  }\rho^{\ast}_{0}\left(
\frac{\varepsilon}{\rho}\right) =-\frac{\partial U}{\partial V}=P
\end{align}
where $V=1/\rho$ is the specific volume, $U=\varepsilon/\rho$, is the specific energy, and $P$ is the pressure.
On the other hand, we find that
\begin{align}
\frac{\partial\mathcal{L}}{\partial\mathbf{v}}  &  =-\frac{\rho^{\ast}_{0}}{\rho
}\frac{\partial\varepsilon}{\partial\rho}\frac{d}{d\mathbf{v}
}\left(  \frac{\rho^{\ast}}{\gamma}\right)
  =\rho^{\ast}_{0}\ \frac{\varepsilon+P}{\rho}\mathbf{v}.
\end{align}
Thus, the action given in Eq.\ (\ref{ActionLagrange}) leads exactly to Eq.\ (\ref{RelEuler}),
\begin{equation}
\frac{d}{d\tau}\left[  \frac{\varepsilon+P}{\rho
}\mathbf{u}\right]  =-\frac{1}{\rho}\nabla P.
\label{EoM-Hydro-Larrange}
\end{equation}

Here we have shown that the relativistic Euler equation can be obtained from the variational principle, where the form of the action (\ref{ModelAction}) was inferred from the Landau definition of the fluid velocity as the time-like eigenvalue of the energy-momentum tensor.
However, once given the action Eq.\ (\ref{ModelAction}), the energy-momentum tensor can be derived as the Noether conserved current associated with the local translational invariance of the Lagrangian density \cite{Elsas:2014kua}.

Finally, we would like to make a comment on the applications of the variational approach on collective flow observables in relativistic heavy ion collisions.
In this formalism, we specify the final state of the collision in terms of macroscopic hydrodynamic variables which correspond to these observables, and we also introduce some model initial condition consistent with the centrality event selection.
Obviously, for a given event, the C-G variables are uniquely defined but the inverse is not true.
There are many events with distinct microscopic configurations that might give the same hydrodynamic final state.
This is also the case for the initial condition.
Furthermore, the C-G procedure can even be extended to include the averaging procedure over different events which are rather loosely classified (see the discussion in Ref.\ \cite{Denicol:2010br}).
This means that, in practice we may further allow a large statistical ensemble contained in the C-G scheme within a certain given resolution of the physical definition of the hydrodynamic event.
If the ensemble of all the microscopic dynamics (including the different events) equivalent within the C-G criteria is large enough, then the macroscopic variables specified by such C-G procedure distribute sharply around their mean-values as a consequence of the central limit theorem.
The variational approach with these C-G variables then results in the hydrodynamic picture of the event averaged flow.
This can be used to explain why smooth initial conditions reproduce the collective behavior of the event averaged collective observables reasonably well as the EbyE fluctuating initial condition hydrodynamics, and also why collective signals are so robust.

\subsubsection{Application of Variational Approach - Effective Hydrodynamics}

Once the variation principle is established for relativistic hydrodynamics, we can apply such an approach to obtain physical insight about the collective behavior by introducing a simplified parametrization of the fluid dynamics and approximately obtain the dynamical behavior of the solution.
For example, the equation for the dynamics of a cavitation bubble, known as Rayleigh-Plesst equation and its relativistic form can be derived from the variational approach \cite{Elze:1999kc}.
Also, many years ago in low energy nuclear physics the description of nuclear collective motions was often formulated in terms of the variational principle.
For instance, the dynamical mixture of Steinwedel-Jensen and Gamow-Teller modes in giant resonance phenomena is discussed using the variational approach and it described the behavior of the observed data very well \cite{Myers:1977zz}.
Such models can be considered as examples of the C-G procedure to extract the global collective dynamics.

\paragraph{SPH Formalism}

The same line of thinking can also be applied to obtain a numerical method to solve the hydrodynamic equations, called Smoothed Particle Method (SPH).
The SPH algorithm was first introduced in astrophysical applications \cite{Lucy:1977zz,Kitsionas:2002sp} and in Refs.\ \cite{Aguiar:2000hw,Hama:2004rr} this numerical method was applied in relativistic heavy-ion collisions using the variational approach discussed in the preceding subsection.
The basic idea of the SPH method is to parametrize the continuous density distribution of any extensive physical quantity in terms of a discrete sum of base functions.
This is somewhat the inverse way of the coarse-graining procedure for particle systems.
Let $a(\mathbf{r},t)$ be the density distribution of a conserved extensive quantity.
Then we parametrize
\begin{equation}
a(\mathbf{r},t)\rightarrow a_{SPH}(\mathbf{r},t)=\sum_{i}^{N}A_{i}
W(\mathbf{r-r}_{i};h),\label{aSPH}
\end{equation}
where as before, $W$ is normalized, $\int W(\mathbf{r-r}^{\prime};h)d^{3}\mathbf{r}^{\prime}=1\,,$ and having the property of finite support, $W(\mathbf{r-r}^{\prime};h)\rightarrow0,\ \mathrm{for}\ |\mathbf{r-r}^{\prime }|>h$, and the weight $A_{i}$ should be chosen appropriately to minimize the difference between $a(\mathbf{r},t)$ and $a_{SPH}(\mathbf{r},t)$ everywhere.
The above expression means that we are representing the continuous density as sum of finite number of unit distributions (kernel) carrying the quantity $A_{i}$.
These unit density distributions are centered at the position $\mathbf{r}_{i}\left(  t\right)  \,$.
Due to the normalization of the kernel $W$, we have
\[
\int a_{SPH}(\mathbf{r},t)d^{3}\mathbf{r}=\sum_{i}^{N}A_{i}\,,
\]
which should be the total value of the extensive quantity $A$ of the system.
\smallskip

For the application in hydrodynamics, we can use these parameters as the variational variables in Eq.\ (\ref{ModelAction}) by representing the conserved density, for example the entropy density $s,$ in the SPH form as
\begin{equation}
s_{SHP}(\mathbf{r},t)=\sum_{i}^{N}\nu_{i}W(\mathbf{r-r}_{i}(t);h)\,,
\label{rho_SPH}
\end{equation}
where the weight $\mathbf{\nu}_{i}$'s are constant in time.
Another extensive quantity, say $A$, is calculated as in Eq.\ (\ref{aSPH}) with weights, $A_{i}=\left(  a/s\right)_{i}\nu_{i}$.
In this scheme, the conservation of $s_{SPH}$ is automatically satisfied by defining
\begin{equation}
\mathbf{j}_{SPH}(\mathbf{r},t)=\sum_{i}\mathbf{\dot{r}}_{i}\,\nu
_{i}W(\mathbf{r-r}_{i}(t))\,.
\end{equation}

The effective Lagrangian Eq.\ (\ref{ModelAction}), written in the SPH representation, is
\begin{equation}
L_{SPH}\left(  \{\mathbf{r}_{i},\mathbf{{\dot{r}}}_{i}\}\right)  =-\sum_{i}%
\nu_{i}(\varepsilon/s)_{i}=-\sum_{i}\left(  \frac{E}{\gamma}\right)
_{i}, \label{L_SPH}%
\end{equation}
where $E_{i}=\left(  \varepsilon^{\ast}/s^{\ast}\right)  _{i}$ is the ``rest energy'' of the $i$-th ``particle'' which carries the amount of entropy $\nu_{i}$.
The equations of motion are obtained from the variational procedure as
\begin{equation}
\frac{d}{dt}\left(  \nu_{i}\frac{P_{i}+\varepsilon_{i}}{s_{i}}\,\gamma_{i}\,\mathbf{\dot{r}}_{i}\right) + \sum_{j}\nu_{i}\nu_{j}\bigg[\frac{P_{i}}{{s_{i}^{\ast}}^{2}}+\frac{P_{j}}{{s_{j}^{\ast}}^{2}}\bigg]\,\mathbf{\nabla}_{i}W(\mathbf{r}_{\,i}-\mathbf{r}_{\,j};h)=0\,. 
\label{SPH_eq}
\end{equation}
One of the advantages of the SPH approach is that we do not introduce numerical spatial derivatives using a difference method.
In fact, in SPH spatial derivatives can be calculated analytically in terms of the kernel $W$.
Another nice feature of SPH is its flexibility since it can be used in problems with initial conditions without any symmetry.
Furthermore, due to the Lagrangian nature of the method, it is suitable for explosive processes such as those found in relativistic heavy ion collisions and it is also very easy to construct the freeze-out surface in this method.
In addition, the variational approach guarantees that the SPH equations (\ref{SPH_eq}) give the optimal equation of motions for the parameters $\{\mathbf{r}_{i}(t)\}$ within the given total number of \textquotedblleft particles\textquotedblright.
In this approach, no numerical instabilities hardly occur since the whole system is a Lagrangian system and the total energy of the system is in principle conserved.

\paragraph{Chiral Field and Hydrodynamics}

The variational formulation \cite{Aguiar:2003pp} can be used to obtain the equations of motion of a relativistic plasma of quarks interacting with the mass generating chiral mean field at finite temperature \cite{Csernai:1995zn,Mishustin:1998yc,Scavenius:1999zc,Scavenius:2000bb,Paech:2003fe,Paech:2005cx,Nahrgang:2011vn,Nahrgang:2011mg,Herold:2013qda}.
The effective Lagrangian density for the four-component isovector chiral field, $\phi=\left( \sigma, \vec{\pi} \right)$, where $\sigma$ is a scalar field and $\pi^i$ are pseudoscalar fields playing the role of the pions, in the presence of the quark thermal bath which is at rest is expressed as
\begin{equation}
\mathcal{L}_{eff}^{(\phi)}=\frac{1}{2}\partial_{\mu}\phi\partial^{\mu}%
\phi-\Omega(T,\mu,\phi), \label{Leff-fai}%
\end{equation}
where $\Omega$ is the thermodynamic potential.
On the other hand, the effective Lagrangian in Eulerian form with constraints from the conserved quantities is
\begin{equation}
\mathcal{L}_{eff}^{\left(  fluid\right)  }    =-\epsilon(n,s)-n u^{\mu}\partial_{\mu}\lambda   -s u^{\mu}\partial_{\mu}\zeta+\frac{1}{2}w\left(  u^{\mu}u_{\mu}-1\right)  \;,
\label{L_eff}
\end{equation}
where $n$ and $s$ are the conserved proper number and entropy densities and $\lambda,$ $\zeta$ and $w$ are Lagrangian multipliers.
It can be shown that the Lagrangian density evaluated in the proper comoving frame of the fluid is equal to \cite{Aguiar:2003pp}
\begin{equation}
\mathcal{L}_{eff}^{\left(  fluid\right)  }=-\epsilon(n,s)+\mu n+Ts=p\;=-\Omega
. \label{L_eff_prop}%
\end{equation}
Our coupled system is then described by an effective Lagrangian
\begin{equation}
\mathcal{L}_{eff}^{\left(  \phi+fluid\right)  }    =\frac{1}{2}(\partial_{\mu}\phi)(\partial^{\mu}\phi)-\epsilon(n,s,\phi)-nu^{\mu}\partial_{\mu}\lambda -su^{\mu}\partial_{\mu}\zeta+\frac{1}{2}w\left(  u^{\mu}u_{\mu}-1\right) \;,
\label{L-coupled}
\end{equation}
and the corresponding equation of motions are
\begin{equation}
\Box\Phi=-R,\;
\qquad
\partial^{\nu}T_{\nu\mu}^{(fluid)}=R\partial_{\mu}\phi\;,
\label{divTmunu2}
\end{equation}
where $R$ is given as
\begin{equation}
R=\frac{\partial V(\phi)}{\partial\phi}+g\phi\rho(T,\mu,\phi)\;,
\end{equation}
with
\begin{equation}
\rho=\nu_{q}\int\frac{d^{3}k}{(2\pi)^{3}}\frac{1/E_{k}(\phi)}{e^{[E_{k}%
(\phi)-\mu_{q}]/T}+1}+(\mu_{q}\rightarrow-\mu_{q})
\end{equation}
is a scalar density for quarks.
Here $\nu_{q}$ stands for the color-spin-isospin degeneracy of the quarks, $E_{k}(\phi)=(\vec{k}^{2}+m_{q}^{2}(\phi))^{1/2}$, and $m_{q}(\phi)=(g^{2}\phi^{2})^{1/2}=g(\sigma^{2}+\vec{\pi}^{2})^{1/2}$ plays the role of an effective mass for the quarks.

\paragraph{Viscosity and Variational Approach}

When the fluid is not ideal, the above mentioned variational method requires the introduction of some other components to take into account the violation of time-reversal symmetry since the standard variational method is always time-reversible.
One way to accomplish this is to introduce a Rayleigh dissipation function and extend the variation of the internal energy to include dissipative effects.
In the context of general relativity, the Lagrange approach for viscous hydrodynamics has been developed \cite{Jiang:2014uxa,Andersson:2006nr}.
There, heat conduction and multiple component fluids are discussed.
The SPH approach can also be extended to include viscosities by introducing the time variation of the weight factor $\left\{  \nu_{i}\right\}  $ and equations of motion for viscous tensors \cite{Denicol:2009am,Denicol:2010tr,Noronha-Hostler:2013hsa}.

Another way to introduce dissipative phenomena in the variational formalism was developed by K. Yasue by replacing classical variables by stochastic ones \cite{Yasue1981327} in the so-called Stochastic Variational Method (SVM).
It has been shown that the non-relativistic NSF equation can be derived from the SVM \cite{Koide:2011sa} and this method also opens up an interesting possibility for the interpretation of the origin of quantum mechanics \cite{Koide:2011sa}.
However, unfortunately its relativistic form is not yet known and, in the pragmatic side, an application of the SVM in solving viscous hydrodynamics equations requires technically non-trivial procedures.

\subsection{Known Analytic Solutions}

\label{sec:analytic}

Several analytic solutions of relativistic hydrodynamics are known.
Naturally, they describe some specific situations but they often offer important physics insights and in some cases they serve as a powerful tool to study realistic situations even at the quantitative level.
Furthermore, analytic solutions can be used to check numerical codes, which is of fundamental importance in numerical calculations of relativistic hydrodynamics since they employ many subtle technical details to avoid numerical instabilities.

The first analytic solution of ideal relativistic hydrodynamics in (1+1)D was given by Khalatnikov \cite{Khalatnikov} for the massless ideal gas initially at rest and confined within a finite spatial interval with thickness, say $2\ell$.
This initial condition was used in Landau's model for the multiple pion production \cite{Landau:1953gs,Belenkij:1956cd}.
By changing the independent variables $\left(  t,z\right)  $ to the hydrodynamic variables $\left(  \alpha,\zeta\right)  ,$ where $\alpha$ is the fluid rapidity, and $\zeta=\ln\left(  T/T_{0}\right)  ,$ the light-cone variables, $\left(  t\pm z\right)  /\ell$ are expressed in terms of integrals of modified Bessel function $I_{0}\ $ containing dimensionless variables $\alpha$ and $\zeta$.
Therefore, to obtain the distributions of thermodynamic quantities and the fluid velocity as function of space and time, we still need to invert the resulting coupled transcendental equations.
In spite of this complexity, the Khalatnikov's analytical solution is useful to extract physical insights of the flow.
Landau predicted that the rapidity distribution of the final produced particles was approximately Gaussian with a width that was a function of the incident energy.
Further studies and applications of this model have been extensively done, including the generalization of the EoS from a gas of massless particles to an arbitrary, but constant speed of sound $c_{s}^{2}$ \cite{Shuryak:1972zq,Carruthers1974,Carruthers:1973ws,Denicol:2008zz,Wong:2008ex}.
The observed Gaussian rapidity distribution of charged particles in A--A collisions has been discussed within the scope of the Landau model \cite{Steinberg:2004vy,Netrakanti:2005iy,Wong:2008ex,Petersen:2006mp}.
For a more recent review on the Landau model and the Khalatnikov solution see \cite{Wong:2014sda}.
As an application of the Landau model in the case of constituent quarks, the system size dependence of the multiplicity and the rapidity distribution are discussed in Ref.\ \cite{Sarkisyan:2015gca}.

The Landau model and its analytic solution become very simple in the high energy asymptotic regime and at very high energy the central rapidity regime becomes boost invariant \cite{Hwa:1974gn,Chiu:1975hx,Chiu:1975hw}.
In this case, the rapidity of the fluid element is given exactly by
\begin{equation}
y=\eta\equiv\frac{1}{2}\ln\frac{t+z}{t-z}. \label{BoostInv}%
\end{equation}
As mentioned in Sec.\ \ref{sec:INIC}, Bjorken argued physically that such a situation should occur in relativistic heavy ion collisions at ultra-relativistic energies and derived a simple formula to estimate the energy density in the initial state \cite{Bjorken:1982qr}.
Although now we know that the real situation does not exactly satisfy those scaling conditions, this has been used in many (2+1)D calculations in the central rapidity domain.
It is also emphasized in \cite{Sen:2014pfa} that, within realistic values of parameters in A--A collisions, the time scale for which the longitudinal flow approaches the Bjorken scaling limit starting from the full-stopping initial condition is too large even at LHC energies, which suggests the necessity of some longitudinal flow profile already in the initial state.

In order to take into account the non-boost invariant nature of heavy ion collisions, Bialas, Janik, and Peschanski \cite{Bialas:2007iu} generalized Eq.\ (\ref{BoostInv}) as $2y=\ln f_{+}(t+z)-\ln f_{-}\left(  t-z\right)  ,$ where $f_{\pm}$ are functions to be determined by the hydrodynamic equations.
These functions obey an identical simple first order differential equation, and from this they have obtained a one-parameter family of exact solutions for relativistic hydrodynamics in (1+1) dimensions, which interpolates between the two limiting solutions of a Hwa-Bjorken boost invariant case and the Landau-Khalatnikov full-stopping initial condition \cite{Bialas:2007iu}.

The Budapest group (Cs\"{o}rg\H{o}, Nagy, Csan\'{a}d) found a simple family of accelerating solutions in ideal relativistic hydrodynamics that are exact, explicit and analytic \cite{Csorgo:2006ax,Csorgo:2007ea,Nagy:2007xn,Csanad:2007iv}, by assuming that the solution depends only on the Rindler coordinates, $t=\tau\cosh\eta,~\ r=\tau\sinh\eta$.
Their solutions are expressed as
\begin{equation}
v=\tanh(\lambda\eta),\ \ \ \frac{p}{p_{0}}=\left(  \frac{\tau_{0}}{\tau
}\right)  ^{\lambda d\left(  \kappa+1\right)  /\kappa}\frac{1}{\cosh
^{\phi_{\lambda}\left(  d-1\right)  }\left(  \frac{\eta}{2}\right)  }
\end{equation}
where $v\left(  \tau,\eta\right)  $ and $p\left(  \tau,\eta\right)  $ are the velocity field and pressure, respectively, and $d$ is the dimension of the space.
It is important to note that the coordinate $r$ for $d>1$ is the radius of a $d$-dimensional sphere.
The constants parameters $\lambda,\kappa$ and $\phi_{\lambda}$ specify the type of physical solutions (in particular $\kappa^{-1}$ is the velocity of sound squared, specifying the EoS as before), but not all of these constants are independent and they are constrained for given value of $d$.
Depending on the constraint, some of these solutions reduce to already known solutions \cite{Hwa:1974gn,Bjorken:1982qr,Biro:1999eh,Biro:2000nj,Csorgo:2003rt,Csorgo:2003ry} while others are new and accelerating ones.
The accelerating solution can be used, for example, to estimate the initial energy density from the experimental observed rapidity distribution, and they showed that the Bjorken formula significantly underestimates the real value at RHIC A--A collisions since it neglects the effects of work done by acceleration.
For a more detailed discussion we refer the reader to Refs.\ \cite{Nagy:2007xn,Csanad:2007iv}.

Pratt developed a method to solve the equation of motion of relativistic hydrodynamics in the co-moving (Lagrangian) coordinates and obtained an explicit solution for the (1+1)D dimensional case with an arbitrary value of the speed of sound (the parameter $\kappa$ above) for the initial Gaussian density profile.
He analyzed the longitudinal acceleration of this model, in particular with respect to the HBT measurements of longitudinal size \cite{Pratt:2008jj}.

All of above mentioned analytic solutions are essentially (1+1)D solutions due to symmetry even if the space has $d$ spatial dimensions.
The first (3+1)D simple, explicit analytic solution for cylindrically symmetric case was found by Bir\'{o} \cite{Biro:1999eh}, where the transverse expansion of the fluid involved the equation of motion corresponding to a mixed phase with a first order phase transition, $P=const$.
The solution was also extended to include non-zero initial velocity profiles \cite{Biro:2000nj}.
In this case the solution is still analytic but not anymore explicit.
Sinyukov and Karpenko \cite{Sinyukov:2004am} studied the same EoS to extend the ellipsoidal case to study elliptic flow in ideal hydrodynamics.

Khalatnikov's method to obtain transverse flow with longitudinal boost invariance was carried out by Peschanski and Saridakis \cite{Peschanski:2009tg} assuming that the transverse flow is quasi stationary and driven by the cooling of the temperature due to the longitudinal expansion.
Under this condition, the transverse flow pattern was calculated analytically by a similar method used in the Khalatnikov solution.
Identifying the entropy flow as particle flow, they studied the relation between the initial eccentricity and the final elliptic flow coefficient.

A different class of cylindrically symmetric analytic solutions was constructed by Lin and Liao \cite{Lin:2009kv} by embedding a transversal Hubble flow in the (3+1)D equation with an EoS of the type $\varepsilon=\kappa p$ and then constructing the longitudinal flow equation.
In addition to the 2D and 3D pure Hubble solutions, they obtained a set of non-trivial explicit solutions.

The method based on self-similar flow initially introduced to find solutions for non-relativistic cases known as Buda-Lund approach \cite{Bondorf:1978kz,Nucl.Phys.A305.226,Csorgo:1994fg,Helgesson:1997zz,Csizmadia:1998ef,Csorgo:1998yk,Akkelin:2000ex,Csorgo:2001xm,Csorgo:2001ru,Csorgo:1995bi,Csorgo:1995vf} has shown to be very powerful in studying the flow dynamics of non-relativistic hydrodynamics.
Recently this approach was extended to include (rigid) rotation, and explicit solutions with rotation were found \cite{Nagy:2009eq,Csorgo:2013ksa}.
In this line, a class of analytic solutions for truly 3-dimensional solutions for ideal hydrodynamics were obtained by generalizing the Bunda-Lund self-similar flow method to the relativistic cases \cite{Csorgo:2003rt,Csorgo:2003ry}.

The key point of this approach is that the continuity equation in $d$-dimensions can be solved for some specific flow profile, $\rho\left(  \mathbf{r},t\right)  =g\left(  t\right)  f\left( \mathbf{r}^{T}A^{-2}\mathbf{r}\right)  $, where $A$ is a time dependent diagonal $d\times d$ matrix $A=\mathrm{diag}\left(  X_{1},..,X_{d}\right)  ,$ and $f\left(  s\right)  $ is an arbitrary smooth function of the scaling variable $s=\mathbf{r}^{T}A^{-2}\mathbf{r}$ with normalization $\int_{0}^{\infty}s^{d-1}ds\ f\left(  s\right)  =1$.
The velocity profile is then given by $\mathbf{v}\mathbf{=}A^{-1}\dot{A}\left(  t\right)  \ \mathbf{r}$.
It is easy to verify that these ans\"{a}tze for $\rho$ and $\mathbf{v}$ satisfy the continuity equation for any $\left\{  X_{1}\left(  t\right),..,X_{d}\left(  t\right)  \right\}  $.
Taking the EoS as that of an ideal gas of massive quanta,
\begin{equation}
\varepsilon=mn+\varepsilon_{in},\ \ p=nT,
\end{equation}
accounts for the presence of mass flow.
The solutions given in Refs.\ \cite{Csorgo:2003rt,Csorgo:2003ry} have a simple, explicit analytic form and correspond to truly (3+1)D cases that contain the Hwa-Bjorken solution as a special case.
In this scenario, it was shown that the final observables are directly related to the initial distribution function $f$ and solutions of the Buda-Lund type predict a universal feature of the observables.
Such aspects were discussed in Ref.\ \cite{Csanad:2005gv}.
Recently, the Budapest group further extended these solutions to incorporate general, realistic equations of state \cite{Csanad:2012hr} and also multipolarity of the initial distribution to analyze the experimental data of higher order flow coefficients \cite{Csanad:2014dpa}.

These studies show that the choice of certain symmetry aspects of the solution play a fundamental role in constructing analytic solutions.
In the limit of conformal symmetry, a class of analytic solutions of the transverse flow with longitudinally boost invariant background was first shown by Ref.\ \cite{Gubser:2010ze}.
The conformal symmetry assumption also allowed to include first order viscous corrections \cite{Gubser:2010ze}.
This approach has been developed offering a more general theoretical scheme to study a variety of physical situations, including angular momentum and anisotropic flow in a perturbative form \cite{Gubser:2010ui}.

The generalization of Gubser's work to Israel-Stewart theory (IS) was carried out in Marrochio \textit{et al.} \cite{Marrochio:2013wla} and the first analytic (and semi-analytic) viscous solutions were obtained for the boost invariant transversely expanding flow.
These IS solutions \cite{Marrochio:2013wla} have well defined temperature and shear stress tensor profiles and do not display negative temperatures as in the case of the first order theory originally considered in Ref.\ \cite{Gubser:2010ui}.
In fact, in Ref.\ \cite{Marrochio:2013wla}a comparison between the first order case and IS was performed and the importance of the relaxation time was discussed.
More recently, Hatta, Noronha, and Xiao and collaborators in their series of works \cite{Hatta:2014gqa,Hatta:2014gga,Hatta:2014jva,Pang:2014ipa} explored various aspects of conformal symmetry, finding many new analytic solutions for viscous relativistic hydrodynamics in different situations and their properties were discussed.
In particular, in Ref.\ \cite{Hatta:2014gga}, they showed that several new solutions for second order viscous hydrodynamics can be obtained using different background geometries, which also include vorticity effects.
Furthermore, Ref.\ \cite{Hatta:2014jva} calculated analytically the harmonic flow parameters as perturbations around radial flow and obtained explicitly the $n$ dependence of $v_{n}$ and discussed also the effects of shear viscosity (a semi-phenomenological discussion on bulk viscosity was also included).

As we have seen, these solutions have played an important role in the sense that they offer many checking points of the physical assumptions assumed in hydrodynamic modeling.
In analytic approaches, obviously the real physical event by event fluctuating observables are not incorporated by definition.
However, some well-chosen observables reflect the symmetry of the system so that the analytic solutions which meet these symmetry requirements can be discussed in the framework of analytic approaches.
For such purposes, explicit and analytic solutions from which physical observables can be calculated also analytically are extremely useful.
For example, the Buda-Lund type of solutions clearly exhibit the observed scaling laws \cite{Csanad:2012hr}, such as the transverse energy dependence of HBT radii, mass dependences of the effective temperature of single particle spectra, and etc.
They also serve to study the fundamental problems intrinsic to the hydrodynamic approach, such as the onset of instabilities in relativistic viscous hydrodynamics for a given equation of state and transport coefficients.

\subsection{Miscellaneous Aspects of Relativistic Hydro Modeling}

Hydrodynamics represents the macroscopic motion of a fluid close to local thermal equilibrium.
If the system is far from equilibrium, it is not easy to describe the dynamics correctly only with macroscopic dynamical variables.
In fact, in numerical simulations with large values of $\eta/s$ or $\zeta/s$ the corresponding viscous correction to the single particle distribution function $\delta f$ becomes easily as large as the thermal equilibrium function $f_{eq}$.
This means that the basic assumption behind the kinetic derivations of dissipative hydrodynamics does not hold in these extreme cases.
However, in some specific situations we may expect that the macroscopic degrees of freedom can still cover the principal features of the dynamics of the system without resorting to a full microscopic description of the non-equilibrium dynamics.

This is the case of the very early stage of the hydrodynamical evolution where the there is still a clear distinction between the transverse and longitudinal directions as we mentioned before in Sec.\ \ref{sec:INIC} and also will be discussed in Sec.\ \ref{sec:DiscussionAndFuture}.
In such a situation, the single particle distribution in momentum space will not be isotropic in contrast to the case of LTE.
The extreme case where the longitudinal direction is free-streaming while the transverse plane is thermally equilibrated was first proposed in Ref.\ \cite{Heinz:2002rs}.
See also Ref.\ \cite{Bialas:2007gn}.
Later, a theoretical scheme of hydrodynamics which takes into account such an anisotropy in the momentum space including dissipative effects has been developed in Refs.\ \cite{Ryblewski:2010tn,Martinez:2010sc,Florkowski:2010cf}.
For comprehensive reviews see Refs.\ \cite{Strickland:2014pga,Strickland:2014eua} and the recent review of relativistic hydrodynamics by Jeon and Heinz where detailed discussions on this approach \cite{Jeon:2015dfa} are given.
Longitudinal/transverse difference in momentum space can also occur when the incident flow of the colliding matter is incorporated and such a situation is generally related to the discussion of rapidity dynamics of baryon rich matter.
To handle such a situation, multicomponent fluid models have been studied \cite{Mishustin:1988mj,Brachmann:1997bq,Ivanov:2014ioa}.
These models are considered to be appropriate in the description of the rapidity dependence of the direct flow parameter $v_1$ in the energy regime relevant for NICA and FAIR experiments.

A quite different view of the origin of the deviation of the single particle spectrum from that of Boltzmann comes from the so-called non-extensive statistics.
It has been long known that the single particle spectra of particles in high energy collisions decay exponentially as function of the transverse energy for lower values of $\sqrt{s}$ but, at higher values of $\sqrt{s}$, they have a power law tail.
In view of QCD, such power law tail of hard components can be well understood in terms of perturbative processes, and the soft exponential part can be understood in terms of canonical statistic effects.
Phenomenologically, both behaviors can be interpolated by the so-called L\'{e}vy-Tsallis distribution \cite{Abelev:2006cs,Adare:2011vy,Aamodt:2011zj,Khachatryan:2011tm,Aad:2010ac,Sogaard:2015tna}.
Tsallis derived this distribution postulating a non-extensive nature of the entropy \cite{GellMannTsallis}.
Some authors have discussed the possibility to incorporate such a formalism in the construction of dissipative hydrodynamics \cite{Osada:2008cn,Biro:2012ix}.
Osada argued that such a situation can be considered as a consequence of the existence of long-range correlations which introduce non-extensive behavior in the local thermodynamic relation \cite{Osada:2009cc} and also proposed a generalized matching condition including such a situation with the presence of particle and anti-particle distribution functions \cite{Osada:2011gx}.

An important point that has been less extensively discussed is the effect of angular momentum and the possible emergence of instabilities in fluids with low viscosity.
In the usual hydrodynamic description of relativistic heavy ion collisions, this subject has attracted little attention since it requires a high resolution calculation with a very small numerical viscosity.
However, if we look for genuine hydrodynamic signals, the nonlinear response which leads to instabilities (for example, Kelvin-Helmholtz instability) may offer important additional sets of observables as advocated in Ref.\ \cite{Csernai:2011qq}.
The study of relativistic hydrodynamics with spin was developed by Becattini \cite{Becattini:2011zz}.
The use of the effects of rotation and associated phenomena such as vortices and turbulences as new tools for precise understanding of hydrodynamic modes has recently been discussed in Ref.\ \cite{Csernai:2014nva}.
Recently, exact solutions of the relativistic Boltzmann equation in the relaxation time approximation (RTA) form have been constructed and used to derive dissipative hydrodynamic equations, including anisotropic hydrodynamics \cite{Florkowski:2013lya}.
Later, solutions of the RTA Boltzmann equation corresponding to Gubser flow were constructed and the dissipative hydrodynamics was derived \cite{Denicol:2014xca,Denicol:2014tha}.
These interesting subjects are quickly developing and are unfortunately beyond the scope of the present review (see, for example, Refs.\ \cite{Noronha:2015jia,Hatta:2015kia}).

\section{Dynamical Effects of Coarse-Graining in Hydrodynamics}

\label{sec:CoarseGraining}

So far we have seen that, in spite of the great success of the hydrodynamic description of relativistic heavy ion collisions and the general picture of the Little Bang being established as some kind of ``standard model'' of heavy ion collisions, the proper physical foundation of the validity of hydrodynamics in relativistic heavy ion collisions is yet far from being established.
The main quantitative question involves the validity of the LTE on an EbyE basis and the size of the coarse-graining (C-G) scale associated with it.
As we have mentioned, the physical observables at the moment are yet not sufficient to determine uniquely the hydrodynamic picture.
Furthermore, as suggested in Sec.\ \ref{sec:vp}, hydrodynamics may also correspond to an ensemble averaged dynamics, in accordance with the observables from the experimental measurements, so it is crucial to understand how precisely one is able to probe the system's profile.
If the observables investigated do not have enough resolution of the macroscopic dynamics, then the properties of the matter deduced from them can be just an effective description, e.g. corresponding to an effective EoS and transport coefficients.
Therefore, one needs to be careful with the assumptions required for the validity of the hydrodynamic description and whether they are supported by experimental observables.
A precise analysis of the profile of the system's dynamics may provide essential information about the properties of the matter as well as the initial condition of the collision (cf. discussions in Ref.\ \cite{Berges:2012ks}).
However, using only the available data, it is a difficult task to reconstruct the event profile evolution.

On the other hand, there exist several transport models which exhibit collective behavior similar to the matter created in heavy ion collisions.
In particular, the PHSD model developed by the FIAS/Frankfurt and Giesen Groups can reproduce well the observed collective signals (see for example, Refs.\ \cite{Cassing:2008sv,Cassing:2009vt,Cassing:2008nn,Peshier:2005pp,Cassing:2007yg,Cassing:2007nb}), but the LTE is not obviously being explicitly implemented in this approach.
Therefore, it is interesting to study the hydrodynamic aspects of such a model by introducing the hydrodynamic variables through the C-G procedure described in the Sec.\ \ref{sec:vp}.
In this way, we expect to test the correlation between the observed collective flow parameters and the concept of LTE used in the hydrodynamic approach.

In the PHSD model, the transport dynamics are described in terms of quasi-particle modes of the Kadanoff-Baym equation \cite{Kadanoff1962}, and the effects of many-body interactions are taken into account in the form of a mean field, which in turn, generates the mass of the quasi-particles.
Therefore, this model is particularly suitable for our purposes since it represents the quantum mechanical microscopic transport dynamics in terms of effective (resummed) strongly interacting dynamical quasi-particles and off-shell hadrons as degrees of freedom \cite{Cassing:2008nn,Cassing:2008sv,Cassing:2009vt,Bratkovskaya:2011wp}.
In this section, we report shortly on the hydrodynamic representation of the PHSD dynamics and see how much the LTE is attained dynamically.

\subsection{Implementation of Coarse-Graining in PHSD}

A single event in the PHSD simulation provides the evolution of coordinates and momenta corresponding to the quasi-particles.
Therefore, it is straightforward to apply the C-G procedure similar to Sec.\ \ref{sec:vp} to obtain the macroscopic energy and momentum tensor,
\begin{equation*}
T^{\mu \nu }\left( x\right) =\sum_{i}\frac{p_{i}^{\mu }p_{i}^{\nu }}{p_{i}^{0}}W\left( \mathbf{r}-\mathbf{r}_{i}(t);h\right) ,
\end{equation*}
where $W$ is a normalized smoothing function (here we take a Gaussian with width $h$) and the sum is taken over all the quasi-particles of the event.
However, as mentioned in Sec.\ \ref{sec:vp}, we can also include the different ``events'' in this sum.

Roughly speaking, an event of the PHSD simulation is a kind of particle observation for a given quantum wave function of the system.
However, a single event does not carry information about the interaction among them.
Therefore, we need to introduce an ensemble of such events (referred as ``parallel events'') to take into account the dynamics of the wave-function which contains the information of many-body interactions.
Such information is determined by the particles of a number of parallel events starting from the same initial condition.
This number, referred here as NUM, represents the number of ensembles of parallel events to generate the mean-field effects.

By increasing NUM, the effect of the potentials shows up in the dynamics of the system's evolution.
On the other hand, if we take numbers that are too large, the corresponding energy density profile looses its particle nature, that is, we loose the resolution in the classical sense.
Thus, NUM also constitutes a parameter of the C-G procedure in the present approach.

\begin{figure}[tbh]
\begin{center}
\includegraphics[width=0.48\textwidth]{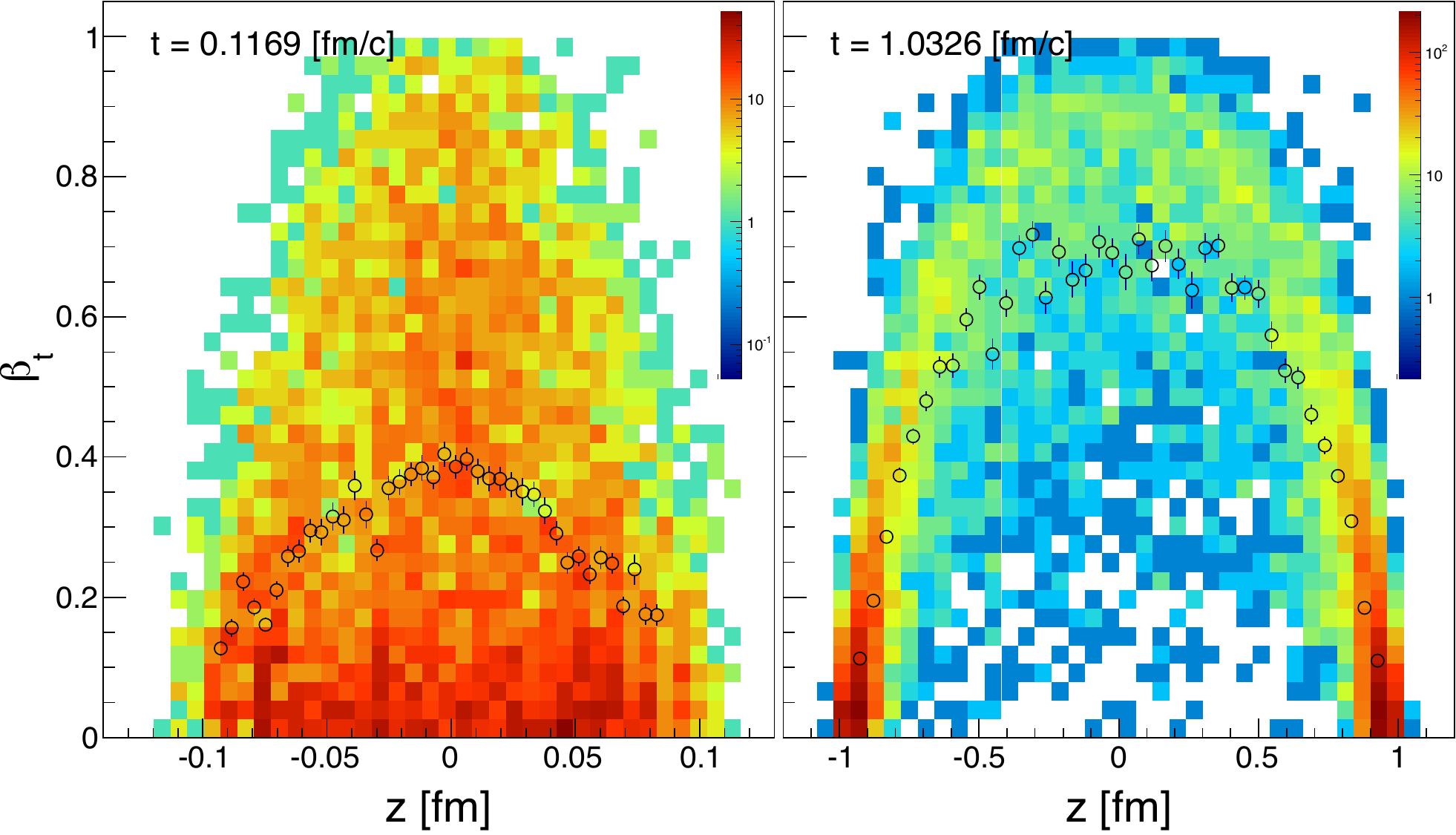}
\includegraphics[width=0.48\textwidth]{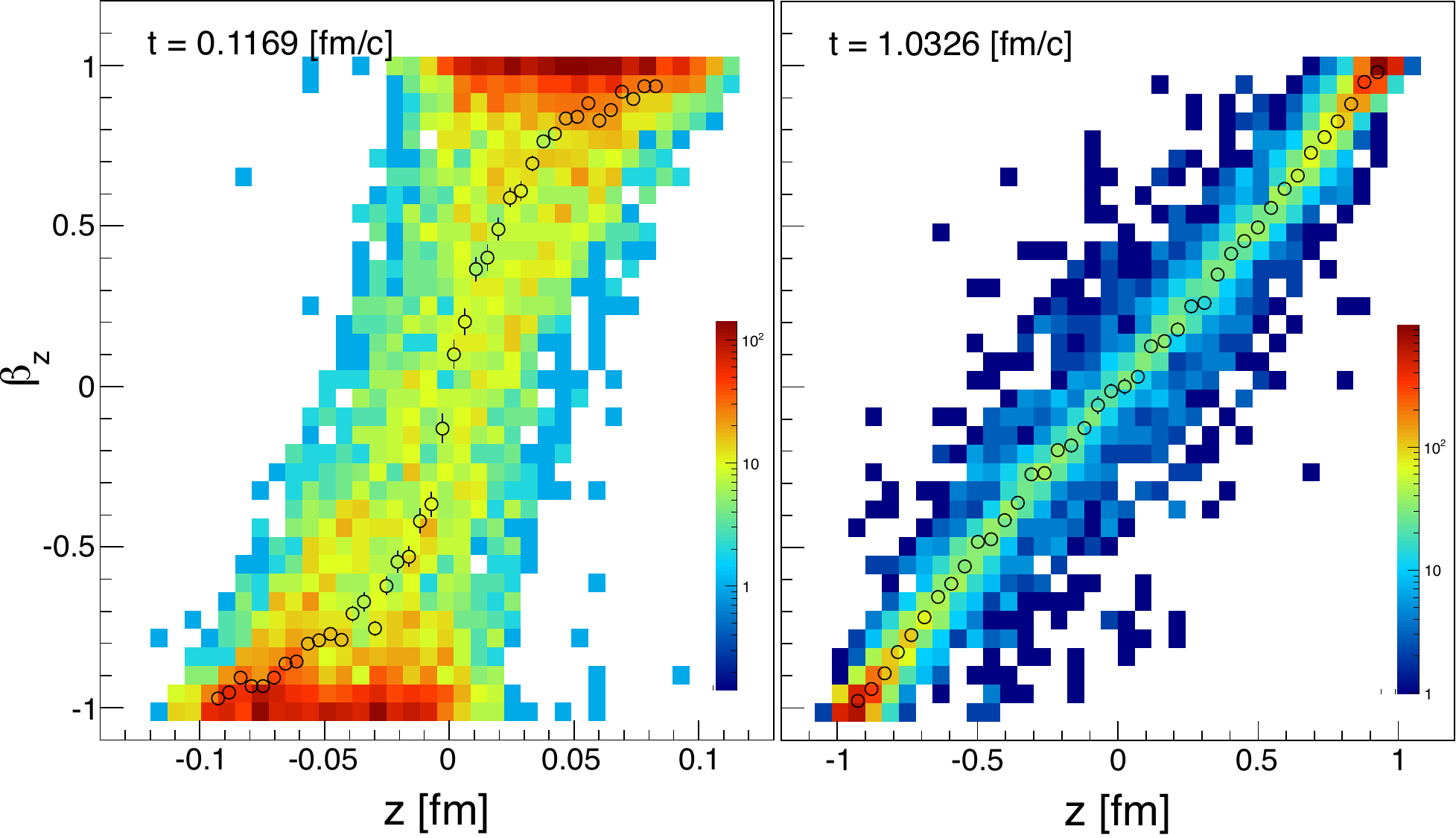}
\end{center}
\par
\vspace{-0.5cm}\caption{(color online) Snapshots of the longitudinal profile of the particle velocity in PHSD for two different time-steps of the evolution for a central Au--Au collision at $\sqrt{s_{NN}}$ = 200 GeV. The color scale in each panel represents the number of particles.}
\label{fig:phsd_snapshots}
\end{figure}

We have performed several tests using a simulated sample of events produced with PHSD in order to check the hydrodynamic properties extracted from the C-G procedure described above.
Here, we summarize the results presented in Ref.\ \cite{IWoC2014}.
The plots in Fig.\ \ref{fig:phsd_snapshots} show the longitudinal profiles of the components of the particle velocity for two snapshots of the evolution (the open circles represent the mean values of the distributions) for one single PHSD event of a Au--Au collision at $\sqrt{s_{NN}}$ = 200 GeV.

\begin{figure}[tbh]
\begin{center}
\includegraphics[width=0.4\textwidth]{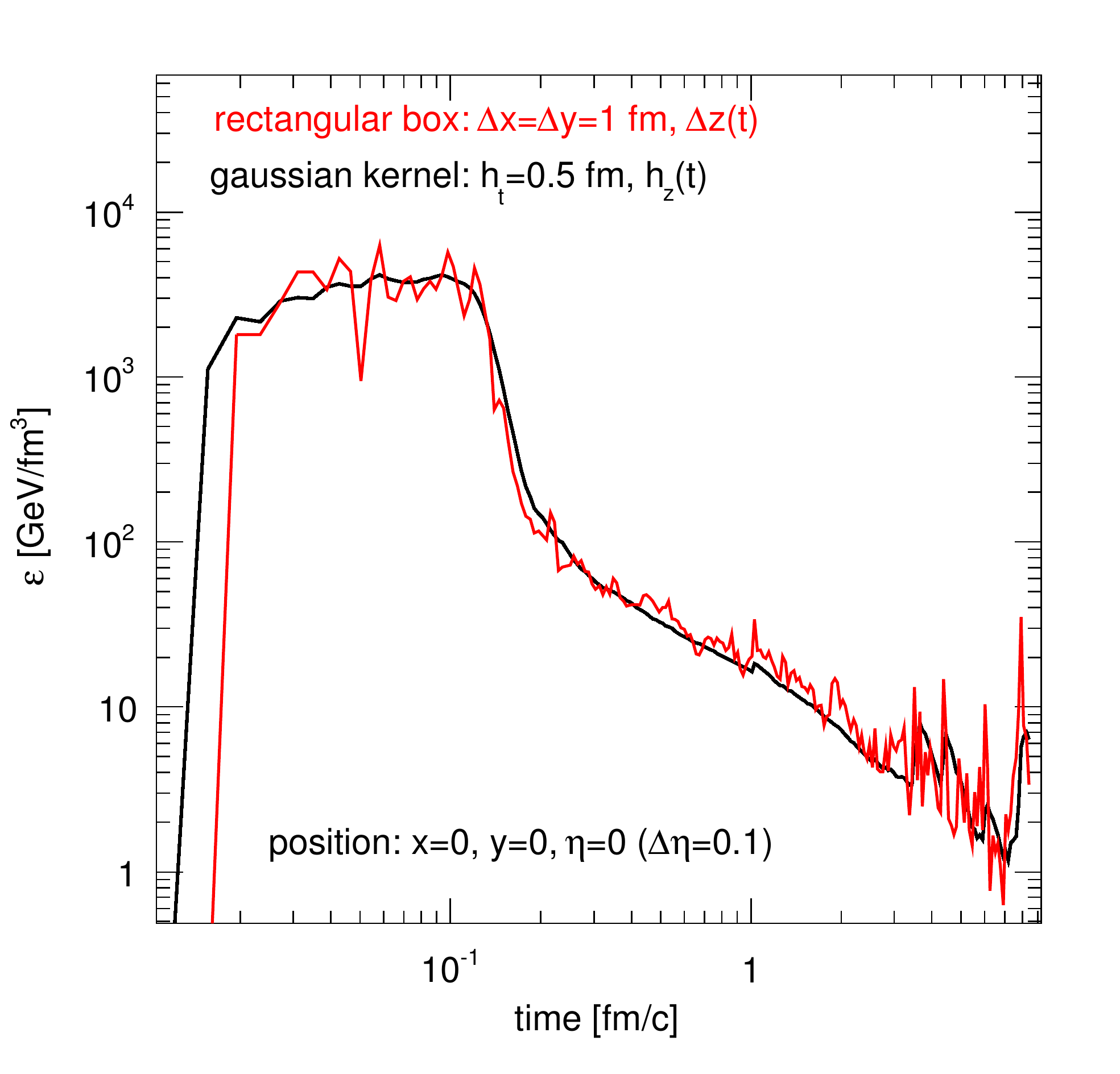}
\includegraphics[width=0.4\textwidth]{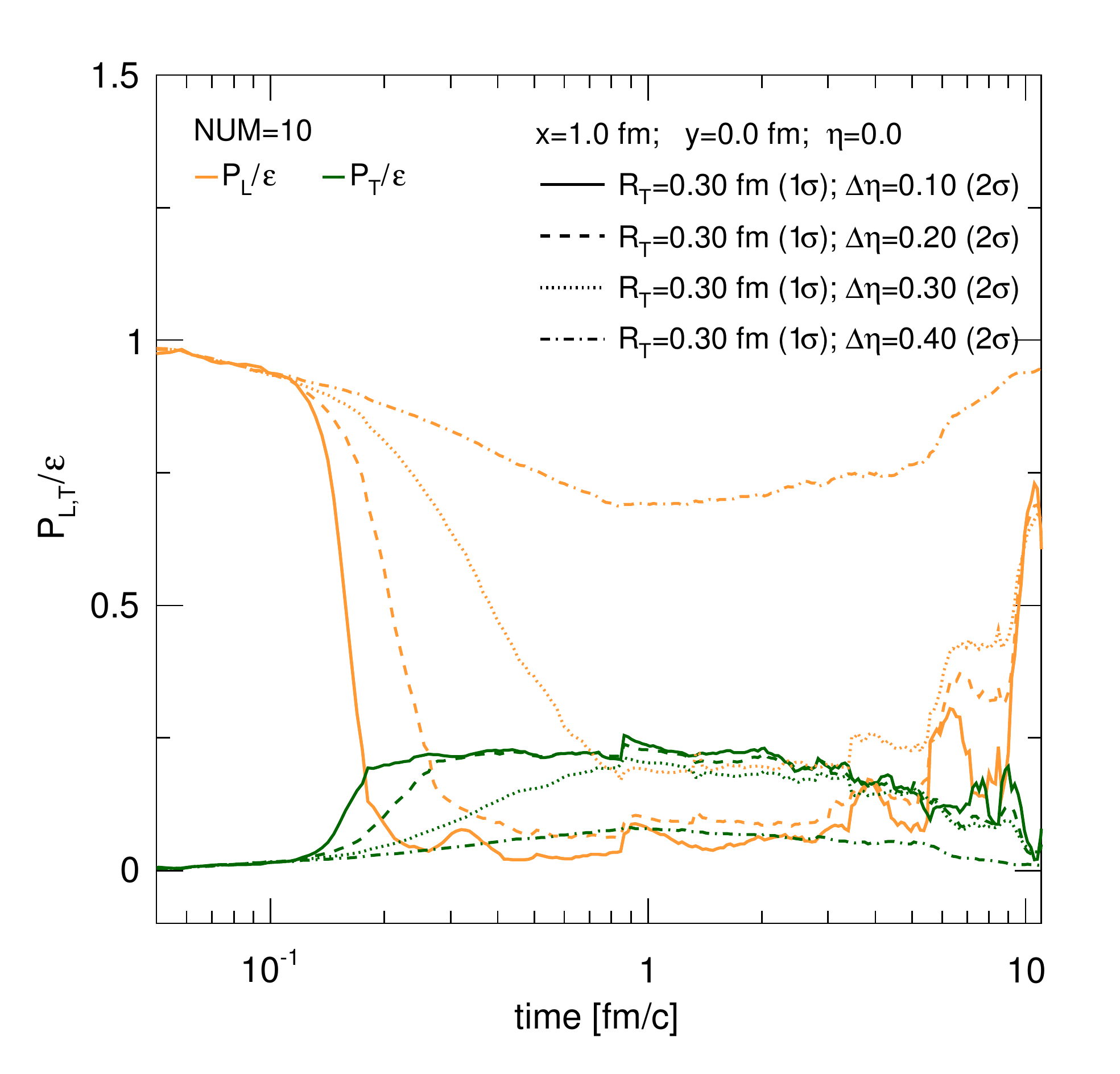}
\end{center}
\par
\caption{(color online) Dependence of the local energy density $\epsilon$ with the time evolution for two different types of kernel functions (left plot); dependence of the longitudinal and transverse pressure components with the system evolution considering different characteristic lengths in the longitudinal direction (right plot).}
\label{fig:kernel_and_pressure}
\end{figure}

We tested two types of kernel functions: (a) a simple rectangular box in which $W$=$1/(\Delta x\Delta y\Delta z)$ if the particle is inside the box and $W$=$0$ if it is outside, with $\Delta x$=$\Delta y$ being the transverse lengths and $\Delta z(t)$ the longitudinal length of the box; and (b) a 3D gaussian function with characteristic lengths defined by the widths $h_{t}$ in the transverse direction and $h_{z}(t)$ in the longitudinal direction.
The left plot in Fig.\ \ref{fig:kernel_and_pressure} shows the dependence of the local energy density as a function of time for the rectangular box (red curve) and the Gaussian kernel (black curve).
The energy-momentum tensor in this case was calculated in the neighborhood of the origin ($x$=$0$, $y$=$0$, $z$=$\eta$=$0$), considering a rapidity window $\Delta\eta$=$0.1$.
We tried to choose the sizes of each kernel to be roughly of the same order though they are not exactly equivalent.
As expected, both cases presented similar behavior with the Gaussian kernel being smoother and more strongly correlated.
The plot on the right hand side of Fig.\ \ref{fig:kernel_and_pressure} shows the dependence of the ratios of the longitudinal ($P_{L}$) and transverse ($P_{T}$) components of the pressure with respect to the proper energy density $\epsilon$ as a function of the time evolution of the system.
In this case we computed the $T^{\mu\nu}$ in the neighborhood of the point $(x,y,z)$=$(1,0,0)$ fm (avoiding the divergence at the origin) and four different sizes of the longitudinal length of the Gaussian kernel were tested, which are displayed in the legend of the plot by $\Delta\eta$ and corresponding to the range of $-\sigma$ to $+\sigma$ around the Gaussian peak.
The transverse length was fixed as $0.3$ fm (which corresponds to $1\sigma$ of the Gaussian kernel).
For all configurations tested we noticed that the transverse components of the pressure always start very low and then increase as the system evolves, while the longitudinal component always starts very high and then decrease.
For $\Delta\eta$=$0.3$ (dotted lines) the pressure components become isotropic and last for an interval of about $2-3$ fm/c, starting at $t$$\sim$$1.0$ fm/c.
However, with a further increase of the longitudinal length of the Gaussian kernel, the isotropy of the pressure components is not maintained.
Actually, the large size of $\Delta \eta$ introduces an artifact due to the mixing the collective longitudinal expansion.
On the other hand, in the infinitesimal limit of $\Delta \eta$, EbyE fluctuations become too large.
Different sizes for the transverse length of the kernel function were also tested without producing any important changes in the observed behavior.
Finally, we varied the parameter NUM (1, 2, 5, 10, 20 and 30) in order to check the effects due to the mean-field potential and we found that the general behavior converges for NUM$>$$5$, with the fluctuations being very large otherwise.

\subsection{Collectivity in the evolution of PHSD events}

In order to check whether the initial spatial anisotropy in the PHSD events is converted into a final momentum anisotropy as one would expect in a scenario with collectivity, we evaluated the time evolution of the eccentricities $\varepsilon_n$ of the particle distribution in the transverse plane as well as the flow coefficients $v_n$ computed with respect to the corresponding eccentricity phase angle: 
\begin{align}
\varepsilon_{n} &= \dfrac{\langle r^{n} \cos\left(  n \left[  \phi- \Phi_{n} \right]  \right)  \rangle}{r^{n}}, & \Phi_{n} &= \dfrac{1}{n} \arctan\left(  \dfrac{\langle r^{n} \sin(n\phi) \rangle }{\langle r^{n} \cos(n\phi) \rangle} \right), \\
v_{n} &= \langle\cos\left(  n \left[  \psi- \Psi_{n} \right]  \right)  \rangle, & \Psi_{n} &= \Phi_{n} + \pi/n,
\end{align}
where $r=\sqrt{x^2+y^2}$ is the distance of a given particle to the center of mass of the distribution in the transverse plane, $\phi= \arctan\left(  y/x \right)$ is the angle of its position vector with respect to the $x$-axis, and $\psi= \arctan\left(  p_{y}/p_{x} \right)$ is the angle of its momentum vector with respect to the $x$-axis.

\begin{figure}[htb]
\begin{center}
\includegraphics[width=0.4\textwidth]{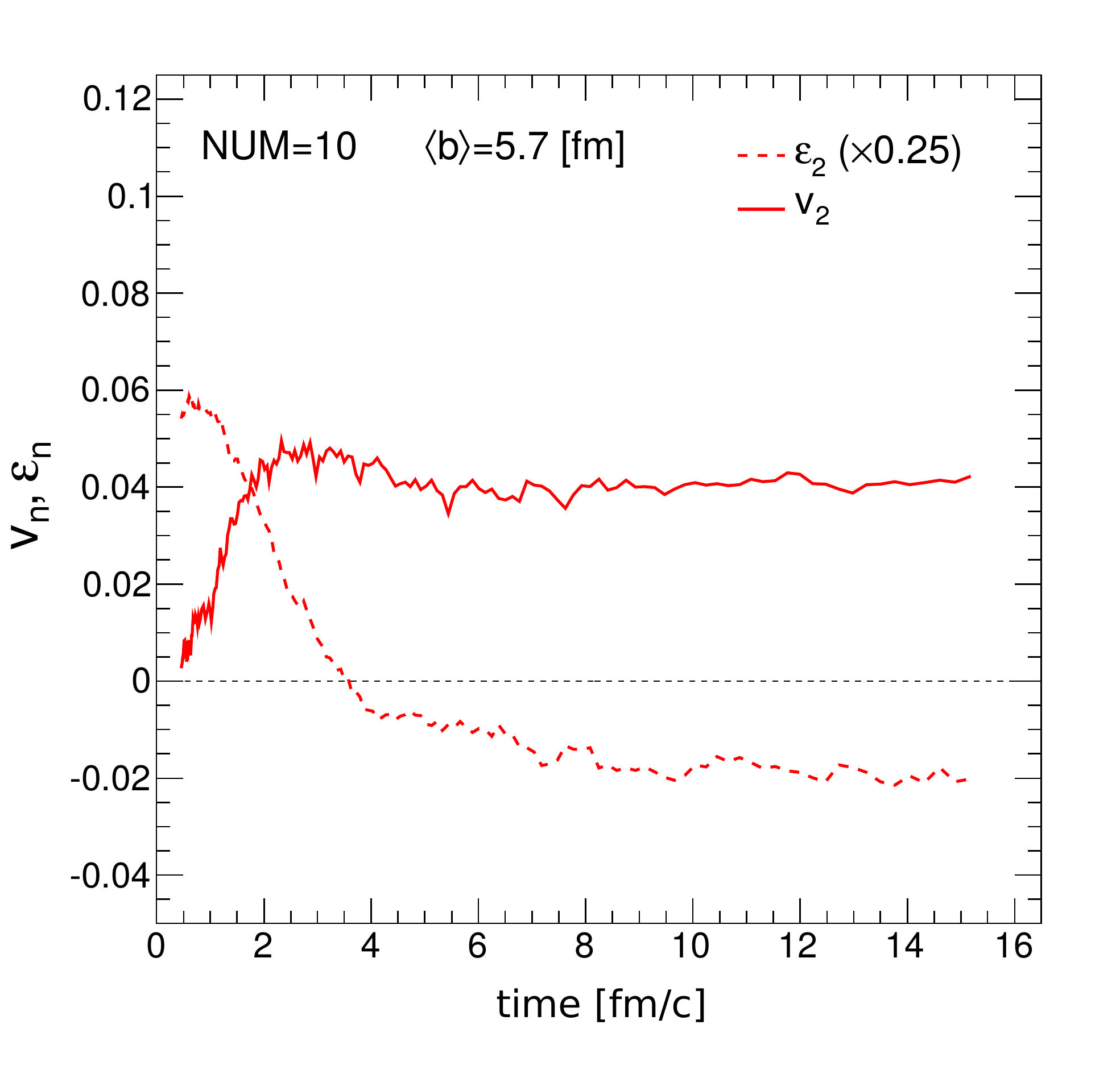}
\includegraphics[width=0.4\textwidth]{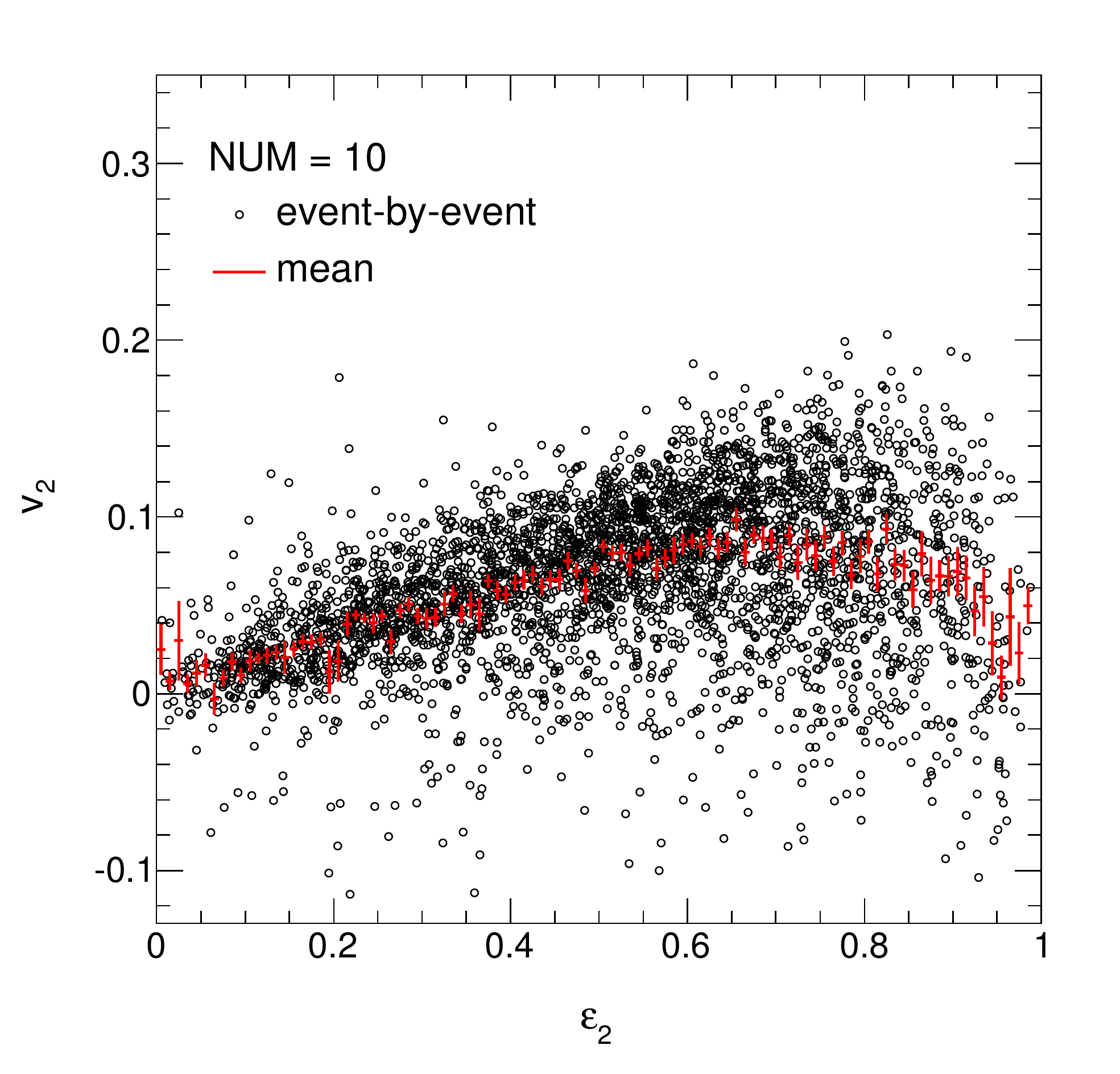}
\end{center}
\caption{Time evolution of the spatial eccentricity $\varepsilon_{2}$ and the elliptic flow coefficient $v_{2}$ for a single event (left plot); the distribution of $\varepsilon_{2}$ vs $v_{2}$ (right plot) for many events. The red line represents the average $\langle v_{2} \rangle$. }
\label{fig:v2_ecc2}
\end{figure}

The left plot in Fig.\ \ref{fig:v2_ecc2} shows the behavior of the coefficients $\varepsilon_{2}$ and $v_{2}$ as a function of the evolution of the system for a single PHSD event. 
The plot on the right shows the distribution of the final $v_{2}$ versus the initial $\varepsilon_{2}$ coefficients for a sample of many events.
The positive correlation observed is an indication of collective evolution of the system.
Moreover, the relevant transfer of the anisotropy from the initial to the final states occurs during the initial (partonic) stages.

Higher harmonics show similar behavior for central and intermediate cases, but the ``quality'' of the transfer becomes poor for the most peripheral cases.
Very similar results have been observed in ideal hydrodynamic calculations as discussed in Refs.\ \cite{Gardim:2011xv,Gardim:2012dc}, which demonstrates that even if complete LTE is not present, the usual observables that signal the system's collective behavior are similar and not sensitive to deviations from equilibrium.
Therefore, the development of new observables that are sensitive to deviations from equilibrium of the matter during the time evolution is of great interest in order to understand the collective properties of heavy ion collisions and the evolution of collectivity.

\section{Discussion and Future Problems}

\label{sec:DiscussionAndFuture}

The hydrodynamic picture has shown to be very successful in describing many aspects of the dynamics of the matter produced in relativistic heavy ion collisions.
Such a picture is consistent with the Little Bang picture of the Quark-Gluon Plasma.
In this work we have reviewed, from a pedagogical point of view, the basic structure and physical meaning of the relativistic hydrodynamic approach with a short historical overview.
We mention also the most recent developments based on the LHC and RHIC observations.
In particular, we emphasized the meaning of coarse-graining implicitly contained in the hydrodynamic description.
Although the hydrodynamic description achieved an overwhelming success in reproducing the experimental data, there still remain some important questions associated with the potential existence of an unknown C-G scale, which should be clarified.

For example, C-G is often loosely quantified in terms of the Knudsen number in kinetic derivations but reality is much more subtle because the kinetic approach itself contains a kind of C-G from the beginning.
At first sight, it seems that the Boltzmann approach is somewhat conflicting with the idea of a strongly interacting and dense sQGP fluid because of the assumption of dilute gas, which is the starting point of the Stossahlansatz.
In spite of this, the Boltzmann equation is considered to be very useful to discuss relativistic hydrodynamics in particular to study the structure of relativistically covariant dissipative terms.
This is so because, due to the assumptions of local binary collisions and the absence of memory effects, relativistic covariance is easy to maintain.
In the presence of non-local interactions and correlations due to three (or more) body collisions, it is quite difficult to satisfy relativistic covariance, as is known in the traditional cascade calculations \cite{Kodama:1983yk}.
In any case, in the derivation of hydrodynamics from a microscopic theory, the extraction of macroscopic variables and the associated C-G of space-time scales are the fundamental procedures.
In addition, to close the system of equations in these macroscopic variables, we introduce the thermodynamic/statistical properties (usually) implicitly depending on C-G.
Unfortunately, there is no established method to perform a C-G procedure which is applicable to any general situation in relativistic heavy ion collisions, and even a general method may not exist at all.
In fact, the C-G procedure is not known even for non-relativistic systems far from equilibrium.
Several approaches discussed in Sec.\ \ref{sec:DifferentFormulations} are examples of these C-G procedures in different situations.
In this sense, the formulation of relativistic hydrodynamics within the context of applications in the description of relativistic heavy ion collisions is not yet completely established and further investigations are necessary.

As discussed in Sec.\ \ref{sec:vp}, due to C-G one hydrodynamic event represents an extremely large number of microscopically distinct physical events.
The larger the C-G scale is, the easier a macroscopic description such as hydrodynamics becomes applicable.
However if a much larger C-G scale is used, we loose the required resolution in the space-time recognition.
In fact, we cannot observe inhomogeneities with smaller wavelength than the C-G scale.
This affects directly the class of observables that the model can describe.

Even though, we do not know a priori what kind of observables are insensitive to the C-G scale.
As an example, we recall that the success of hydrodynamics was first established by the analyzes of the elliptic flow coefficient using smooth initial conditions (s-IC) and, subsequently, EbyE analyzes with fluctuating initial conditions (f-IC) confirmed this hydrodynamic interpretation for $v_{2}$.
But this seems to be somewhat amazing because the hydrodynamic equation couples the fluid flow and the corresponding driving force in a very complicated nonlinear form.
If elliptic flow is due to genuine non-linear signature aspects of hydrodynamics, one might expect that the two analyses should give appreciable differences.
For example, if we consider an observable $O$ as functional of the initial condition $IC,$ in general we expect for a non-linear response, 
\[
\left\langle O\left[  IC\right]  \right\rangle \neq O\left[  \left\langle
IC\right\rangle \right]  ,
\]
due to the hydrodynamic evolution, where $\left\langle {}\right\rangle $ represents the average on different IC's.
In the equation above, on the right-hand side we identify s-IC as an average over all f-IC fixing a parameter such as centrality.

\begin{figure}[h]
\centering
\includegraphics[height=4.5cm,keepaspectratio]{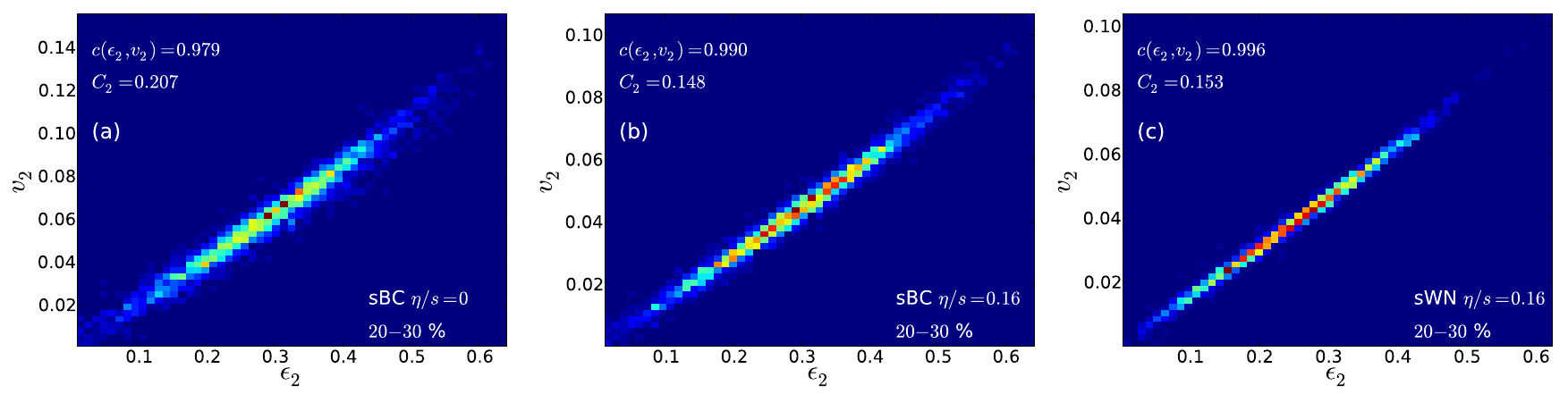}
\caption{Correlation $v_{2}$ vs $\varepsilon_{2}$ in fluctuating initial conditions. Figure\protect\footnotemarkB\ taken from Ref.\ \cite{Niemi:2012aj}.}
\label{v2-e2}
\end{figure}

In fact, some differences between the left and the right-hand sides have been observed for some observables such as the $p_{T}$ dependence of $v_{2}$, particle spectra, and HBT radii, but there is almost no difference in the integrated $v_{2},$ except for very central collisions.
These differences can easily be absorbed in the systematic uncertainties of the hydrodynamic modeling. This fact can be understood since for $v_{2}$ the effects from the initial pressure gradient, which is related directly to $\varepsilon_{2}$, is overwhelming over the small inhomogeneity in the initial conditions.
This picture was quantitatively demonstrated in Ref.\ \cite{Niemi:2012aj} using Monte-Carlo Glauber type initial conditions.
As shown in Fig.\ \ref{v2-e2} we first observe that these plots indicate that the average elliptic flow is very much insensitive to the fluctuations in the initial conditions since the width of the distribution of events around the average is very narrow for a given value of $\varepsilon_{2}$.
Second, the event-averaged $v_{2}$ has a very keen linear dependence on $\varepsilon_{2}$.
Therefore, any nonlinear aspect of genuine nonlinear hydrodynamic response is not contained in $v_{2}$.
In other words, the information of the detailed IC through $v_{2}$ is limited up to the resolution specified only by the initial eccentricity $\varepsilon_{2}$, and any other inhomogeneity is hindered.
In fact, if we use the integrated $v_{2}$ as a filter for initial condition selection, the corresponding C-G is too large and the hydrodynamic information obtained only from this observable is rather imprecise.
The fact that $v_2$ is almost linear in $\epsilon_2$ and does not show significant EbyE fluctuations indicate strongly that $v_2$ is almost uniquely determined by the initial geometry.
As is pointed out in Ref.\ \cite{Torrieri:2012wz}, several puzzling ``scaling properties'' are observed, in particular the quasi-independence of $v_2(p_T )/\epsilon_2$  with energy and system size, and the very good scaling of $\langle v_2 \rangle \sim \langle p_T \rangle \sim  (1/S)(dN/dy) $.
It is interesting to investigate whether these observations can be understood in hydrodynamics, and the role of  C-G
\footnotetextB{Reprinted with permission from H.\ Niemi \textit{et al.}, \textit{Event-by-event distributions of azimuthal asymmetries in ultra-relativistic heavy-ion collisions}, \textit{Phys.\ Rev}.\ C87 (2013) 054901. Copyright \textcopyright\ 2013 by the American Physical Society.}

The higher order harmonics are more sensitive to the effects of inhomogeneities, as seen from the fluctuation measure in scattered plots of $v_{n}$ vs $\varepsilon_{n}$ \cite{Niemi:2012aj}.
Of course, a more detailed structure of $v_{2}$ may also give a larger fluctuation, for example at large $p_{T}$, especially if the initial state dynamics is more peculiar than that of the Monte-Carlo Glauber one generating very sharp hot spots. 

As mentioned in Sec.\ref{sec:Overview}, in LHC experiments the anisotropic flow parameters can be determined in EbyE basis and their distributions can be measured, which were not possible in RHIC experiments.
Due to the strong correlations between $\left\{  v_{n}\right\}  $ and $\left\{  \varepsilon_{n}\right\}  $ in hydrodynamic calculations, the LHC data on flow parameters provide unique insight into the initial state of the matter produced in relativistic heavy ion collisions.
In fact, one can classify the event geometry class more accurately than only using centrality. In particular, not only the distributions, but also the correlations between $v_{n}$ and $v_{m}$ or their higher ones and event-plane correlations are considered to be useful to map the final state to the initial condition quite precisely \cite{Gale:2012rq,Niemi:2012aj,Yan:2015jma,Mota:2012qv}.
The opportunity of observing collective signals for one single event opens several remarkable possibilities associated with genuine nonlinear hydrodynamic behavior.
In particular, one needs to construct observables which are sensitive to the traditional genuine hydro-signals, such as shock waves \cite{Betz:2010qh,Mota:2010zz,Bouras:2010nt} and Rayleigh-Helmholtz instabilities \cite{LaszloBook} in terms of correlations of flow parameters in one single event.

Most of the discussions about the hydrodynamic behavior of the matter produced at RHIC and LHC until now are concentrated on the mid-rapidity region.
In this case, the boost-invariant aspects have been emphasized.
This is consistent with the CGC-Glasma type initial conditions where the classical gluon field is first generated (boost invariant) and then is subsequently converted into partons.
As mentioned in Sec.\ \ref{sec:INIC}, the pre-equilibrium energy-momentum tensor possesses anisotropic pressure components and the transverse pressure is dominant over the longitudinal one.
On the other hand, as shown in Fig.\ \ref{fig:kernel_and_pressure}, in the initial condition based on the PHSD model the longitudinal pressure in the corresponding stage is much larger than the transverse one.
This difference comes from the fact that in the PHSD approach the energy-momentum tensor is composed of particle-like objects (partons and strings).
The huge longitudinal pressure is generated by nearly free-streaming particles going in opposite directions.
This picture is not boost invariant at least in the very early stage where the usual hydrodynamic picture is not valid in the sense that the system is not in LTE.
However, as discussed in Ref.\ \cite{Strickland:2014pga} such a large value of the longitudinal pressure decreases very quickly due to the rapid longitudinal expansion and the transverse components dominate over the longitudinal one at $t\sim0.2fm$.
At this time, the transverse components become more or less isotropic indicating that the anisotropic hydrodynamic picture may hold.
On the other hand, in the CGC-Glasma picture, the energy-momentum tensor comes from the partons generated from a boost invariant classical gluon field.
These partons have mainly transverse momenta.
These apparently opposite behaviors represent a fundamental difference between the initial conditions for hydrodynamics constructed from transport and the CGC-Glasma pictures.
It should be noted that the PHSD and CGC-Glasma approaches are based on QCD to obtain the dynamics of the quantum wave function of the collisions, which possesses the wave-particle duality.
In our oppinion, it seems that the PHSD approach emphasizes the particle nature in this dynamics while the CGC-Glasma focuses on the wave nature (classical field).
Although this point represents an interesting question, any clear observable which distinguishes qualitatively the difference between the two pictures has not yet been proposed to the best of the authors' knowledge.

In the above discussion, boost invariance plays a fundamental role to distinguish the two pictures.
On the other hand, this refers only to the matter produced from the vacuum in the central rapidity region.
In fact, thermal model analyzes show that the baryonic chemical potential is close to null at RHIC and LHC, which shows that the particles come dominantly form the excitation of the vacuum.
However, for $\left\vert y\right\vert \gtrsim$ $1$, the rapidity distributions of charged particles are clearly deviated from the boost invariant form even at RHIC energies \cite{Steinberg:2007iv}.
Furthermore, for the low energy heavy ion collisions such as BES at RHIC and the CBM experiment where one expects a large stopping power of colliding nuclei, the effects of finite baryonic chemical potential will not be negligible and the boost invariant postulate is not adequate. In this aspect, insight from analytic approaches such as those in Refs.\ \cite{Bialas:2007iu,Nagy:2007xn} will be useful as mentioned in Sec.\ \ref{sec:analytic}.

The determination of the flow parameters based on an EbyE analysis, even for photons and leptons in LHC experiments, also opens the possibility of investigating the time evolution of the collective profile realized in relativistic heavy ion collisions.
As we have discussed, the proper formalism of relativistic hydrodynamics is still an open issue.
One does not even know if a consistent framework exists that allows for the derivation of relativistic dissipative hydrodynamics for strongly interacting matter because, in this case, the concepts of the theory of relativity and thermodynamics require two different conflicting conditions: strict locality vs.\ thermodynamic limit.
Studies of such observables related to the time evolution of the sytem will certainly offer a stringent check for the hydrodynamic picture and contribute
to the determination of the transport properties of the fluid in question.

In this review, we have explored several questions associated with relativistic hydrodynamics applied to relativistic heavy ion physics.
In spite of the overwhelming successes of the hydrodynamic description of the experimental data from SPS-RHIC to LHC energies, we pointed out that there exist some disturbing questions from the conceptual point of view, mainly related to the role of coarse-graining and its intrinsic scale.
There are some intriguing phenomena, for example, the observed flow patterns in p--p and p--A collisions and the associated ridge structure in two-particle correlations.
If they really fit in the hydrodynamic picture, then the associated coarse-graining scale should really be small.
If this is so, the realization of local thermal equilibrium occurs almost at the microscopic scale compared with the hadronic size, which would be rather amazing.
In this sense, it is important to investigate in detail whether the thermodynamic relations and transport coefficients which quantitatively reproduce the data in such small systems remain exactly the same as those used in A--A collisions on an EbyE basis.

Additionally, we pointed out that the boost-invariant approximation will break down when the nuclear stopping power becomes more effective.
In this case, the incident baryonic flow becomes more dominant and we expect that some new features may appear compared to the dynamics of the matter directly created from the QCD vacuum.

This field of physics is presently very quickly developing and new ideas and techniques are being actively proposed.
We believe that the several questions raised in this paper will be crucial to improve our understanding of the physics of relativistic heavy ion collisions and QCD dynamics in extreme conditions.

\bigskip

The authors express their sincere thanks to G. Denicol, J. Noronha, J. Noronha-Hostler, E. Fraga, G. Torrieri, T. Cs\"{o}rg\H{o}, Y. Sinyukov, H. Petersen, T. Osada, for invaluable suggestions and in particular to G. Denicol, J. Noronha, J. Noronha-Hostler, G. Torrieri and T. Cs\"{o}rg\H{o} for critical readings of the manuscript.
We also acknowledge E. Bratkovskaya and W. Cassing for providing the PHSD code and as well as the support concerning the physics of the model.
This work has been supported by CNPq, FAPERJ, CAPES, PRONEX and grant 2014/09167-8, S\~{a}o Paulo Research Foundation (FAPESP).

\setlength{\bibsep}{3pt}


\begin{thebibliography}{100}

\bibitem{ODarrigol.WorldsofFlow}
O.~Darrigol, Worlds of Flow: A History of Hydrodynamics from the Bernoullis to
  Prandtl  (Oxford University Press 2005), 1st ed.

\bibitem{LDLandau.FluidMechanics}
L.~Landau and E.~Lifshitz, Fluid Mechanics, Second Edition: Volume 6 (Course of
  Theoretical Physics)  (Butterworth-Heinemann 1987), 2nd ed.

\bibitem{WDMcComb.ThePhysicsofFluidTurbulence}
W.~D. McComb, The Physics of Fluid Turbulence (Oxford Engineering Science)
  (Oxford University Press 1992)

\bibitem{Hohne:2013pqa}
C.~{H\"{o}hne}, \emph{J. Phys. Conf. Ser.} 420 (2013) 012016

\bibitem{Schmidt:2014lva}
H.~R. Schmidt, \emph{J. Phys. Conf. Ser.} 509 (2014) 012084

\bibitem{Kolb:2003dz}
P.~F. Kolb and U.~W. Heinz, \emph{ar{X}iv:nucl-th/0305084}  (2003)

\bibitem{Heinz:2009xj}
U.~W. Heinz, \emph{Landolt-Bornstein} 23 (2010) 240

\bibitem{Ollitrault:2008zz}
J.-Y. Ollitrault, \emph{Eur. J. Phys.} 29 (2008) 275

\bibitem{Huovinen:2006jp}
P.~Huovinen and P.~Ruuskanen, \emph{Ann. Rev. Nucl. Part. Sci.} 56 (2006) 163

\bibitem{Sorensen:2009cz}
P.~Sorensen, \emph{ar{X}iv:0905.0174}  (2009)

\bibitem{Teaney:2009qa}
D.~A. Teaney, \emph{ar{X}iv:0905.2433}  (2009)

\bibitem{Hama:2004rr}
Y.~Hama, T.~Kodama and J.~Socolowski, O., \emph{Braz. J. Phys.} 35 (2005) 24

\bibitem{WFlorkowski.PhenomenologyOfURHIC}
W.~Florkowski, Phenomenology of Ultra-relativistic Heavy-ion Collisions  (World
  Scientific Publishing Company 2010)

\bibitem{Sarkar2009}
S.~Sarkar, H.~Satz and B.~Sinha (Eds.) The Physics of the Quark-Gluon Plasma:
  Introductory Lectures (Lecture Notes in Physics 785)  (Springer 2009)

\bibitem{Hirano:2012kj}
T.~Hirano \emph{et~al.}, \emph{Prog. Part. Nucl. Phys.} 70 (2013) 108

\bibitem{Int.J.Mod.Phys.A28.11.1340011}
C.~Gale, S.~Jeon and B.~Schenke, \emph{Int. J. Mod. Phys. A} 28 (2013) 1340011

\bibitem{Heinz:2013th}
U.~Heinz and R.~Snellings, \emph{Ann. Rev. Nucl. Part. Sci.} 63 (2013) 123

\bibitem{Jia:2014jca}
J.~Jia, \emph{J. Phys.} G41 (2014) 124003

\bibitem{Snellings:2014kwa}
R.~Snellings, \emph{J. Phys.} G41 (2014) 124007

\bibitem{Petersen:2014yqa}
H.~Petersen, \emph{J. Phys.} G41 (2014) 124005

\bibitem{Fermi:1950jd}
E.~Fermi, \emph{Prog. Theor. Phys.} 5 (1950) 570

\bibitem{Landau:1953gs}
L.~Landau, \emph{Izv. Akad. Nauk Ser. Fiz.} 17 (1953) 51

\bibitem{Belenkij:1956cd}
S.~Belenkij and L.~Landau, \emph{Nuovo Cim. Suppl.} 3S10 (1956) 15

\bibitem{Khalatnikov}
I.~Khalatnikov, \emph{Zh. Eksp. Teor. Fiz.} 27 (1954)

\bibitem{Hagedorn:1965st}
R.~Hagedorn, \emph{Nuovo Cim. Suppl.} 3 (1965) 147

\bibitem{Cabibbo:1975ig}
N.~Cabibbo and G.~Parisi, \emph{Phys. Lett.} B59 (1975) 67

\bibitem{Ericson:2003ya}
T.~E.~O. Ericson and J.~Rafelski, \emph{CERN Cour.} 43N7 (2003) 30

\bibitem{NoronhaHostler:2007jf}
J.~Noronha-Hostler, C.~Greiner and I.~Shovkovy, \emph{Phys. Rev. Lett.} 100
  (2008) 252301

\bibitem{NoronhaHostler:2008ju}
J.~Noronha-Hostler, J.~Noronha and C.~Greiner, \emph{Phys. Rev. Lett.} 103
  (2009) 172302

\bibitem{NoronhaHostler:2012ug}
J.~Noronha-Hostler, J.~Noronha and C.~Greiner, \emph{Phys. Rev.} C86 (2012)
  024913

\bibitem{Dashen:1969ep}
R.~Dashen, S.-K. Ma and H.~J. Bernstein, \emph{Phys. Rev.} 187 (1969) 345

\bibitem{Carruthers}
P.~Carruthers, \emph{Annals N. Y. Aca. Sci.} 229 (1974) 91

\bibitem{Cooper:1974mv}
F.~Cooper and G.~Frye, \emph{Phys. Rev.} D10 (1974) 186

\bibitem{Lee:1974ma}
T.~Lee and G.~Wick, \emph{Phys. Rev.} D9 (1974) 2291

\bibitem{Phys.Rev.Lett.21.1479}
W.~Scheid, R.~Ligensa and W.~Greiner, \emph{Phys. Rev. Lett.} 21 (1968) 1479

\bibitem{Scheid:1974zz}
W.~Scheid, H.~Muller and W.~Greiner, \emph{Phys. Rev. Lett.} 32 (1974) 741

\bibitem{Relativistic:1730719}
\emph{CERN Courier} 21 (1981) 393

\bibitem{Hewish:1968bj}
A.~Hewish \emph{et~al.}, \emph{Nature} 217 (1968) 709

\bibitem{Pilkington:1968bk}
J.~Pilkington \emph{et~al.}, \emph{Nature} 218 (1968) 126

\bibitem{Landau-1932}
L.~D. Landau, \emph{Phys. Z. Sowjetunion} 1 (1932) 285

\bibitem{Baade-Zwicky-PR-46-76-1934}
W.~Baade and F.~Zwicky, \emph{Phys. Rev.} 46 (1934) 76

\bibitem{NKGlendenning.CompactStars}
N.~K. Glendenning, Compact Stars: Nuclear Physics, Particle Physics, and
  General Relativity (Astronomy and Astrophysics Library)  (Springer 2000), 2nd
  ed.

\bibitem{Itoh:1970uw}
N.~Itoh, \emph{Prog. Theor. Phys.} 44 (1970) 291

\bibitem{LarryTDLee}
L.~McLerran and N.~Samios, T.D. Lee: Relativistic Heavy Ion Collisions and the
  RIKEN Brookhaven Center. ( 2006)

\bibitem{Baym:2001in}
G.~Baym, \emph{Nucl. Phys.} A698 (2002) XXIII

\bibitem{Fritzsch:1972jv}
H.~Fritzsch and M.~Gell-Mann, \emph{eConf} C720906V2 (1972) 135

\bibitem{Fritzsch:1973pi}
H.~Fritzsch, M.~Gell-Mann and H.~Leutwyler, \emph{Phys. Lett.} B47 (1973) 365

\bibitem{Fritsch:2012sw}
H.~Fritsch, \emph{CERN Cour.} 52N8 (2012) 21

\bibitem{Gross:1973id}
D.~J. Gross and F.~Wilczek, \emph{Phys. Rev. Lett.} 30 (1973) 1343

\bibitem{Politzer:1973fx}
H.~D. Politzer, \emph{Phys. Rev. Lett.} 30 (1973) 1346

\bibitem{Hanson:1975fe}
G.~Hanson \emph{et~al.}, \emph{Phys. Rev. Lett.} 35 (1975) 1609

\bibitem{Hanson:1981em}
G.~Hanson \emph{et~al.}, \emph{Phys. Rev.} D26 (1982) 991

\bibitem{Shuryak:1978ij}
E.~V. Shuryak, \emph{Phys. Lett.} B78 (1978) 150

\bibitem{Cugnon:1982qw}
J.~Cugnon, \emph{Nucl. Phys.} A387 (1982) 191C

\bibitem{Kodama:1983yk}
T.~Kodama \emph{et~al.}, \emph{Phys. Rev.} C29 (1984) 2146

\bibitem{Stocker:1981zz}
H.~Stocker \emph{et~al.}, \emph{Phys. Rev. Lett.} 47 (1981) 1807

\bibitem{Rafelski:1982pu}
J.~Rafelski and B.~Muller, \emph{Phys. Rev. Lett.} 48 (1982) 1066

\bibitem{Matsui:1986dk}
T.~Matsui and H.~Satz, \emph{Phys. Lett.} B178 (1986) 416

\bibitem{Rapp:1999ej}
R.~Rapp and J.~Wambach, \emph{Adv.Nucl. Phys.} 25 (2000) 1

\bibitem{Bjorken:1982tu}
J.~Bjorken, \emph{{FERMILAB-PUB-82-059-THY}}  (1982)

\bibitem{Yagi2005}
K.~Yagi, Quark-Gluon plasma : from big bang to little bang  (Cambridge
  University Press, Cambridge 2005)

\bibitem{Letessier2002}
J.~Letessier, Hadrons and Quark-Gluon Plasma  (Cambridge University Press,
  Cambridge 2002)

\bibitem{Nix:1979tc}
J.~Nix, \emph{Prog. Part. Nucl. Phys.} 2 (1979) 237

\bibitem{Prog.Part.Nucl.Phys.4.133-195}
H.~{St\"{o}cker} \emph{et~al.}, \emph{Progress in Particle and Nuclear Physics}
  4 (1980) 133

\bibitem{Kapusta:1982va}
J.~I. Kapusta, \emph{Phys. Rept.} 88 (1982) 365

\bibitem{Nagamiya:1982kn}
S.~Nagamiya and M.~Gyulassy, \emph{Adv.Nucl. Phys.} 13 (1984) 201

\bibitem{Harris:1996zx}
J.~W. Harris and B.~Muller, \emph{Ann. Rev. Nucl. Part. Sci.} 46 (1996) 71

\bibitem{Wong:1995jf}
C.~Wong, Introduction to high-energy heavy-ion collisions  (World Scientific,
  Singapore River Edge, NJ 1994)

\bibitem{LaszloBook}
L.~Csernai, Introduction to Relativistic Heavy Ion Collisions  (Wiley 1994)

\bibitem{Ritter:2014uca}
H.~G. Ritter and R.~Stock, \emph{J. Phys.} G41 (2014) 124002

\bibitem{Csernai:2014cwa}
L.~Csernai and H.~{St\"{o}cker}, \emph{J. Phys.} G41 (2014) 124001

\bibitem{Fukushima:2013rx}
K.~Fukushima and C.~Sasaki, \emph{Prog. Part. Nucl. Phys.} 72 (2013) 99

\bibitem{Aoki:2006we}
Y.~Aoki \emph{et~al.}, \emph{Nature} 443 (2006) 675

\bibitem{Asakawa:1989bq}
M.~Asakawa and K.~Yazaki, \emph{Nucl. Phys.} A504 (1989) 668

\bibitem{Fodor:2004nz}
Z.~Fodor and S.~Katz, \emph{JHEP} 0404 (2004) 050

\bibitem{Borsanyi:2012cr}
S.~Borsanyi \emph{et~al.}, \emph{JHEP} 1208 (2012) 053

\bibitem{Philipsen:2012nu}
O.~Philipsen, \emph{Prog. Part. Nucl. Phys.} 70 (2013) 55

\bibitem{McLerran:2007qj}
L.~McLerran and R.~D. Pisarski, \emph{Nucl. Phys.} A796 (2007) 83

\bibitem{deForcrand:2010ys}
P.~de~Forcrand, \emph{PoS} LAT2009 (2009) 010

\bibitem{Fukushima:2010bq}
K.~Fukushima and T.~Hatsuda, \emph{Rept. Prog. Phys.} 74 (2011) 014001

\bibitem{Tawfik:2014eba}
A.~N. Tawfik, \emph{Int. J. Mod. Phys.} A29 (2014) 1430021

\bibitem{BraunMunzinger:1995bp}
P.~Braun-Munzinger \emph{et~al.}, \emph{Phys. Lett.} B365 (1996) 1

\bibitem{Becattini:1997uf}
F.~Becattini, \emph{J. Phys.} G23 (1997) 1933

\bibitem{BraunMunzinger:2003zd}
P.~Braun-Munzinger, K.~Redlich and J.~Stachel, \emph{ar{X}iv:nucl-th/0304013}
  (2003)

\bibitem{Becattini:2002en}
F.~Becattini, \emph{J. Phys.} G28 (2002) 1553

\bibitem{Xu:2001zj}
N.~Xu and M.~Kaneta, \emph{Nucl. Phys.} A698 (2002) 306

\bibitem{Torrieri:2004zz}
G.~Torrieri \emph{et~al.}, \emph{Comput. Phys. Commun.} 167 (2005) 229

\bibitem{Wheaton:2004qb}
S.~Wheaton and J.~Cleymans, \emph{Comput. Phys. Commun.} 180 (2009) 84

\bibitem{Rafelski:1995vp}
J.~Rafelski, J.~Letessier and A.~Tounsi, \emph{Phys. Lett.} B390 (1997) 363

\bibitem{Stachel:2013zma}
J.~Stachel \emph{et~al.}, \emph{J. Phys. Conf. Ser.} 509 (2014) 012019

\bibitem{Becattini:1995if}
F.~Becattini, \emph{Z.Phys.} C69 (1996) 485

\bibitem{Becattini:1996gy}
F.~Becattini, \emph{ar{X}iv:hep-ph/9701275}  (1996)

\bibitem{LandauLifshitzBook}
L.~D. Landau and E.~M. Lifshitz, Fluid Mechanics (Course of Theoretical
  Physics)  (Pergamon Press 1959)

\bibitem{Koide:2013nia}
T.~Koide, R.~O. Ramos and G.~S. Vicente, \emph{Braz. J. Phys.} 45 (2015) 102

\bibitem{Kodama:2001qv}
T.~Kodama \emph{et~al.}, \emph{J. Phys.} G27 (2001) 557

\bibitem{Eckart:1940te}
C.~Eckart, \emph{Phys. Rev.} 58 (1940) 919

\bibitem{Becattini:2014yxa}
F.~Becattini \emph{et~al.}, \emph{Eur. Phys. J.} C75 (2015) 191

\bibitem{groot1962non}
S.~Groot and P.~Mazur, Non-equilibrium thermodynamics, North-Holland series in
  physics  (North-Holland Pub. Co. 1962)

\bibitem{Israel:1976tn}
W.~Israel, \emph{Annals Phys.} 100 (1976) 310

\bibitem{Israel:1979wp}
W.~Israel and J.~Stewart, \emph{Annals Phys.} 118 (1979) 341

\bibitem{Koide:2006ef}
T.~Koide \emph{et~al.}, \emph{Phys. Rev.} C75 (2007) 034909

\bibitem{Koide:2010wt}
T.~Koide, \emph{AIP Conf. Proc.} 1312 (2010) 27

\bibitem{Denicol:2009zz}
G.~Denicol \emph{et~al.}, \emph{J. Phys.} G36 (2009) 035103

\bibitem{Denicol:2008ha}
G.~Denicol \emph{et~al.}, \emph{J. Phys.} G35 (2008) 115102

\bibitem{Pu:2009fj}
S.~Pu, T.~Koide and D.~H. Rischke, \emph{Phys. Rev.} D81 (2010) 114039

\bibitem{Prakash:1993bt}
M.~Prakash \emph{et~al.}, \emph{Phys. Rept.} 227 (1993) 321

\bibitem{Muronga:2001zk}
A.~Muronga, \emph{Phys. Rev. Lett.} 88 (2002) 062302. \textit{Ibid.} 89 (2002)
  159901

\bibitem{Muronga:2006zw}
A.~Muronga, \emph{Phys. Rev.} C76 (2007) 014909

\bibitem{Hiscock:1985zz}
W.~A. Hiscock and L.~Lindblom, \emph{Phys. Rev.} D31 (1985) 725

\bibitem{Hiscock:1987zz}
W.~A. Hiscock and L.~Lindblom, \emph{Phys. Rev.} D35 (1987) 3723

\bibitem{Kouno:1989ps}
H.~Kouno \emph{et~al.}, \emph{Phys. Rev.} D41 (1990) 2903

\bibitem{Van:2011yn}
P.~Van and T.~Bir\'{o}, \emph{Phys. Lett.} B709 (2012) 106

\bibitem{GarciaPerciante:2008ui}
A.~Garcia-Perciante, L.~Garcia-Colin and A.~Sandoval-Villalbazo, \emph{Gen.
  Rel. Grav.} 41 (2009) 1645

\bibitem{Stoecker:1986ci}
H.~Stoecker and W.~Greiner, \emph{Phys. Rept.} 137 (1986) 277

\bibitem{Clare:1986qj}
R.~Clare and D.~Strottman, \emph{Phys. Rept.} 141 (1986) 177

\bibitem{Bjorken:1982qr}
J.~Bjorken, \emph{Phys. Rev.} D27 (1983) 140

\bibitem{Baym:1984sr}
G.~Baym \emph{et~al.}, \emph{Nucl. Phys.} A407 (1983) 541

\bibitem{Ornik:1989jp}
U.~Ornik, F.~Pottag and R.~Weiner, \emph{Phys. Rev. Lett.} 63 (1989) 2641

\bibitem{Rischke:1995ir}
D.~H. Rischke, S.~Bernard and J.~A. Maruhn, \emph{Nucl. Phys.} A595 (1995) 346

\bibitem{Hirano:2008hy}
T.~Hirano, N.~van~der Kolk and A.~Bilandzic, \emph{Lect.Notes Phys.} 785 (2010)
  139

\bibitem{Wang:1996yf}
X.-N. Wang, \emph{Phys. Rept.} 280 (1997) 287

\bibitem{Sjostrand:2006za}
T.~Sjostrand, S.~Mrenna and P.~Z. Skands, \emph{JHEP} 0605 (2006) 026

\bibitem{Drescher:2000ha}
H.~Drescher \emph{et~al.}, \emph{Phys. Rept.} 350 (2001) 93

\bibitem{Bleicher:1999xi}
M.~Bleicher \emph{et~al.}, \emph{J. Phys.} G25 (1999) 1859

\bibitem{Werner:2005jf}
K.~Werner, F.-M. Liu and T.~Pierog, \emph{Phys. Rev.} C74 (2006) 044902

\bibitem{Pang:2012he}
L.~Pang, Q.~Wang and X.-N. Wang, \emph{Phys. Rev.} C86 (2012) 024911

\bibitem{CGC}
L.~McLerran \emph{et~al.}, Relativistic Heavy Ion Physics
  (Landolt-{B\"{o}rnstein}: Numerical Data and Functional Relationships in
  Science and Technology - New Series / Elementary Particles, Nuclei and
  Atoms), vol.~23  (Springer 2010), 334+

\bibitem{Albacete:2014fwa}
J.~L. Albacete and C.~Marquet, \emph{Prog. Part. Nucl. Phys.} 76 (2014) 1

\bibitem{Paatelainen:2013eea}
R.~Paatelainen \emph{et~al.}, \emph{Phys. Lett.} B731 (2014) 126

\bibitem{Niemi:2015qia}
H.~Niemi, K.~J. Eskola and R.~Paatelainen, \emph{ar{X}iv:1505.02677}  (2015)

\bibitem{Gyulassy:1996br}
M.~Gyulassy, D.~H. Rischke and B.~Zhang, \emph{Nucl. Phys.} A613 (1997) 397

\bibitem{Osada:2001hw}
T.~Osada \emph{et~al.}, \emph{ar{X}iv:nucl-th/0102011}  (2001)

\bibitem{Aguiar:2000hw}
C.~Aguiar \emph{et~al.}, \emph{J. Phys.} G27 (2001) 75

\bibitem{Drescher:2000ec}
H.~Drescher \emph{et~al.}, \emph{Phys. Rev.} C65 (2002) 054902

\bibitem{Socolowski:2004hw}
J.~Socolowski, O. \emph{et~al.}, \emph{Phys. Rev. Lett.} 93 (2004) 182301

\bibitem{Andrade:2006yh}
R.~Andrade \emph{et~al.}, \emph{Phys. Rev. Lett.} 97 (2006) 202302

\bibitem{Alver:2010gr}
B.~Alver and G.~Roland, \emph{Phys. Rev.} C81 (2010) 054905

\bibitem{Bass:1998ca}
S.~Bass \emph{et~al.}, \emph{Prog. Part. Nucl. Phys.} 41 (1998) 255

\bibitem{Petersen:2008dd}
H.~Petersen \emph{et~al.}, \emph{Phys. Rev.} C78 (2008) 044901

\bibitem{Cassing:2008nn}
W.~Cassing, \emph{Eur. Phys. J.ST} 168 (2009) 3

\bibitem{Cassing:2008sv}
W.~Cassing and E.~Bratkovskaya, \emph{Phys. Rev.} C78 (2008) 034919

\bibitem{Cassing:2009vt}
W.~Cassing and E.~Bratkovskaya, \emph{Nucl. Phys.} A831 (2009) 215

\bibitem{Bratkovskaya:2011wp}
E.~Bratkovskaya \emph{et~al.}, \emph{Nucl. Phys.} A856 (2011) 162

\bibitem{Drescher:2006ca}
H.-J. Drescher and Y.~Nara, \emph{Phys. Rev.} C75 (2007) 034905

\bibitem{Drescher:2007ax}
H.-J. Drescher and Y.~Nara, \emph{Phys. Rev.} C76 (2007) 041903

\bibitem{Kharzeev:2004if}
D.~Kharzeev, E.~Levin and M.~Nardi, \emph{Nucl. Phys.} A747 (2005) 609

\bibitem{Bartels:2002cj}
J.~Bartels, K.~J. Golec-Biernat and H.~Kowalski, \emph{Phys. Rev.} D66 (2002)
  014001

\bibitem{Kowalski:2003hm}
H.~Kowalski and D.~Teaney, \emph{Phys. Rev.} D68 (2003) 114005

\bibitem{Schenke:2012wb}
B.~Schenke, P.~Tribedy and R.~Venugopalan, \emph{Phys. Rev. Lett.} 108 (2012)
  252301

\bibitem{Dumitru:2012yr}
A.~Dumitru and Y.~Nara, \emph{Phys. Rev.} C85 (2012) 034907

\bibitem{Karsch:2006sm}
F.~Karsch, \emph{AIP Conf. Proc.} 842 (2006) 20

\bibitem{Durr:2010vn}
S.~Durr \emph{et~al.}, \emph{Phys. Lett.} B701 (2011) 265

\bibitem{Borsanyi:2014rza}
S.~Borsanyi \emph{et~al.}, \emph{ar{X}iv:1410.7917}  (2014)

\bibitem{Huovinen:2009yb}
P.~Huovinen and P.~Petreczky, \emph{Nucl. Phys.} A837 (2010) 26

\bibitem{Petreczky:2012rq}
P.~Petreczky, \emph{J. Phys.} G39 (2012) 093002

\bibitem{Danielewicz:1984ww}
P.~Danielewicz and M.~Gyulassy, \emph{Phys. Rev.} D31 (1985) 53

\bibitem{Kovtun:2004de}
P.~Kovtun, D.~T. Son and A.~O. Starinets, \emph{Phys. Rev. Lett.} 94 (2005)
  111601

\bibitem{Song:2009rh}
H.~Song and U.~W. Heinz, \emph{Phys. Rev.} C81 (2010) 024905

\bibitem{Ryu:2015vwa}
S.~Ryu \emph{et~al.}, \emph{ar{X}iv:1502.01675}  (2015)

\bibitem{Gale:2012rq}
C.~Gale \emph{et~al.}, \emph{Phys. Rev. Lett.} 110 (2013) 012302

\bibitem{Molnar:2014zha}
E.~Molnar \emph{et~al.}, \emph{Phys. Rev.} C90 (2014) 044904

\bibitem{Denicol:2009am}
G.~Denicol \emph{et~al.}, \emph{Phys. Rev.} C80 (2009) 064901

\bibitem{Monnai:2009ad}
A.~Monnai and T.~Hirano, \emph{Phys. Rev.} C80 (2009) 054906

\bibitem{Noronha-Hostler:2013gga}
J.~Noronha-Hostler \emph{et~al.}, \emph{Phys. Rev.} C88 (2013) 044916

\bibitem{Noronha-Hostler:2014dqa}
J.~Noronha-Hostler, J.~Noronha and F.~Grassi, \emph{Phys. Rev.} C90 (2014)
  034907

\bibitem{Gardim:2014tya}
F.~G. Gardim \emph{et~al.}, \emph{Phys. Rev.} C91 (2015) 034902

\bibitem{Dusling:2011fd}
K.~Dusling and T.~{Sch\"{a}fer}, \emph{Phys. Rev.} C85 (2012) 044909

\bibitem{Bozek:2009dw}
P.~Bozek, \emph{Phys. Rev.} C81 (2010) 034909

\bibitem{Huovinen:2007xh}
P.~Huovinen, \emph{Eur. Phys. J.} A37 (2008) 121

\bibitem{Bondorf:1978kz}
J.~Bondorf, S.~Garpman and J.~Zim\'anyi), \emph{Nucl. Phys.} A296 (1978) 320

\bibitem{Navarra:1992ej}
F.~Navarra \emph{et~al.}, \emph{Phys. Rev.} C45 (1992) 2552

\bibitem{Schnedermann:1994gc}
E.~Schnedermann and U.~W. Heinz, \emph{Phys. Rev.} C50 (1994) 1675

\bibitem{Bondorf:2000pm}
J.~Bondorf \emph{et~al.}, \emph{Phys. Rev.} C65 (2002) 017601

\bibitem{Eskola:2007zc}
K.~Eskola, H.~Niemi and P.~Ruuskanen, \emph{Phys. Rev.} C77 (2008) 044907

\bibitem{Schafer:2009dj}
T.~{Sch\"{a}fer} and D.~Teaney, \emph{Rept. Prog. Phys.} 72 (2009) 126001

\bibitem{Holopainen:2013jna}
H.~Holopainen and P.~Huovinen, \emph{J. Phys. Conf. Ser.} 509 (2014) 012114

\bibitem{Niemi:2014wta}
H.~Niemi and G.~Denicol, \emph{ar{X}iv:1404.7327}  (2014)

\bibitem{Arbex:2001vx}
N.~Arbex \emph{et~al.}, \emph{Phys. Rev.} C64 (2001) 064906

\bibitem{Hirano:2002ds}
T.~Hirano and K.~Tsuda, \emph{Phys. Rev.} C66 (2002) 054905

\bibitem{Kisiel:2005hn}
A.~Kisiel \emph{et~al.}, \emph{Comput. Phys. Commun.} 174 (2006) 669

\bibitem{Rybczynski:2013yba}
M.~Rybczynski \emph{et~al.}, \emph{Comput. Phys. Commun.} 185 (2014) 1759

\bibitem{Shen:2014vra}
C.~Shen \emph{et~al.}, \emph{ar{X}iv:1409.8164}  (2014)

\bibitem{Grassi:1994nf}
F.~Grassi, Y.~Hama and T.~Kodama, \emph{Phys. Lett.} B355 (1995) 9

\bibitem{Grassi:2004dz}
F.~Grassi, \emph{Braz. J. Phys.} 35 (2005) 52

\bibitem{Akkelin:2004cp}
S.~Akkelin, M.~Borysova and Y.~Sinyukov, \emph{Acta Phys. Hung..} A22 (2005)
  165

\bibitem{Csernai:2005ht}
L.~Csernai \emph{et~al.}, \emph{Eur. Phys. J.} A25 (2005) 65

\bibitem{Akkelin:2008eh}
S.~Akkelin \emph{et~al.}, \emph{Phys. Rev.} C78 (2008) 034906

\bibitem{Sinyukov:2002if}
Y.~Sinyukov, S.~Akkelin and Y.~Hama, \emph{Phys. Rev. Lett.} 89 (2002) 052301

\bibitem{Karpenko:2009wf}
I.~Karpenko and Y.~Sinyukov, \emph{Phys. Lett.} B688 (2010) 50

\bibitem{Karpenko:2010te}
I.~Karpenko and Y.~Sinyukov, \emph{Phys. Rev.} C81 (2010) 054903

\bibitem{Shapoval:2014wya}
V.~Shapoval \emph{et~al.}, \emph{Nucl. Phys.} A929 (2014) 1

\bibitem{Shapoval:2013bga}
V.~Shapoval, Y.~M. Sinyukov and I.~A. Karpenko, \emph{Phys. Rev.} C88 (2013)
  064904

\bibitem{Karpenko:2012yf}
I.~Karpenko, Y.~Sinyukov and K.~Werner, \emph{Phys. Rev.} C87 (2013) 024914

\bibitem{Shapoval:2013jca}
V.~Shapoval \emph{et~al.}, \emph{Phys. Lett.} B725 (2013) 139

\bibitem{Sinyukov:2013zna}
Y.~Sinyukov \emph{et~al.}, \emph{Adv. High Energy Phys.} 2013 (2013) 198928

\bibitem{Noronha-Hostler:2013ria}
J.~Noronha-Hostler \emph{et~al.}, \emph{Phys. Rev.} C89 (2014) 054904

\bibitem{Ollitrault:1992bk}
J.-Y. Ollitrault, \emph{Phys. Rev.} D46 (1992) 229

\bibitem{Poskanzer:1998yz}
A.~M. Poskanzer and S.~Voloshin, \emph{Phys. Rev.} C58 (1998) 1671

\bibitem{Voloshin:2008dg}
S.~A. Voloshin, A.~M. Poskanzer and R.~Snellings, \emph{ar{X}iv:0809.2949}
  (2008)

\bibitem{Borghini:2001vi}
N.~Borghini, P.~M. Dinh and J.-Y. Ollitrault, \emph{Phys. Rev.} C64 (2001)
  054901

\bibitem{Bilandzic:2010jr}
A.~Bilandzic, R.~Snellings and S.~Voloshin, \emph{Phys. Rev.} C83 (2011) 044913

\bibitem{Qiu:2012uy}
Z.~Qiu and U.~Heinz, \emph{Phys. Lett.} B717 (2012) 261

\bibitem{Philos.Mag.45.1954.663}
R.~Hanbury-Brown and R.~Twiss, \emph{Philos. Mag.} 45 (1954) 663

\bibitem{Goldhaber:1960sf}
G.~Goldhaber \emph{et~al.}, \emph{Phys. Rev.} 120 (1960) 300

\bibitem{Baym:1997ce}
G.~Baym, \emph{Acta Phys. Polon.} B29 (1998) 1839

\bibitem{Wiedemann:1999qn}
U.~A. Wiedemann and U.~W. Heinz, \emph{Phys. Rept.} 319 (1999) 145

\bibitem{Weiner:1999th}
R.~Weiner, \emph{Phys. Rept.} 327 (2000) 249

\bibitem{Heinz:1999rw}
U.~W. Heinz and B.~V. Jacak, \emph{Ann. Rev. Nucl. Part. Sci.} 49 (1999) 529

\bibitem{Padula:2004ba}
S.~S. Padula, \emph{Braz. J. Phys.} 35 (2005) 70

\bibitem{Csorgo:2005gd}
T.~{Cs\"{o}rg\H{o}}, \emph{J. Phys. Conf. Ser.} 50 (2006) 259

\bibitem{Humanic:2005ye}
T.~J. Humanic, \emph{Int. J. Mod. Phys.} E15 (2006) 197

\bibitem{Lisa:2005dd}
M.~A. Lisa \emph{et~al.}, \emph{Ann. Rev. Nucl. Part. Sci.} 55 (2005) 357

\bibitem{Pratt:2008qv}
S.~Pratt, \emph{Phys. Rev. Lett.} 102 (2009) 232301

\bibitem{TeaneyQGP4}
D.~A. Teaney, Viscous Hydrodynamics and the Quark Gluon Plasma, chap.~4,
  207--266

\bibitem{BNLNews}
{RHIC Scientists Serve Up ``Perfect'' Liquid} (2005).
  http://www.bnl.gov/newsroom/news.php?a=1303

\bibitem{Hirano:2005wx}
T.~Hirano and M.~Gyulassy, \emph{Nucl. Phys.} A769 (2006) 71

\bibitem{Gardim:2011xv}
F.~G. Gardim \emph{et~al.}, \emph{Phys. Rev.} C85 (2012) 024908

\bibitem{Luzum:2013yya}
M.~Luzum and H.~Petersen, \emph{J. Phys.} G41 (2014) 063102

\bibitem{Huo:2013qma}
P.~Huo, J.~Jia and S.~Mohapatra, \emph{Phys. Rev.} C90 (2014) 024910

\bibitem{Gale:2013da}
C.~Gale, S.~Jeon and B.~Schenke, \emph{Int. J. Mod. Phys.} A28 (2013) 1340011

\bibitem{Huovinen:2013wma}
P.~Huovinen, \emph{Int. J. Mod. Phys.} E22 (2013) 1330029

\bibitem{Bucher:2015eia}
M.~Bucher, \emph{Int. J. Mod. Phys.} D24 (2015) 1530004

\bibitem{Wang:1991qh}
S.~Wang \emph{et~al.}, \emph{Phys. Rev.} C44 (1991) 1091

\bibitem{Adler:2002pu}
C.~Adler \emph{et~al.}, \emph{Phys. Rev.} C66 (2002) 034904

\bibitem{Bhalerao:2003xf}
R.~Bhalerao, N.~Borghini and J.~Ollitrault, \emph{Nucl. Phys.} A727 (2003) 373

\bibitem{Selyuzhenkov:2007zi}
I.~Selyuzhenkov and S.~Voloshin, \emph{Phys. Rev.} C77 (2008) 034904

\bibitem{Aad:2014vba}
G.~Aad \emph{et~al.}, \emph{Eur. Phys. J.} C74 (2014) 3157

\bibitem{Abelev:2007qg}
B.~Abelev \emph{et~al.}, \emph{Phys. Rev.} C75 (2007) 054906

\bibitem{Adare:2010ux}
A.~Adare \emph{et~al.}, \emph{Phys. Rev. Lett.} 105 (2010) 062301

\bibitem{Alver:2008zza}
B.~Alver \emph{et~al.}, \emph{Phys. Rev.} C77 (2008) 014906

\bibitem{Qiu:2011iv}
Z.~Qiu and U.~W. Heinz, \emph{Phys. Rev.} C84 (2011) 024911

\bibitem{ALICE:2011ab}
K.~Aamodt \emph{et~al.}, \emph{Phys. Rev. Lett.} 107 (2011) 032301

\bibitem{Adamczyk:2013waa}
L.~Adamczyk \emph{et~al.}, \emph{Phys. Rev.} C88 (2013) 014904

\bibitem{Adamczyk:2013gw}
L.~Adamczyk \emph{et~al.}, \emph{Phys. Rev.} C88 (2013) 014902

\bibitem{Abelev:2010tr}
B.~Abelev \emph{et~al.}, \emph{Phys. Rev.} C81 (2010) 044902

\bibitem{Agakishiev:2011eq}
G.~Agakishiev \emph{et~al.}, \emph{Phys. Rev.} C86 (2012) 014904

\bibitem{Adamczyk:2012ku}
L.~Adamczyk \emph{et~al.}, \emph{Phys. Rev.} C86 (2012) 054908

\bibitem{Adare:2011tg}
A.~Adare \emph{et~al.}, \emph{Phys. Rev. Lett.} 107 (2011) 252301

\bibitem{CMS:2013bza}
S.~Chatrchyan \emph{et~al.}, \emph{JHEP} 1402 (2014) 088

\bibitem{Chatrchyan:2013kba}
S.~Chatrchyan \emph{et~al.}, \emph{Phys. Rev.} C89 (2014) 044906

\bibitem{Chatrchyan:2012ta}
S.~Chatrchyan \emph{et~al.}, \emph{Phys. Rev.} C87 (2013) 014902

\bibitem{Chatrchyan:2012xq}
S.~Chatrchyan \emph{et~al.}, \emph{Phys. Rev. Lett.} 109 (2012) 022301

\bibitem{Chatrchyan:2012vqa}
S.~Chatrchyan \emph{et~al.}, \emph{Phys. Rev. Lett.} 110 (2013) 042301

\bibitem{Aad:2013xma}
G.~Aad \emph{et~al.}, \emph{JHEP} 1311 (2013) 183

\bibitem{ATLAS:2012at}
G.~Aad \emph{et~al.}, \emph{Phys. Rev.} C86 (2012) 014907

\bibitem{Aad:2014eoa}
G.~Aad \emph{et~al.}, \emph{Eur. Phys. J.} C74 (2014) 2982

\bibitem{Abelev:2012di}
B.~Abelev \emph{et~al.}, \emph{Phys. Lett.} B719 (2013) 18

\bibitem{Aamodt:2010pa}
K.~Aamodt \emph{et~al.}, \emph{Phys. Rev. Lett.} 105 (2010) 252302

\bibitem{Abelev:2014pua}
B.~B. Abelev \emph{et~al.}, \emph{ar{X}iv:1405.4632}  (2014)

\bibitem{Miller:2003kd}
M.~Miller and R.~Snellings, \emph{ar{X}iv:nucl-ex/0312008}  (2003)

\bibitem{Alver:2006wh}
B.~Alver \emph{et~al.}, \emph{Phys. Rev. Lett.} 98 (2007) 242302

\bibitem{Flensburg:2011wx}
C.~Flensburg, \emph{ar{X}iv:1108.4862}  (2011)

\bibitem{Ollitrault:2009ie}
J.-Y. Ollitrault, A.~M. Poskanzer and S.~A. Voloshin, \emph{Phys. Rev.} C80
  (2009) 014904

\bibitem{Alver:2010rt}
B.~Alver \emph{et~al.}, \emph{Phys. Rev.} C81 (2010) 034915

\bibitem{Voloshin:2007pc}
S.~A. Voloshin \emph{et~al.}, \emph{Phys. Lett.} B659 (2008) 537

\bibitem{Abelev:2014mda}
B.~B. Abelev \emph{et~al.}, \emph{Phys. Rev.} C90 (2014) 054901

\bibitem{Yan:2013laa}
L.~Yan and J.-Y. Ollitrault, \emph{Phys. Rev. Lett.} 112 (2014) 082301

\bibitem{Alver:2007qw}
B.~Alver \emph{et~al.}, \emph{Phys. Rev. Lett.} 104 (2010) 142301

\bibitem{DerradideSouza:2011rp}
R.~Derradi~de Souza \emph{et~al.}, \emph{Phys. Rev.} C85 (2012) 054909

\bibitem{Collaboration:2011yba}
R.~Snellings, \emph{J. Phys.} G38 (2011) 124013

\bibitem{Miller:2007ri}
M.~L. Miller \emph{et~al.}, \emph{Ann. Rev. Nucl. Part. Sci.} 57 (2007) 205

\bibitem{Drescher:2006pi}
H.-J. Drescher \emph{et~al.}, \emph{Phys. Rev.} C74 (2006) 044905

\bibitem{Gardim:2012im}
F.~G. Gardim \emph{et~al.}, \emph{Phys. Rev.} C87 (2013) 031901

\bibitem{Heinz:2013bua}
U.~Heinz, Z.~Qiu and C.~Shen, \emph{Phys. Rev.} C87 (2013) 034913

\bibitem{Zhou:2014bba}
Y.~Zhou, \emph{Nucl. Phys.} A931 (2014) 949

\bibitem{Borghini:2000sa}
N.~Borghini, P.~M. Dinh and J.-Y. Ollitrault, \emph{Phys. Rev.} C63 (2001)
  054906

\bibitem{Bhalerao:2014mua}
R.~S. Bhalerao \emph{et~al.}, \emph{Phys. Rev. Lett.} 114 (2015) 152301

\bibitem{Mazeliauskas:2015vea}
A.~Mazeliauskas and D.~Teaney, \emph{Phys. Rev.} C91 (2015) 044902

\bibitem{Aamodt:2011by}
K.~Aamodt \emph{et~al.}, \emph{Phys. Lett.} B708 (2012) 249

\bibitem{Abelev:2009af}
B.~Abelev \emph{et~al.}, \emph{Phys. Rev.} C80 (2009) 064912

\bibitem{Alver:2009id}
B.~Alver \emph{et~al.}, \emph{Phys. Rev. Lett.} 104 (2010) 062301

\bibitem{CasalderreySolana:2004qm}
J.~Casalderrey-Solana, E.~Shuryak and D.~Teaney, \emph{J. Phys. Conf. Ser.} 27
  (2005) 22

\bibitem{Armesto:2004pt}
N.~Armesto, C.~A. Salgado and U.~A. Wiedemann, \emph{Phys. Rev. Lett.} 93
  (2004) 242301

\bibitem{Shuryak:2007fu}
E.~Shuryak, \emph{Phys. Rev.} C76 (2007) 047901

\bibitem{Stoecker:2004qu}
H.~Stoecker, \emph{Nucl. Phys.} A750 (2005) 121

\bibitem{Dumitru:2008wn}
A.~Dumitru \emph{et~al.}, \emph{Nucl. Phys.} A810 (2008) 91

\bibitem{Wong:2008yh}
C.-Y. Wong, \emph{Phys. Rev.} C78 (2008) 064905

\bibitem{Takahashi:2009na}
J.~Takahashi \emph{et~al.}, \emph{Phys. Rev. Lett.} 103 (2009) 242301

\bibitem{Luzum:2010sp}
M.~Luzum, \emph{Phys. Lett.} B696 (2011) 499

\bibitem{Teaney:2010vd}
D.~Teaney and L.~Yan, \emph{Phys. Rev.} C83 (2011) 064904

\bibitem{Sorensen:2010zq}
P.~Sorensen, \emph{J. Phys.} G37 (2010) 094011

\bibitem{Adams:2003im}
J.~Adams \emph{et~al.}, \emph{Phys. Rev. Lett.} 91 (2003) 072304

\bibitem{CMS:2012qk}
S.~Chatrchyan \emph{et~al.}, \emph{Phys. Lett.} B718 (2013) 795

\bibitem{Aad:2012gla}
G.~Aad \emph{et~al.}, \emph{Phys. Rev. Lett.} 110 (2013) 182302

\bibitem{ABELEV:2013wsa}
B.~B. Abelev \emph{et~al.}, \emph{Phys. Lett.} B726 (2013) 164

\bibitem{Khachatryan:2010gv}
V.~Khachatryan \emph{et~al.}, \emph{JHEP} 1009 (2010) 091

\bibitem{ATLAS:2012ap}
G.~Aad \emph{et~al.}, \emph{JHEP} 1205 (2012) 157

\bibitem{ATLAS:2012as}
G.~Aad \emph{et~al.}, \emph{JHEP} 1207 (2012) 019

\bibitem{Aad:2014lta}
G.~Aad \emph{et~al.}, \emph{Phys. Rev.} C90 (2014) 044906

\bibitem{Basar:2013hea}
G.~{Ba\c{s}ar} and D.~Teaney, \emph{Phys. Rev.} C90 (2014) 054903

\bibitem{Adare:2014keg}
A.~Adare \emph{et~al.}, \emph{Phys. Rev. Lett.} 114 (2015) 192301

\bibitem{Khachatryan:2014jra}
V.~Khachatryan \emph{et~al.}, \emph{Phys. Lett.} B742 (2015) 200

\bibitem{Adams:2004bi}
J.~Adams \emph{et~al.}, \emph{Phys. Rev.} C72 (2005) 014904

\bibitem{Adare:2006ti}
A.~Adare \emph{et~al.}, \emph{Phys. Rev. Lett.} 98 (2007) 162301

\bibitem{Shuryak:2013ke}
E.~Shuryak and I.~Zahed, \emph{Phys. Rev.} C88 (2013) 044915

\bibitem{Conesa:2007gr}
G.~Conesa \emph{et~al.}, \emph{Nucl. Phys.} A782 (2007) 356

\bibitem{Wilde:2012wc}
M.~Wilde, \emph{Nucl. Phys.} A904-905 (2013) 573c

\bibitem{Shen:2013vja}
C.~Shen \emph{et~al.}, \emph{Phys.Rev.} C89 (2014) 044910

\bibitem{Adare:2014fwh}
A.~Adare \emph{et~al.}, \emph{ar{X}iv:1405.3940}  (2014)

\bibitem{Adare:2008ab}
A.~Adare \emph{et~al.}, \emph{Phys. Rev. Lett.} 104 (2010) 132301

\bibitem{Lohner:2012ct}
D.~Lohner, \emph{J. Phys. Conf. Ser.} 446 (2013) 012028

\bibitem{Adare:2011zr}
A.~Adare \emph{et~al.}, \emph{Phys. Rev. Lett.} 109 (2012) 122302

\bibitem{Gale:2012xq}
C.~Gale, \emph{Nucl. Phys.} A910-911 (2013) 147

\bibitem{Chatterjee:2013naa}
R.~Chatterjee \emph{et~al.}, \emph{Phys. Rev.} C88 (2013) 034901

\bibitem{Linnyk:2013hta}
O.~Linnyk \emph{et~al.}, \emph{Phys. Rev.} C88 (2013) 034904

\bibitem{Dion:2011pp}
M.~Dion \emph{et~al.}, \emph{Phys. Rev.} C84 (2011) 064901

\bibitem{Shen:2015qba}
C.~Shen \emph{et~al.}, \emph{ar{X}iv:1504.07989}  (2015)

\bibitem{Nature178.1046-1048}
R.~Hanbury~Brown and R.~Twiss, \emph{Nature} 178 (1956) 1046

\bibitem{Adams:2004yc}
J.~Adams \emph{et~al.}, \emph{Phys. Rev.} C71 (2005) 044906

\bibitem{Abelev:2009tp}
B.~Abelev \emph{et~al.}, \emph{Phys. Rev.} C80 (2009) 024905

\bibitem{Abelev:2013pqa}
B.~B. Abelev \emph{et~al.}, \emph{Phys. Rev.} C89 (2014) 024911

\bibitem{Aamodt:2011mr}
K.~Aamodt \emph{et~al.}, \emph{Phys. Lett.} B696 (2011) 328

\bibitem{Adamova:2002ff}
D.~Adamova \emph{et~al.}, \emph{Phys. Rev. Lett.} 90 (2003) 022301

\bibitem{Abelev:2014pja}
B.~B. Abelev \emph{et~al.}, \emph{Phys. Lett.} B739 (2014) 139

\bibitem{Kozlov:2014hya}
I.~Kozlov \emph{et~al.}, \emph{Nucl. Phys.} A931 (2014) 1045

\bibitem{Kozlov:2014fqa}
I.~Kozlov \emph{et~al.}, \emph{ar{X}iv:1405.3976}  (2014)

\bibitem{Werner:2013ipa}
K.~Werner \emph{et~al.}, \emph{Phys.Rev.Lett.} 112 (2014) 232301

\bibitem{Pratt:2015zsa}
S.~Pratt \emph{et~al.}, \emph{Phys. Rev. Lett.} 114 (2015) 202301

\bibitem{Geiger:1992si}
K.~Geiger, \emph{Phys. Rev.} D46 (1992) 4965

\bibitem{Baier:2000sb}
R.~Baier \emph{et~al.}, \emph{Phys. Lett.} B502 (2001) 51

\bibitem{Kharzeev:2005qg}
D.~Kharzeev, \emph{Nucl. Phys.} A774 (2006) 315

\bibitem{Mueller:2005un}
A.~Mueller, A.~Shoshi and S.~Wong, \emph{Phys. Lett.} B632 (2006) 257

\bibitem{Epelbaum:2014xea}
T.~Epelbaum and F.~Gelis, \emph{Nucl. Phys.} A926 (2014) 122

\bibitem{Berges:2013fga}
J.~Berges \emph{et~al.}, \emph{Phys. Rev.} D89 (2014) 114007

\bibitem{Noronha:2015jia}
J.~Noronha and G.~S. Denicol, \emph{ar{X}iv:1502.05892}  (2015)

\bibitem{Rep.Prog.Phys.62.1035}
D.~Jou, J.~Casas-V\'{a}zquez and G.~Lebon, \emph{Reports on Progress in
  Physics} 62 (1999) 1035

\bibitem{Ciancio1991}
V.~Ciancio and J.~Verh{\'{a}}s, \emph{Journal of Non-Equilibrium
  Thermodynamics} 16 (1991)

\bibitem{Geroch:1990bw}
R.~P. Geroch and L.~Lindblom, \emph{Phys. Rev.} D41 (1990) 1855

\bibitem{Heinson:1995xw}
A.~Heinson, A.~Belyaev and E.~Boos, \emph{ar{X}iv:hep-ph/9509274}  (1995)

\bibitem{Calzetta:1997aj}
E.~Calzetta, \emph{Class.Quant.Grav.} 15 (1998) 653

\bibitem{Grmela:1997zz}
M.~Grmela and H.~C. Ottinger, \emph{Phys. Rev.} E56 (1997) 6620

\bibitem{Ottinger2005}
H.~C. \"{O}ttinger, Beyond Equilibrium Thermodynamics  (Wiley-Interscience
  2005)

\bibitem{Denicol:2014loa}
G.~Denicol, \emph{J. Phys.} G41 (2014) 124004

\bibitem{Cercignani2002}
C.~Cercignani and G.~M. Kremer, The Relativistic Boltzmann Equation: Theory and
  Applications  (Birkh\"{a}user Basel 2002)

\bibitem{Reichl2009}
L.~E. Reichl, A Modern Course in Statistical Physics  (Wiley-VCH 2009)

\bibitem{Liboff2003}
R.~Liboff, Kinetic Theory: Classical, Quantum, and Relativistic Descriptions
  (Graduate Texts in Contemporary Physics)  (Springer 2003)

\bibitem{chapman1970mathematical}
S.~Chapman and T.~Cowling, The Mathematical Theory of Non-uniform Gases: An
  Account of the Kinetic Theory of Viscosity, Thermal Conduction and Diffusion
  in Gases, Cambridge Mathematical Library  (Cambridge University Press 1970)

\bibitem{Phys.Rev.Lett.18.988}
B.~J. Alder and T.~E. Wainwright, \emph{Phys. Rev. Lett.} 18 (1967) 988

\bibitem{cond-mat/9707146}
M.~H. {Ernst}, \emph{eprint arXiv:cond-mat/9707146}  (1997)

\bibitem{Dorfman198177}
J.~Dorfman, \emph{Physica A: Statistical Mechanics and its Applications} 106
  (1981) 77

\bibitem{vanHove1954517}
L.~van Hove, \emph{Physica} 21 (1954) 517

\bibitem{DeMasi1984}
A.~D. Masi \emph{et~al.}, Studies in Statistical Mechanics: Non-equilibrium
  phenomena II, from Stochastics to Hydrodynamics, vol.~II  (North-Holland,
  Amsterdan 1984)

\bibitem{Tsumura:2012kp}
K.~Tsumura and T.~Kunihiro, \emph{Eur. Phys. J.} A48 (2012) 162

\bibitem{Grad1949}
H.~Grad, \emph{Communications on Pure and Applied Mathematics} 2 (1949) 331

\bibitem{Denicol:2012cn}
G.~Denicol \emph{et~al.}, \emph{Phys. Rev.} D85 (2012) 114047

\bibitem{Hoover1991}
W.~G. Hoover, Computational Statistical Mechanics  (Elsevier Science 1991)

\bibitem{Garcia2008149}
L.~Garc\'{i}a-Col\'{i}n, R.~Velasco and F.~Uribe, \emph{Physics Reports} 465
  (2008) 149

\bibitem{Denicol:2010xn}
G.~Denicol, T.~Koide and D.~Rischke, \emph{Phys. Rev. Lett.} 105 (2010) 162501

\bibitem{Jaiswal:2013fc}
A.~Jaiswal, R.~S. Bhalerao and S.~Pal, \emph{Phys. Rev.} C87 (2013) 021901

\bibitem{Zwanzig2001}
R.~Zwanzig, Nonequilibrium Statistical Mechanics  (Oxford University Press
  2001)

\bibitem{Breuer2002}
H.-P. Breuer and F.~Petruccione, The Theory of Open Quantum Systems  (Oxford
  University Press 2002)

\bibitem{Kubo1983}
M.~Toda \emph{et~al.}, Statistical physics, no. v.2 in Springer series in
  solid-state sciences  (Springer-Verlag 1983)

\bibitem{Koide:2008nw}
T.~Koide and T.~Kodama, \emph{Phys. Rev.} E78 (2008) 051107

\bibitem{Koide:2007ja}
T.~Koide, \emph{Phys. Rev.} E75 (2007) 060103

\bibitem{Huang:2010sa}
X.-G. Huang \emph{et~al.}, \emph{Phys. Rev.} C83 (2011) 024906

\bibitem{Huang:2011ez}
X.-G. Huang and T.~Koide, \emph{Nucl. Phys.} A889 (2012) 73

\bibitem{Koide:2005qb}
T.~Koide, \emph{Phys. Rev.} E72 (2005) 026135

\bibitem{Kapusta:2011gt}
J.~I. Kapusta, B.~Muller and M.~Stephanov, \emph{Phys. Rev.} C85 (2012) 054906

\bibitem{fox1978gaussian}
R.~F. Fox, \emph{Physics Reports} 48 (1978) 179

\bibitem{Fick2012}
E.~Fick and G.~Sauermann, The Quantum Statistics of Dynamic Processes (Springer
  Series in Solid-State Sciences)  (Springer 2012)

\bibitem{Minami:2012hs}
Y.~Minami and Y.~Hidaka, \emph{Phys. Rev.} E87 (2013) 023007

\bibitem{Koide:2009sy}
T.~Koide, E.~Nakano and T.~Kodama, \emph{Phys. Rev. Lett.} 103 (2009) 052301

\bibitem{Ann.Phys.154.229}
A.~Hosoya, M.~aki Sakagami and M.~Takao, \emph{Annals of Physics} 154 (1984)
  229

\bibitem{Denicol:2010br}
G.~S. Denicol \emph{et~al.}, \emph{Phys. Lett.} B708 (2012) 174

\bibitem{Denicol:2011fa}
G.~S. Denicol \emph{et~al.}, \emph{Phys. Rev.} D83 (2011) 074019

\bibitem{Zwanzig1965}
R.~Zwanzig, \emph{Annu. Rev. Phys. Chem.} 16 (1965) 67

\bibitem{Koide:2004yn}
T.~Koide and M.~Maruyama, \emph{Nucl. Phys.} A742 (2004) 95

\bibitem{Baier:2007ix}
R.~Baier \emph{et~al.}, \emph{JHEP} 0804 (2008) 100

\bibitem{Finazzo:2014cna}
S.~I. Finazzo \emph{et~al.}, \emph{JHEP} 1502 (2015) 051

\bibitem{Bhattacharyya:2008jc}
S.~Bhattacharyya \emph{et~al.}, \emph{JHEP} 0802 (2008) 045

\bibitem{Policastro:2001yc}
G.~Policastro, D.~T. Son and A.~O. Starinets, \emph{Phys. Rev. Lett.} 87 (2001)
  081601

\bibitem{Kanitscheider:2009as}
I.~Kanitscheider and K.~Skenderis, \emph{JHEP} 0904 (2009) 062

\bibitem{Buchel:2007mf}
A.~Buchel, \emph{Phys. Lett.} B663 (2008) 286

\bibitem{Bigazzi:2010ku}
F.~Bigazzi and A.~L. Cotrone, \emph{JHEP} 1008 (2010) 128

\bibitem{Taub:1954zz}
A.~Taub, \emph{Phys. Rev.} 94 (1954) 1468

\bibitem{Andersson:2006nr}
N.~Andersson and G.~Comer, \emph{Living Rev.Rel.} 10 (2007) 1

\bibitem{Mota:2012qv}
P.~Mota \emph{et~al.}, \emph{Eur. Phys. J.} A48 (2012) 165

\bibitem{Elze:1999kc}
H.-T. Elze \emph{et~al.}, \emph{J. Phys.} G25 (1999) 1935

\bibitem{Elsas:2014kua}
J.~Gaspar~Elsas, T.~Koide and T.~Kodama, \emph{Braz. J. Phys.} 45 (2015) 334

\bibitem{Myers:1977zz}
W.~Myers \emph{et~al.}, \emph{Phys. Rev.} C15 (1977) 2032

\bibitem{Lucy:1977zz}
L.~Lucy, \emph{Astron.J.} 82 (1977) 1013

\bibitem{Kitsionas:2002sp}
S.~Kitsionas and A.~Whitworth, \emph{Mon.Not.Roy.Astron.Soc.} 330 (2002) 129

\bibitem{Aguiar:2003pp}
C.~Aguiar, E.~Fraga and T.~Kodama, \emph{J. Phys.} G32 (2006) 179

\bibitem{Csernai:1995zn}
L.~Csernai and I.~Mishustin, \emph{Phys. Rev. Lett.} 74 (1995) 5005

\bibitem{Mishustin:1998yc}
I.~Mishustin and O.~Scavenius, \emph{Phys. Rev. Lett.} 83 (1999) 3134

\bibitem{Scavenius:1999zc}
O.~Scavenius and A.~Dumitru, \emph{Phys. Rev. Lett.} 83 (1999) 4697

\bibitem{Scavenius:2000bb}
O.~Scavenius \emph{et~al.}, \emph{Phys. Rev.} D63 (2001) 116003

\bibitem{Paech:2003fe}
K.~Paech, H.~Stoecker and A.~Dumitru, \emph{Phys. Rev.} C68 (2003) 044907

\bibitem{Paech:2005cx}
K.~Paech and A.~Dumitru, \emph{Phys. Lett.} B623 (2005) 200

\bibitem{Nahrgang:2011vn}
M.~Nahrgang \emph{et~al.}, \emph{J. Phys.} G40 (2013) 055108

\bibitem{Nahrgang:2011mg}
M.~Nahrgang \emph{et~al.}, \emph{Phys. Rev.} C84 (2011) 024912

\bibitem{Herold:2013qda}
C.~Herold \emph{et~al.}, \emph{Nucl. Phys.} A925 (2014) 14

\bibitem{Jiang:2014uxa}
B.-F. Jiang, D.-F. Hou and J.-R. Li, \emph{Int. J. Mod. Phys.Conf.Ser.} 29
  (2014) 1460217

\bibitem{Denicol:2010tr}
G.~Denicol, T.~Kodama and T.~Koide, \emph{J. Phys.} G37 (2010) 094040

\bibitem{Noronha-Hostler:2013hsa}
J.~Noronha-Hostler \emph{et~al.}, \emph{J. Phys. Conf. Ser.} 458 (2013) 012018

\bibitem{Yasue1981327}
K.~Yasue, \emph{Journal of Functional Analysis} 41 (1981) 327

\bibitem{Koide:2011sa}
T.~Koide and T.~Kodama, \emph{J. Phys.} A45 (2012) 255204

\bibitem{Shuryak:1972zq}
E.~V. Shuryak, \emph{Yad.Fiz.} 16 (1972) 395

\bibitem{Carruthers1974}
P.~{Carruthers}, \emph{Annals of the New York Academy of Sciences} 229 (1974)
  91

\bibitem{Carruthers:1973ws}
P.~Carruthers and M.~Doung-van, \emph{Phys. Rev.} D8 (1973) 859

\bibitem{Denicol:2008zz}
G.~Denicol \emph{et~al.}, \emph{J. Phys.} G35 (2008) 104130

\bibitem{Wong:2008ex}
C.-Y. Wong, \emph{Phys. Rev.} C78 (2008) 054902

\bibitem{Steinberg:2004vy}
P.~Steinberg, \emph{Acta Phys. Hung..} A24 (2005) 51

\bibitem{Netrakanti:2005iy}
P.~K. Netrakanti and B.~Mohanty, \emph{Phys. Rev.} C71 (2005) 047901

\bibitem{Petersen:2006mp}
H.~Petersen and M.~Bleicher, \emph{PoS} CPOD2006 (2006) 025

\bibitem{Wong:2014sda}
C.-Y. Wong \emph{et~al.}, \emph{Phys. Rev.} C90 (2014) 064907

\bibitem{Sarkisyan:2015gca}
E.~K.~G. Sarkisyan \emph{et~al.}, \emph{ar{X}iv:1506.09080}  (2015)

\bibitem{Hwa:1974gn}
R.~C. Hwa, \emph{Phys. Rev.} D10 (1974) 2260

\bibitem{Chiu:1975hx}
C.~B. Chiu and K.-H. Wang, \emph{Phys. Rev.} D12 (1975) 272

\bibitem{Chiu:1975hw}
C.~B. Chiu, E.~Sudarshan and K.-H. Wang, \emph{Phys. Rev.} D12 (1975) 902

\bibitem{Sen:2014pfa}
A.~Sen \emph{et~al.}, \emph{Phys. Rev.} C91 (2015) 024901

\bibitem{Bialas:2007iu}
A.~Bialas, R.~Janik and R.~B. Peschanski, \emph{Phys. Rev.} C76 (2007) 054901

\bibitem{Csorgo:2006ax}
T.~{Cs\"{o}rg\H{o}}, M.~Nagy and M.~Csan\'ad, \emph{Phys. Lett.} B663 (2008)
  306

\bibitem{Csorgo:2007ea}
T.~{Cs\"{o}rg\H{o}}, M.~Nagy and M.~Csan\'ad, \emph{Braz. J. Phys.} 37 (2007)
  723

\bibitem{Nagy:2007xn}
M.~Nagy, T.~{Cs\"{o}rg\H{o}} and M.~Csan\'ad, \emph{Phys. Rev.} C77 (2008)
  024908

\bibitem{Csanad:2007iv}
M.~Csan\'ad, M.~Nagy and T.~{Cs\"{o}rg\H{o}}, \emph{Eur. Phys. J.ST} 155 (2008)
  19

\bibitem{Biro:1999eh}
T.~S. Bir\'{o}, \emph{Phys. Lett.} B474 (2000) 21

\bibitem{Biro:2000nj}
T.~S. Bir\'{o}, \emph{Phys. Lett.} B487 (2000) 133

\bibitem{Csorgo:2003rt}
T.~{Cs\"{o}rg\H{o}} \emph{et~al.}, \emph{Phys. Lett.} B565 (2003) 107

\bibitem{Csorgo:2003ry}
T.~{Cs\"{o}rg\H{o}} \emph{et~al.}, \emph{Heavy Ion Phys.} A21 (2004) 73

\bibitem{Pratt:2008jj}
S.~Pratt, \emph{Phys. Rev.} C75 (2007) 024907

\bibitem{Sinyukov:2004am}
Y.~Sinyukov and I.~Karpenko, \emph{Acta Phys. Hung..} A25 (2006) 141

\bibitem{Peschanski:2009tg}
R.~Peschanski and E.~N. Saridakis, \emph{Phys. Rev.} C80 (2009) 024907

\bibitem{Lin:2009kv}
S.~Lin and J.~Liao, \emph{Nucl. Phys.} A837 (2010) 195

\bibitem{Nucl.Phys.A305.226}
J.~N. {De} \emph{et~al.}, \emph{Nuclear Physics A} 305 (1978) 226

\bibitem{Csorgo:1994fg}
T.~{Cs\"{o}rg\H{o}}, B.~{L\"{o}rstad} and J.~Zim\'anyi), \emph{Phys. Lett.}
  B338 (1994) 134

\bibitem{Helgesson:1997zz}
J.~Helgesson \emph{et~al.}, \emph{Phys. Rev.} C56 (1997) 2626

\bibitem{Csizmadia:1998ef}
P.~Csizmadia, T.~{Cs\"{o}rg\H{o}} and B.~Luk\'{a}cs, \emph{Phys. Lett.} B443
  (1998) 21

\bibitem{Csorgo:1998yk}
T.~{Cs\"{o}rg\H{o}}, \emph{Central Eur. J. Phys.} 2 (2004) 556

\bibitem{Akkelin:2000ex}
S.~Akkelin \emph{et~al.}, \emph{Phys. Lett.} B505 (2001) 64

\bibitem{Csorgo:2001xm}
T.~{Cs\"{o}rg\H{o}} \emph{et~al.}, \emph{Phys. Rev.} C67 (2003) 034904

\bibitem{Csorgo:2001ru}
T.~{Cs\"{o}rg\H{o}}, \emph{Acta Phys. Polon.} B37 (2006) 483

\bibitem{Csorgo:1995bi}
T.~{Cs\"{o}rg\H{o}} and B.~{L\"{o}rstad}, \emph{Phys. Rev.} C54 (1996) 1390

\bibitem{Csorgo:1995vf}
T.~{Cs\"{o}rg\H{o}} and B.~{L\"{o}rstad}, \emph{Nucl. Phys.} A590 (1995) 465C

\bibitem{Nagy:2009eq}
M.~Nagy, \emph{Phys. Rev.} C83 (2011) 054901

\bibitem{Csorgo:2013ksa}
T.~{Cs\"{o}rg\H{o}} and M.~Nagy, \emph{Phys. Rev.} C89 (2014) 044901

\bibitem{Csanad:2005gv}
M.~Csan\'ad \emph{et~al.}, \emph{Eur. Phys. J.} A38 (2008) 363

\bibitem{Csanad:2012hr}
M.~Csan\'{a}d, M.~Nagy and S.~{L\"{o}k\"{o}s}, \emph{Eur. Phys. J.} A48 (2012)
  173

\bibitem{Csanad:2014dpa}
M.~Csan\'ad and A.~Szabo, \emph{Phys. Rev.} C90 (2014) 054911

\bibitem{Gubser:2010ze}
S.~S. Gubser, \emph{Phys. Rev.} D82 (2010) 085027

\bibitem{Gubser:2010ui}
S.~S. Gubser and A.~Yarom, \emph{Nucl. Phys.} B846 (2011) 469

\bibitem{Marrochio:2013wla}
H.~Marrochio \emph{et~al.}, \emph{Phys. Rev.} C91 (2015) 014903

\bibitem{Hatta:2014gqa}
Y.~Hatta, J.~Noronha and B.-W. Xiao, \emph{Phys. Rev.} D89 (2014) 051702

\bibitem{Hatta:2014gga}
Y.~Hatta, J.~Noronha and B.-W. Xiao, \emph{Phys. Rev.} D89 (2014) 114011

\bibitem{Hatta:2014jva}
Y.~Hatta \emph{et~al.}, \emph{Phys. Rev.} D90 (2014) 074026

\bibitem{Pang:2014ipa}
L.-G. Pang \emph{et~al.}, \emph{Phys. Rev.} D91 (2015) 074027

\bibitem{Heinz:2002rs}
U.~W. Heinz and S.~M.~H. Wong, \emph{Phys. Rev.} C66 (2002) 014907

\bibitem{Bialas:2007gn}
A.~Bialas, M.~Chojnacki and W.~Florkowski, \emph{Phys. Lett.} B661 (2008) 325

\bibitem{Ryblewski:2010tn}
R.~Ryblewski and W.~Florkowski, \emph{Phys. Rev.} C82 (2010) 024903

\bibitem{Martinez:2010sc}
M.~Martinez and M.~Strickland, \emph{Nucl. Phys.} A848 (2010) 183

\bibitem{Florkowski:2010cf}
W.~Florkowski and R.~Ryblewski, \emph{Phys. Rev.} C83 (2011) 034907

\bibitem{Strickland:2014pga}
M.~Strickland, \emph{Acta Phys. Polon.} B45 (2014) 2355

\bibitem{Strickland:2014eua}
M.~Strickland, \emph{Nucl. Phys.} A926 (2014) 92

\bibitem{Jeon:2015dfa}
S.~Jeon and U.~Heinz, \emph{ar{X}iv:1503.03931}  (2015)

\bibitem{Mishustin:1988mj}
I.~N. Mishustin, V.~N. Russkikh and L.~M. Satarov, \emph{Sov. J. Nucl. Phys.}
  48 (1988) 454. [Yad. Fiz.48,711(1988)]

\bibitem{Brachmann:1997bq}
J.~Brachmann \emph{et~al.}, \emph{Nucl. Phys.} A619 (1997) 391

\bibitem{Ivanov:2014ioa}
{\relax Yu}.~B. Ivanov and A.~A. Soldatov, \emph{Phys. Rev.} C91 (2015) 024915

\bibitem{Abelev:2006cs}
B.~Abelev \emph{et~al.}, \emph{Phys. Rev.} C75 (2007) 064901

\bibitem{Adare:2011vy}
A.~Adare \emph{et~al.}, \emph{Phys. Rev.} C83 (2011) 064903

\bibitem{Aamodt:2011zj}
K.~Aamodt \emph{et~al.}, \emph{Eur. Phys. J.} C71 (2011) 1655

\bibitem{Khachatryan:2011tm}
V.~Khachatryan \emph{et~al.}, \emph{JHEP} 1105 (2011) 064

\bibitem{Aad:2010ac}
G.~Aad \emph{et~al.}, \emph{New J. Phys.} 13 (2011) 053033

\bibitem{Sogaard:2015tna}
C.~Søgaard and K.~Gulbrandsen, \emph{EPJ Web Conf.} 90 (2015) 02004

\bibitem{GellMannTsallis}
M.~Gell-Mann and C.~Tsallis (Eds.) Nonextensive Entropy : Interdisciplinary
  Applications, Santa Fe Institute Studies on the Sciences of Complexity
  (Oxford University Press, USA 2004)

\bibitem{Osada:2008cn}
T.~Osada and G.~Wilk, \emph{Central Eur. J. Phys.} 7 (2009) 432

\bibitem{Biro:2012ix}
T.~{Bir\'{o}} and E.~{Moln\'{a}r}, \emph{Eur. Phys. J.} A48 (2012) 172

\bibitem{Osada:2009cc}
T.~Osada, \emph{Phys. Rev.} C81 (2010) 024907

\bibitem{Osada:2011gx}
T.~Osada, \emph{Phys. Rev.} C85 (2012) 014906

\bibitem{Csernai:2011qq}
L.~Csernai, D.~Strottman and C.~Anderlik, \emph{Phys. Rev.} C85 (2012) 054901

\bibitem{Becattini:2011zz}
F.~Becattini, \emph{Phys. Part. Nucl. Lett..} 8 (2011) 801

\bibitem{Csernai:2014nva}
L.~Csernai, F.~Becattini and D.~Wang, \emph{J. Phys. Conf. Ser.} 509 (2014)
  012054

\bibitem{Florkowski:2013lya}
W.~Florkowski, R.~Ryblewski and M.~Strickland, \emph{Phys. Rev.} C88 (2013)
  024903

\bibitem{Denicol:2014xca}
G.~S. Denicol \emph{et~al.}, \emph{Phys. Rev. Lett.} 113 (2014) 202301

\bibitem{Denicol:2014tha}
G.~S. Denicol \emph{et~al.}, \emph{Phys. Rev.} D90 (2014) 125026

\bibitem{Hatta:2015kia}
Y.~Hatta, M.~Martinez and B.-W. Xiao, \emph{Phys. Rev.} D91 (2015) 085024

\bibitem{Berges:2012ks}
J.~Berges, J.-P. Blaizot and F.~Gelis, \emph{J. Phys.} G39 (2012) 085115

\bibitem{Peshier:2005pp}
A.~Peshier and W.~Cassing, \emph{Phys. Rev. Lett.} 94 (2005) 172301

\bibitem{Cassing:2007yg}
W.~Cassing, \emph{Nucl. Phys.} A791 (2007) 365

\bibitem{Cassing:2007nb}
W.~Cassing, \emph{Nucl. Phys.} A795 (2007) 70

\bibitem{Kadanoff1962}
L.~Kadanoff and G.~Baym, Quantum statistical mechanics: Green's function
  methods in equilibrium and nonequilibrium problems, Frontiers in physics
  (W.A. Benjamin 1962)

\bibitem{IWoC2014}
R.~Derradi~de Souza \emph{et~al.}, Evaluating the role of coarse graining in
  hydrodynamic modeling of heavy-ion collisions from microscopic dynamics
  (2014). {T}o appear in the Proceedings of the International Workshop on
  Collectivity in Relativistic Heavy Ion Collisions.

\bibitem{Gardim:2012dc}
F.~G. Gardim \emph{et~al.}, \emph{Nucl. Phys.} A904-905 (2013) 503c

\bibitem{Niemi:2012aj}
H.~Niemi \emph{et~al.}, \emph{Phys. Rev.} C87 (2013) 054901

\bibitem{Torrieri:2012wz}
G.~Torrieri, B.~Betz and M.~Gyulassy, \emph{ar{X}iv:1208.5996}  (2012)

\bibitem{Yan:2015jma}
L.~Yan and J.-Y. Ollitrault, \emph{Phys. Lett.} B744 (2015) 82

\bibitem{Betz:2010qh}
B.~Betz \emph{et~al.}, \emph{Phys. Rev. Lett.} 105 (2010) 222301

\bibitem{Mota:2010zz}
P.~Mota and T.~Kodama, \emph{J. Phys.} G37 (2010) 094032

\bibitem{Bouras:2010nt}
I.~Bouras \emph{et~al.}, \emph{J. Phys. Conf. Ser.} 230 (2010) 012045

\bibitem{Steinberg:2007iv}
P.~Steinberg, \emph{PoS} CPOD2006 (2006) 036

\end{thebibliography}
\end{document}